\documentclass[10pt,letterpaper]{article}
\usepackage{epsfig,rotating,setspace,latexsym,amsmath,epsf,amssymb,bm}
\usepackage{cite,graphicx,authblk,color,subfigure,adjustbox,epstopdf}
\title{Retroactive Anti-Jamming for MISO Broadcast Channels\thanks{E-mail: adhiraj@vt.edu, tandonr@vt.edu, rbuehrer@vt.edu, tcc@vt.edu. Parts of this paper will be presented at Asilomar-2013 and at IEEE Globecom 2013.}}
\vspace{3cm}
\author[]{SaiDhiraj Amuru}
\author[]{Ravi Tandon$^\dagger$}
\author[]{R. Michael Buehrer}
\author[]{T. Charles Clancy$^\dagger$}
\affil[]{\normalsize Bradley Department of Electrical and Computer Engineering,\\
$^\dagger$ Hume Center for National Security and Technology,\\
Virginia Tech, Blacksburg, VA, USA.}

\newtheorem{Theo}{Theorem}
\newtheorem{remark}{Remark}

\newcommand{\PP}{\mathsf{PP}}
\newcommand{\PD}{\mathsf{PD}}
\newcommand{\PN}{\mathsf{PN}}
\newcommand{\DP}{\mathsf{DP}}
\newcommand{\DD}{\mathsf{DD}}
\newcommand{\DN}{\mathsf{DN}}
\newcommand{\NP}{\mathsf{NP}}
\newcommand{\ND}{\mathsf{ND}}
\newcommand{\NN}{\mathsf{NN}}

\newcommand{\Perfect}{\mathsf{P}}
\newcommand{\Delayed}{\mathsf{D}}
\newcommand{\None}{\mathsf{N}}

\newcommand{\CSIT}{\mathsf{CSIT}}
\newcommand{\JSIT}{\mathsf{JSIT}}

\newcommand{\CSIR}{\mathsf{CSIR}}
\newcommand{\JSIR}{\mathsf{JSIR}}

\newcommand{\DoF}{\mathsf{DoF}}

\begin{document}
\maketitle
\thispagestyle{empty}
\vspace{-0.5cm}
\begin{abstract}
Jamming attacks can significantly impact the performance of wireless communication systems. In addition to reducing the capacity, such attacks may lead to insurmountable overhead in terms of re-transmissions and increased power consumption. In this paper, we consider the multiple-input single-output (MISO) broadcast channel (BC) in the presence of a jamming attack in which a subset of the receivers can be jammed at any given time. Further, countermeasures for mitigating the effects of such jamming attacks are presented. The effectiveness of these anti-jamming countermeasures is quantified in terms of the degrees-of-freedom ($\DoF$) of the MISO BC under various assumptions regarding the availability of the channel state information ($\CSIT$) and the jammer state information at the transmitter ($\JSIT$). The main contribution of this paper is the characterization of the $\DoF$ region of the two user MISO BC under various assumptions on the availability of $\CSIT$ and $\JSIT$. Partial extensions to the multi-user broadcast channels are also presented.  
\end{abstract}

\section{Introduction}\label{Introduction}
Wireless communication systems have now become ubiquitous and constitute a key component of the fabric of modern day life.
However, the inherent \textit{openness} of the wireless medium makes it susceptible to adversarial attacks.  
The vulnerabilities of the wireless system can be largely classified based on the capability of an adversary-- 

a) \textit{Eavesdropping attack}, in which the eavesdropper (passive adversary) can listen to the wireless channel and try to infer information (which if leaked may severely compromise data integrity). %Therefore, the goal of the legitimate parties in the presence of an eavesdropper is to ensure secure information transmission, while minimizing information leakage to the eavesdropper.
The study of information theoretic security (or communication in presence of eavesdropping attacks) was initiated by Wyner \cite{WynerWiretap},  Csisz\'{a}r and K\"{o}rner \cite{CsiszarKorner}. Recently, there has been a resurgent interest in extending these results to multi-user scenarios. We refer the reader to a comprehensive tutorial \cite{ITsecurity} on this topic and the references therein. 

b) \textit{Jamming attack}, in which the jammer (active adversary) can transmit information  in order to disrupt reliable data transmission or reception.  While there has been some work in studying the impact of jamming on the capacity of point-to-point channels (such as  \cite{Basar-1983, MedardJamming, Kashyap-IT-2004}), the literature on information theoretic analysis of jamming attacks (and associated countermeasures) for multi-user channels is relatively sparse in comparison to the case of eavesdropping attacks. 

In this paper, we focus on a class of \textit{time-varying} jamming attacks over a fast fading multi-user multiple-input single-output (MISO) broadcast channel (BC), in which a transmitter equipped with $K$ transmit antennas intends to send independent messages to $K$ single antenna receivers. 
While several jamming scenarios are plausible, we initiate the study of jamming attacks by focusing on a simple yet harmful jammer. In particular, we consider a jammer equipped with $K$ transmit antennas and at any given time instant, has the capability of jamming a subset of the receivers. 
%In addition, we assume that  the jammer's signal is additive white Gaussian noise (AWGN), with the same power as that of the transmit signal. The jammer's strategy at any given time is random, i.e., it probabilistically selects the receiver(s) to disrupt.
%In addition, it is assumed that the jamming strategy varies in an i.i.d. manner across time. 
We consider a scenario in which the jammers' strategy at any given time is random, i.e., the subset of receivers to be jammed is probabilistically selected.  Furthermore, the jamming strategy varies in an independent and identically distributed (i.i.d.) manner across time\footnote{While we realize that perhaps more sophisticated jamming scenarios may arise in practice, as a first step, it is important  to understand i.i.d jamming scenarios before studying the impact of more complicated attacks (such as time/signal correlated jammer, on-off jamming etc). Even in the i.i.d. jamming scenarios, interesting and non-trivial problems arise that we address in this paper in the context of broadcast channels.}. Such random, time-varying jamming attacks may be inflicted either intentionally by an adversary or unintentionally, in different scenarios. We next highlight some plausible scenarios in which such random time varying jamming attacks could arise.
%\begin{enumerate}%

%\item 
A resource constrained jammer that intentionally jams the receivers may conserve power by selectively jamming a subset (or none) of the receivers based on its available resources. Such a jammer can also choose to jam the receivers when it has information about channel sounding procedures (i.e., when this procedure occurs) and disrupts the communication only during those specific time instants. 
%\item 
Interference from neighboring cells in a cellular system can act as a bottleneck to improve spectral efficiency and be particularly harmful for cell edge users. The interference seen from adjacent cells in such scenarios can be time varying depending on whether the neighboring cells are transmitting on the same frequency or not (which can change with time); and the spatial separation of the users from interfering cells.
%\item 
%Given that spectrum sharing is gaining importance, spectral co-existence of heterogenous systems such as radar and communication systems (such as LTE/WiMAX) has gained significant attention \cite{Shabnam}. However, the high radar signal power can create harmful interference to the communication systems that operate at much lower power levels by comparison. An interfering pulse radar (e.g. weather or beamforming-based radar) that scans the environment periodically, can possibly disrupt the receivers in such a random, time varying manner.
%\item Such jamming scenarios can also occur when a jammer that is unaware of the location of the transmitter and receiver, can disrupt a subset of the receivers in the process of scanning the environment. For instance, such a jamming scenario in a $2$-user system is shown in Fig.~\ref{JammingScenario}, where either none, or one or both the users are jammed. 
%\begin{figure}%
%\centering
%\includegraphics[width=0.6\columnwidth]{Jamming_Scenario.png}%
%\caption{Jamming Scenario}%
%\label{JammingScenario}%
%\end{figure}
%\item 
A frequency-selective jammer can disrupt communication on certain frequencies (carriers) in multi-carrier (for instance OFDM-based) systems. A jammer that has knowledge about the pilot signal-based synchronization procedures, can jam only those sub carriers that carry the pilot symbols in order to disrupt the synchronization procedure of the multi-carrier system \cite{Clancy}. Our analysis in this paper suggests that the transmitter and receivers based on the knowledge of the jammers' strategy, can reduce the effects of these jamming attacks by coding/transmitting across various jamming states (jamming state here can be interpreted as the subset of frequencies/sub-carriers that are jammed at a given time instant). 
%\end{enumerate}

\begin{figure}[t]
  \centering
\includegraphics[width=12.0cm]{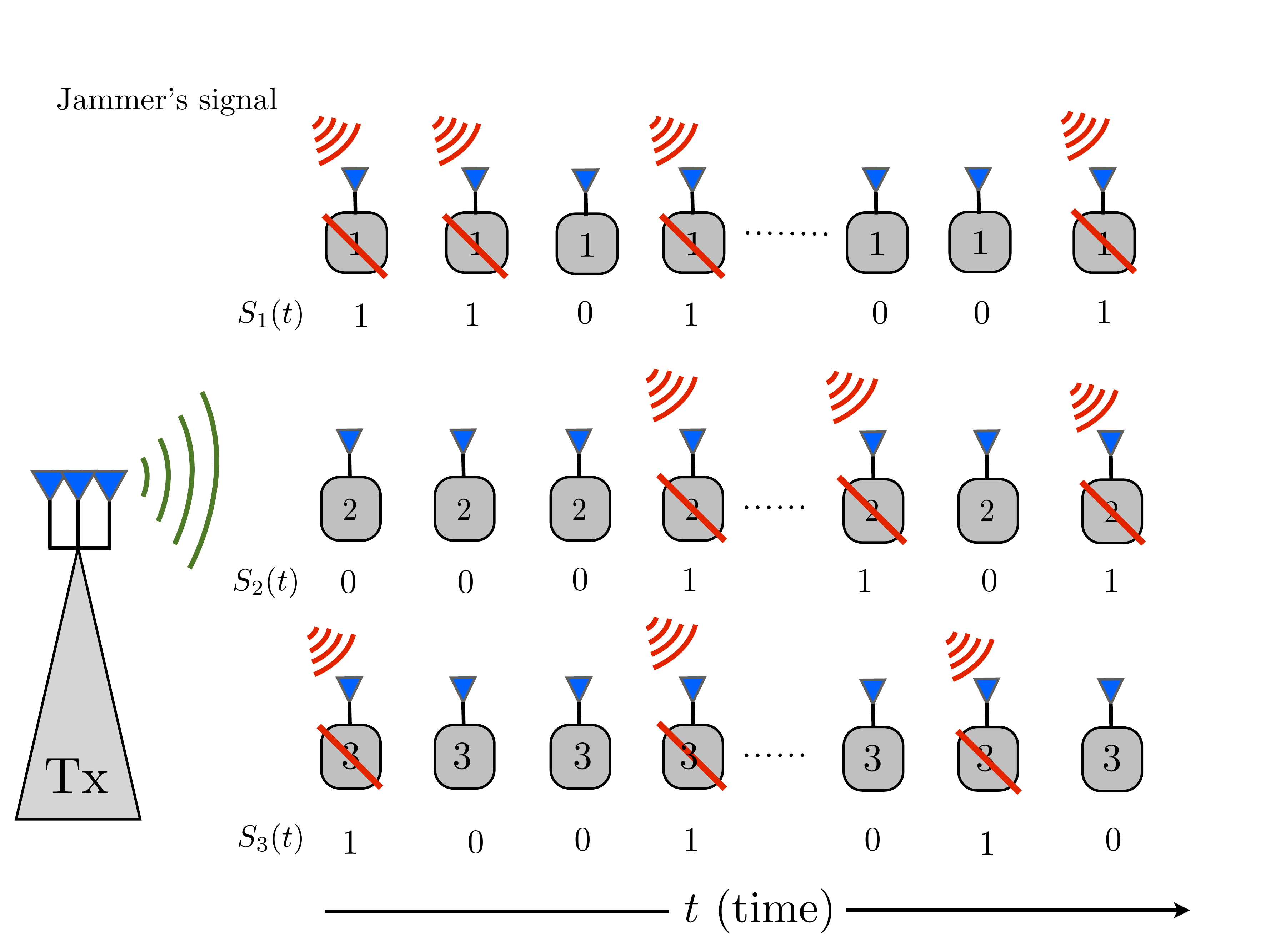}
\vspace{-0.2in}
\caption{Possible jamming scenarios in a $3$-user MISO broadcast channel.}
\label{Fig:FigureModel}
\end{figure}

Interestingly, the MISO BC with a time-varying jamming attack can also be interpreted as a network with a time-varying topology. The  concept of topological interference alignment has been recently introduced in \cite{JafarTopological} (also see \cite{JafarTopological_ISIT}, \cite{AvestimehrTopological}) to understand the effects of time-varying topology on interference mitigation techniques such as interference alignment. In \cite{JafarTopological_ISIT}, the authors characterize the $\DoF$ by studying the interference management problem in such networks using a 1-bit delay-less feedback (obtained from the receivers) indicating the presence or absence of an interference link. The connection between jamming attacks considered in this paper and time-varying network topologies can be noted by observing the following: if at a given time, a receiver is jammed, then its received signal is completely drowned in the jamming signal (assuming jamming power as high as the desired signal) which is analogous to the channel (or link) to the jammed receiver being wiped out. For instance, in a $3$-user MISO BC with a time-varying jamming attack, a total of $2^{3}=8$ \textit{topologies} could arise (see Figure \ref{Fig:FigureModel}) over time: none of the receivers are jammed (one topology), all receivers are jammed (one topology),  only one out of the three receivers is jammed (three topologies), or only two out of three receivers are jammed (i.e., three topologies). Interestingly, the \textit{retroactive anti-jamming} techniques presented in this paper are philosophically related to topological interference alignment with alternating connectivity \cite{JafarTopological_ISIT}. The common theme that emerges is that it is necessary to \textit{code across} multiple jamming states (equivalently, topologies as in \cite{JafarTopological_ISIT}) in order to achieve the optimal performance, which is measured in terms of degrees of freedom (capacity at high SNR). 

The model considered in the paper also bears similarities with broadcast erasure channels studied in \cite{ChihChunWang2012}, \cite{Erasure} etc. The presence of a jamming signal ($J$) at a receiver implies that the information bearing signal ($X$) is un-recoverable from the received signal ($Y= X+ J+ N$) in the context of degrees of freedom (since the pre-log of mutual information between $X$ and $Y$ would be zero as both signal and jamming powers become large). Hence, the presence of a jammer can be interpreted as an ``erasure". In the absence of a jammer (or no ``erasure"), the signal $X$ can be recovered from $Y=X+N$ within noise distortion.  

%Interestingly, the MISO BC with time varying jamming attack could also be interpreted as a network with time-varying topology.
%The novel concept of topological interference alignment has been recently introduced in \cite{JafarTopological} (also see \cite{AvestimehrTopological}) to tackle interference in such time varying networks. 
%Typical interference management techniques are designed for a fixed topology (which does not change over time). 
%It has been shown in  \cite{JafarAlternating} that if there are multiple topologies (each topology corresponds to a network with different interfering/non-interfering links) appearing across time, then significant gains can be achieved by coding across topologies. The connection between the jamming attacks considered in this paper and the time-varying network topologies becomes clear by noting the following: if at a given time, a receiver is jammed, then its received signal is completely drowned in the jamming signal (of power as high as transmit signal) and  as if the link to the jammed receiver is wiped out from degrees of freedom point of view. For example, in a $2$-user MISO BC with a time-varying jamming attack, a total of four topologies could arise over time: both receivers are not jammed, both receivers are jammed, or only one of the receiver is jammed. 

We study the impact of such random time-varying jamming attacks on the degrees-of-freedom (henceforth referred by $\DoF$) region of the MISO BC. The $\DoF$ of a network can be regarded as an approximation of its capacity at high SNR and is also referred to as the pre-log of capacity. Even in the absence of a jammer, it is well known that the $\DoF$ is crucially dependent on the availability of channel state information at the transmitter ($\CSIT$). The $\DoF$ region of the MISO BC has been studied under a variety of assumptions on the availability of $\CSIT$ including full (perfect and instantaneous) $\CSIT$ \cite{MIMOBC}, no $\CSIT$ \cite{CaireShamai, Huang}, delayed $\CSIT$ \cite{MAT2012, VV:DCSI-BC}, compound $\CSIT$ \cite{Weingarten_Shamai_Kramer}, quantized $\CSIT$ \cite{Jindal_BCFB}, mixed (perfect delayed and partial instantaneous) $\CSIT$ \cite{JafarTCBC} and asymmetric $\CSIT$ (perfect $\CSIT$ for one user, delayed $\CSIT$ for the other) \cite{Jafar_corr}.  To note the dependence of $\DoF$ on $\CSIT$, we remark that a sum $\DoF$ of $2$ is achieved in the $2$-user MISO BC when perfect $\CSIT$ information is available \cite{MIMOBC}, while it reduces to $1$ (with statistically equivalent receivers) when no $\CSIT$ is available \cite{Huang}. Interestingly it is shown in \cite{MAT2012} that completely outdated $\CSIT$ in a fast fading channel is still useful and helps increase the $\DoF$ from $1$ to $\frac{4}{3}$. Interesting extensions to the $K$-user case with delayed $\CSIT$ are also presented in \cite{MAT2012}.  
In this paper, we denote the availability of $\CSIT$ (by $\mathsf{CSI}$, we refer to the channel between the transmitter and the receiver, we \emph{do not} assume the knowledge of the jammer's channel at the transmitter or the receivers) through a variable $I_{\CSIT}$, which can take values either $\Perfect$, $\Delayed$ or $\None$; where the state $I_{\CSIT}=\Perfect$ indicates that the transmitter has perfect and instantaneous channel state information at time $t$, the state $I_{\CSIT}=\Delayed$ indicates that the transmitter has perfect but  delayed channel state information (i.e., it has knowledge of the channel realizations of time instants $\{1,2,\ldots, t-1\}$ at time $t$), and the state  $I_{\CSIT}=\None$ indicates that the transmitter has no channel state information. 

As mentioned above, the impact of $\CSIT$ on the $\DoF$ of MISO broadcast channels has been explored for scenarios in which there is no adversarial time-varying interference. The novelty of this work is two fold: a) incorporating adversarial time-varying interference, and b) studying the \textit{joint} impact of $\CSIT$ and the knowledge about the absence/presence of interference at the transmitter (termed $\JSIT$).

As we show in this paper, in the presence of a time-varying jammer, not only the $\CSIT$ availability but also the knowledge of jammer's strategy significantly impacts the $\DoF$. 
Indeed, if the transmitter is non-causally aware of the jamming strategy at time $t$, i.e., if it knows \textit{which} receiver (or receivers) is going to be disrupted at time $t$, the transmitter can utilize this knowledge and adapt its transmission strategy by:
%\begin{itemize}
%\item Transmitting to the receiver which is not jammed (if only one receiver is not jammed); 
either transmitting to a subset of receivers simultaneously (if only a subset of them are jammed/not-jammed) or conserving energy by not transmitting (if all the receivers are jammed).  %\end{itemize}

However, such adaptation may not be feasible if there is delay in learning the jammer's strategy. Feedback delays could arise in practice as the detection of a jamming signal would be done at the receiver (for instance, via a binary hypothesis test \cite{KayDetection} in which the receiver could use energy detection to validate the presence/absence of a jammer in its vicinity). This binary decision could be subsequently fed back to the transmitter. In presence of feedback delays, the standard approach would be to exploit the time correlation in the jammer's strategy to predict the current jammer's strategy from the delayed measurements. 
%\footnote{Detection of a jamming signal can be cast as a binary hypothesis testing problem \cite{KayDetection} in which the receiver can use energy detection to validate the presence/absence of a jammer in its vicinity. This decision is subsequently fed back to the transmitter. For the scope of this paper, we assume that the receivers can correctly detect the presence/absence of a jamming signal. Such assumptions are usually made in spread spectrum-based systems \cite{Proakis} where the receivers detect the presence or absence of the jammer based on the measurements from the adjacent channels. }. 
The predicted jammer state could then be used in place of the true jammer state. However, if the jammer's strategy is completely uncorrelated  across time (which is the case if the jammers' strategy is i.i.d),  delayed feedback reveals no information about the current state, and a predict-then-adapt scheme offers no advantage. A third and perhaps worst case scenario could also arise in which the transmitter only has statistical knowledge of jammer's strategy. This could be the case when the feedback links are unreliable or if the feedback links themselves are susceptible to jamming attacks, i.e., the outputs of feedback links are untrustworthy. 

To take all such plausible scenarios into account, we formally model the jamming strategy via an independent and identically distributed (i.i.d.) random variable $S(t)= (S_1(t), S_{2}(t),\ldots,S_K(t))$; which we call the jammer state information $(\mathsf{JSI})$ at time $t$. Note here that in the context of the paper, the jammers' state only indicates knowledge about the jammers' strategy (i.e., which receivers are jammed) and \emph{not} the channel between the jammer and receiver. At time $t$, if the $k$th component of $S(t)$, i.e., $S_k(t)=1$, it indicates that receiver $k$ is being jammed, and $S_{k}(t)=0$ indicates that receiver $k$ receives a jamming free signal. We denote the availability of jammer state information at the transmitter ($\JSIT$) through a variable $I_{\JSIT}$, which (similar to $I_{\CSIT}$) can take values either $\Perfect$, $\Delayed$ or $\None$; where the state $I_{\JSIT}=\Perfect$ indicates that the transmitter has perfect and instantaneous jammer state information $(S_1(t), S_2(t),\ldots,S_K(t))$ at time $t$, the state $I_{\JSIT}=D$ indicates that the transmitter has delayed jammer state information (i.e., it has access to $\{S_1(i), S_{2}(i),\ldots,S_K(i)\}_{i=1}^{t-1}$ at time $t$), and the state  $I_{\JSIT}=N$ indicates that the transmitter does not have the exact realization of $S(t)$ at its disposal. In all configurations above, it is assumed that the transmitter knows the statistics of $S(t)$. 
%has no jammer state information. For all $\JSIT$ configurations, it is assumed that the transmitter has statistical knowledge of the jammer�s strategy (i.e., statistics of $S(t)$). 
%Note here that the jammer state information only corresponds to the knowledge whether a receiver is jammed or not and does not correspond to the knowledge of the channel between the jammer and the receivers. 

\emph{Summary of Main Results:} 
Depending on the \textit{joint} availability of channel state information ($\CSIT$) and jammer state information ($\JSIT$) at the transmitter,  the variable $I_{\CSIT}I_{\JSIT}$ can take $9$ values and hence a total of $9$ distinct scenarios can arise: $\PP$, $\PD$, $\PN$, $\DP$, $\DD$, $\DN$, $\NP$, $\ND$, and  $\NN$.  The main contributions of this paper are the following.
\begin{enumerate}
\item For the $2$-user scenario, we characterize the exact $\DoF$ region for the $\PP$, $\PD$, $\PN$, $\DP$, $\DD$, $\NP$ and $\NN$ configurations.
\item For the $\DN$ and $\ND$ configurations in a $2$-user MISO BC, we present novel inner bounds to the $\DoF$ regions. 
\item The interplay between $\CSIT$ and $\JSIT$ and the associated impact on the $\DoF$ region in the various configurations is discussed. Specifically, the gain in $\DoF$ by transmitting across various jamming states and the loss in $\DoF$ due to the unavailability of $\mathsf{CSI}$ or $\mathsf{JSI}$ at the transmitter is quantified by the achievable sum $\DoF$. 
%\item The characterization of the degrees-of-freedom ($\DoF$) region of the $2$-user MISO BC for seven out of these $9$ possible scenarios; namely for $\PP$, $\PD$, $\PN$, $\DP$, $\DD$, $\NP$ and $\NN$ configurations. For $\DN$ and $\ND$ configurations, we present novel inner bounds to the $\DoF$ regions. 
\item We extend the analysis in a $2$-user MISO BC to a generic $K$-user MISO BC  with such random time-varying jamming attacks. 
The $\DoF$ region is completely characterized for the $\PP$, $\PD$, $\PN$, $\NP$ and $\NN$ configurations. 
Further, novel inner bounds are presented for the sum $\DoF$ in $\DP$ and $\DD$ configurations. 
These bounds provide insights on the scaling of sum $\DoF$ with the number of receivers $K$. 
\end{enumerate}

The remaining parts of the paper are organized as follows. The system model is introduced in Section~\ref{system_model}. 
The main contributions of the paper i.e., the Theorems describing the $\DoF$ regions in various ($\CSIT$,$\JSIT$) configurations for the $2$-user and $K$-user MISO BC are illustrated in Sections~\ref{Theorems} and \ref{TheoremsKuser} respectively and the corresponding converse proofs are presented in the Appendix.  The coding (transmission) schemes achieving the optimal $\DoF$ regions are described in Sections~\ref{Schemes}, \ref{TheoremsKuser}. Finally, conclusions are drawn in Section~\ref{Conclusions}. 

\section{System Model}\label{system_model}
\begin{figure}[t]
  \centering
\includegraphics[width=12.0cm]{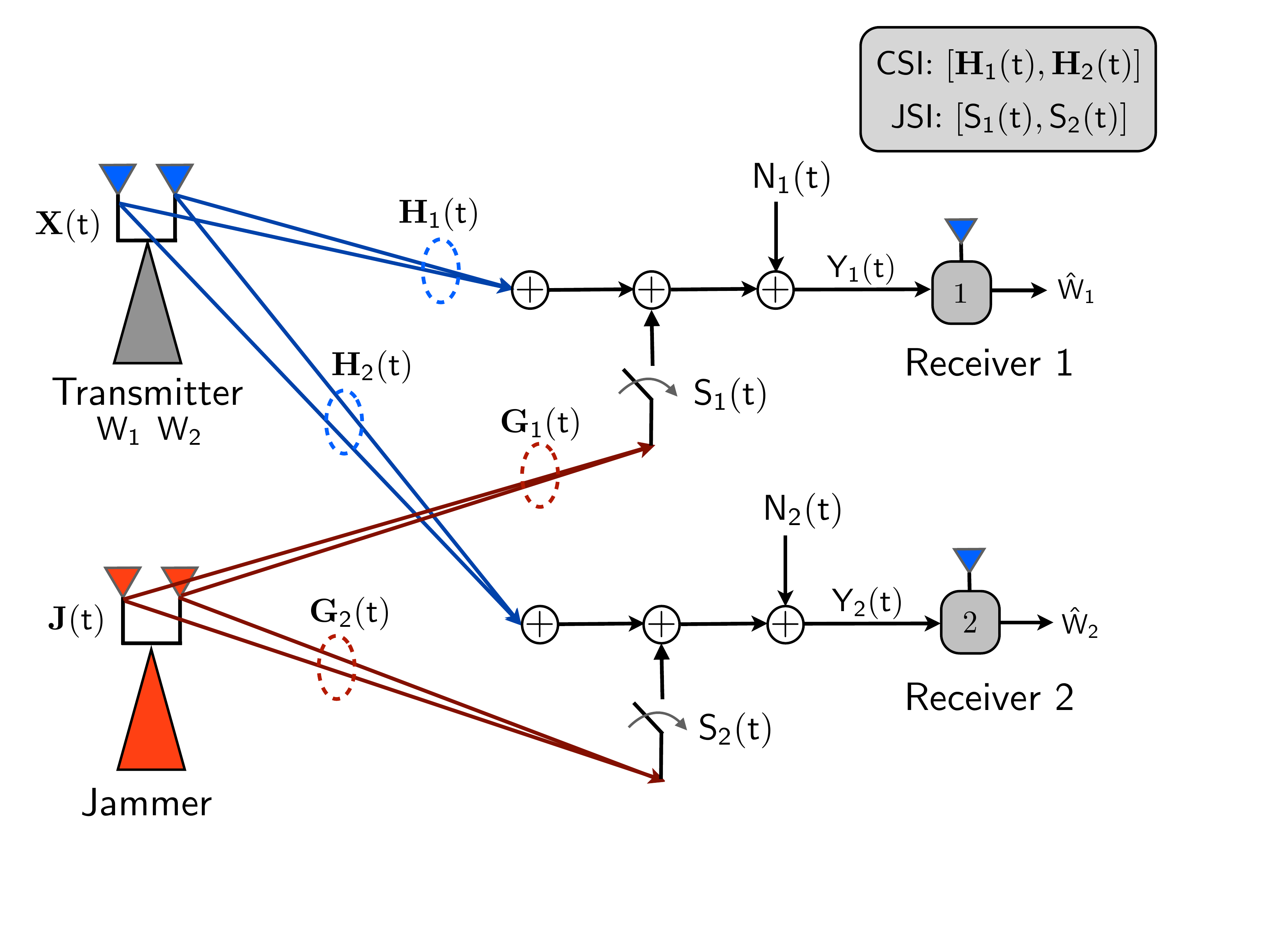}
\vspace{-0.5in}
\caption{System Model for a $2$-user scenario.}
\label{Fig:sys_model}
\end{figure}

A $K$-user MISO broadcast channel with $K$ transmit antennas and $K$ single antenna receivers, is considered in the presence of a random, time-varying jammer. The system model for the $K=2$ user case is shown in Fig.~\ref{Fig:sys_model}. 
The channel output at receiver $k$, for $k=1,2,\ldots,K$ at time $t$ is given as:
\begin{align}\label{system_model_eq}
Y_{k}(t)&= \mathbf{H}_{k}(t)\mathbf{X}(t) + S_k(t)\mathbf{G}_{k}(t)\mathbf{J}(t) + N_{k}(t), 
%Y_{2}(t)&= \mathbf{H}_{2}(t)\mathbf{X}(t) + S_2(t)\mathbf{G}_{2}(t)\mathbf{J}(t) + N_{2}(t),
\end{align}
where $\mathbf{X}(t)$  is the $K \times 1$ channel input vector at time $t$ with
\begin{align}
E\left(|\mathbf{X}(t)|^2\right)\leq P_T,
\end{align}
and $P_T$ is the power constraint on $\mathbf{X}(t)$. In \eqref{system_model_eq}, $\mathbf{H}_{k}(t)=[h_{1k}(t),h_{2k}(t),\ldots,h_{Kk}(t)]$ is the $1\times K$ channel vector from the transmitter to the $k$th receiver at time $t$, $\mathbf{G}_{k}(t)$ is the $1\times K$ channel response from the jammer to receiver $k$ at time $t$ and $\mathbf{J}(t)$ is the $K\times 1$ jammer's channel input at time $t$ (a worst case scenario where the jammer has $K$ degrees-of-freedom to disrupt all $K$ parallel streams of data from the transmitter to the $K$ receivers). Without loss of generality, the channel vectors $\mathbf{H}_k(t)$ and $\mathbf{G}_k(t)$ are assumed to be sampled from any continuous distribution (for instance, Rayleigh) with an identity covariance matrix, and are i.i.d. across time. The additive noise $N_{k}(t)$ is distributed according to $\mathcal{CN}(0,1)$ for $k=1,\ldots,K$ and are assumed to be independent of all other random variables. The random variable $S(t)= \{S_1(t), S_2(t),\ldots,S_K(t)\}$ that denotes the jammer state information $\mathsf{JSI}$ at time $t$, is a $2^K$-valued i.i.d. random variable. 

For example, in the $3$-user MISO BC, the $\mathsf{JSI}$ $S(t)$ is a $8$-ary valued random variable taking values $\{000,001,010,011,100,101,110,111\}$ with probabilities $\{\lambda_{000},\lambda_{001},\lambda_{010},\lambda_{011},\lambda_{100},\lambda_{101},\lambda_{110},\lambda_{111}\}$ respectively,  for arbitrary $\{\lambda_{ijk}\geq 0\}_{i,j,k=0,0,0}^{1,1,1}$ such that $\sum_{i,j,k}\lambda_{ijk}=1$. 
The jammer state $S(t)$ at time $t$ can be interpreted as follows:
\begin{itemize}
\item $S(t)=\left(0,0,0\right)$ : none of the receivers are jammed. This occurs with probability $\lambda_{000}$.
\item $S(t)=\{\left(1,0,0\right)/\left(0,1,0\right)/\left(0,0,1\right)\}$ : only one receiver is jammed. This scenario occurs with probability $\lambda_{100}/\lambda_{010}/\lambda_{001}$ respectively. $S(t)=\left(1,0,0\right)$ indicates that the $1$st receiver is jammed while the receivers $2$ and $3$ are not jammed. 
\item $S(t)=\{\left(1,1,0\right)/\left(1,0,1\right)/\left(0,1,1\right)\}$ : any two out of the three receivers are jammed. This happens with probability $\lambda_{110}/\lambda_{101}/\lambda_{011}$ respectively.
\item $S(t)=\left(1,1,1\right)$ : all the receivers are jammed with probability $\lambda_{111}$.
\end{itemize}
Using the probability vector $\{\lambda_{000}, \lambda_{001}, \lambda_{010}, \lambda_{100}, \lambda_{011}, \lambda_{110}, \lambda_{101}, \lambda_{111} \}$, 
we define the marginal probabilities 
\begin{align}\label{lambda_receiver_1}
\lambda_1&=\lambda_{000}+\lambda_{001}+\lambda_{010}+\lambda_{011}, \nonumber \\
\lambda_2&=\lambda_{000}+\lambda_{001}+\lambda_{100}+\lambda_{101}, \nonumber \\ 
\lambda_3&=\lambda_{000}+\lambda_{010}+\lambda_{100}+\lambda_{110},
\end{align}
%Using these, we can write the marginal probabilities
%\begin{align}
%\lambda_{1}\triangleq \lambda_{00}+\lambda_{01},\ \ \lambda_{2}\triangleq \lambda_{00}+\lambda_{10},
%\end{align}
where $\lambda_{k}\in [0,1]$ denotes the \textit{total} probability with which receiver $k$ is \textit{not} jammed. For example, in the $3$-user scenario, $\lambda_1$ indicates the total probability with which the $1$st receiver is not jammed which happens when any one of the following events happen 1) none of the receivers are jammed with probability $\lambda_{000}$, 2) only the $2$nd receiver is jammed with probability $\lambda_{010}$, 3) only $3$rd receiver is jammed with probability $\lambda_{001}$ or 4) both the $2$nd and $3$rd receivers are jammed with probability $\lambda_{011}$. Similar definitions hold for the $K$-user MISO BC. In general, $S(t)$ is a $K\times 1$ vector where a $1 (0)$ in the $k$th position indicates that the $k$th receiver is jammed (not-jammed). 

%It is assumed that the elements of $\mathbf{J}(t)$ are distributed i.i.d. as $\mathcal{CN}(0,P_T)$, i.e., the jammer transmits i.i.d. AWGN noise at each time instant, and with power equal to the transmit signal power $P_T$. 
It is assumed that the jammer sends a signal with power equal to $P_T$ (the transmit signal power). This formulation attempts to capture the performance of the system in a time-varying interference (here jammer) limited scenario where the received interference power is as high as the transmit signal power $P_T$ (a worst case scenario where the receiver by no means can recover the symbol from the received signal). Furthermore, it is assumed that $\{\mathbf{J}(t)\}_{t=1}^{n}$ is independent of $\{S(t)\}_{t=1}^{n}$. We denote the global channel state information (between transmitter and receivers) at time $t$ by $\mathbf{H}(t)\triangleq \{\mathbf{H}_{1}(t), \mathbf{H}_{2}(t),\ldots,\mathbf{H}_{K}(t)\}$. In all analysis that follows, we assume that both the receivers have complete knowledge of global channel vectors $\{\mathbf{H}(t)\}_{t=1}^{n}$ and also of the jammer's strategy $\{S(t)\}_{t=1}^{n}$, i.e., full $\CSIR$ and full $\JSIR$ (similar assumptions were made in earlier works, see \cite{MAT2012}, \cite{ACSIT-ISIT}, \cite{ACSIT2012} and references therein). 

\paragraph{Assumptions:} The following are the list of assumptions made in this paper.
\begin{itemize}
%\item The transmitter receives feedback from the receivers regarding the channel $\mathbf{H}(t)$ (in other words termed as $\CSIT$)  and about the jammers' strategy i.e., $S(t)$ ($\JSIT$) at any given time $t$. The transmitter does not require knowledge of the channels between the jammer and the receiver i.e., $\mathbf{G}(t)=\{\mathbf{G}_{1}(t),\ldots,\mathbf{G}_{K}(t)\}$. 
\item If $\CSIT$ exists (i.e., when $I_{\CSIT}=\mathsf{P}$ or $\mathsf{D}$), the transmitter receives either instantaneous or delayed feedback from the receivers regarding the channel $\mathbf{H}(t)$. In either scenario, neither the transmitter nor the receivers require knowledge of $\mathbf{G}(t)=\{\mathbf{G}_{1}(t),\ldots,\mathbf{G}_{K}(t)\}$ i.e., the channel between the jammer and the receivers. 
\item If $\JSIT$ exists (i.e., when $I_{\JSIT}=\mathsf{P}$ or $\mathsf{D}$), then the transmitter receives either instantaneous or delayed feedback about the jammers' strategy i.e., $S(t)$.
\item Irrespective of the availability/ un-availability of $\CSIT$ and $\JSIT$, it is assumed that the transmitter has statistical knowledge of the jammer's strategy (i.e., statistics of $S(t)$) which is assumed to be constant across time (these assumptions form the basis for future studies that deal with time varying statistics of a jammer). 
\item While the achievability schemes presented in Sections~\ref{Schemes}, \ref{TheoremsKuser} hold for arbitrary correlations between the random variables 
$S(t)$, $\mathbf{J}(t)$, and $\mathbf{G}(t)$, the converse proofs provided in the Appendix hold under the assumption that these random variables are mutually independent and when the elements of $\mathbf{J}(t)$ are distributed i.i.d. as $\mathcal{CN}(0,P_T)$. 
\item The theorems, achievability schemes and the converse proofs presented in Sections~\ref{Theorems}--\ref{TheoremsKuser} and the Appendix hold true for any continuous distributions that $\mathbf{H}(t)$ and $\mathbf{G}(t)$ may assume. While these achievability schemes are valid for any distribution of the jammers' signal $\mathbf{J}(t)$, the converse proofs are presented for the case in which the jammers' signal is Gaussian distributed. 
\end{itemize}

For the $K$-user MISO BC, a rate tuple $(R_{1},R_{2},\ldots,R_K)$, with $R_{k}= \log(|W_{k}|)/n$, where $n$ is the number of channel uses, $W_k$ denotes the message for the $k$th receiver and $|W_k|$ represents the cardinality of $W_k$, is achievable if there exist a sequence of encoding functions $f^{(n)}$ and decoding functions $g^{(n)}_{k}\left(Y_k^n,\mathbf{H}^n,\mathbf{S}^n\right)$ (one for each receiver) such that for all $k=1,2,\ldots,K$,
\begin{equation}\label{error_convergence}
P\left(W_k\neq g_k^n\left(Y_k^n,\mathbf{H}^n,\mathbf{S}^n\right)\right)\leq n\epsilon_{kn},
\end{equation} 
where
\begin{equation}
\epsilon_{kn}\longrightarrow 0 \quad \mbox{as} \quad n\longrightarrow \infty,%\ \forall k=1,2,\ldots,K,
\end{equation}
i.e, the probability of incorrectly decoding the message $W_k$ from the signal received at user $k$ converges to zero asymptotically. In \eqref{error_convergence}, we have used the following shorthand notations $Y_k^n=\left(Y_k(1),\ldots,Y_k(n)\right)$, $\mathbf{H}^{n}=\left(\mathbf{H}_1(1),..,\mathbf{H}_K(1),..,\mathbf{H}_1(n),..,\mathbf{H}_K(n)\right)$ and  $\mathbf{S}^{n}=\left(S(1),S(2),\ldots,S(n)\right)$. 
%the probability of decoding error for message $W_{k}$, for $k=1,\ldots,K$ can be made arbitrarily small for sufficiently large $n$. 
We are specifically interested in the degrees-of-freedom region $\mathcal{D}$, defined as the set of all achievable pairs $(d_{1},d_{2},\ldots,d_{K})$ with $d_{k}=\lim_{P_{T}\rightarrow \infty} \frac{R_{k}}{\log(P_{T})}$. The encoding functions $f^{(n)}$ that achieve the $\DoF$ described in Sections~\ref{Theorems} and \ref{TheoremsKuser} depend on the availability of $\CSIT$ and $\JSIT$ i.e, on the variable $I_{\CSIT}I_{\JSIT}$. For example, in the $\DD$ (delayed $\CSIT$, delayed $\JSIT$) configuration, the encoding function takes the following form; 
\begin{equation}
\mathbf{X}(n)=f^{(n)}\left(W_{1},W_{2},\ldots,W_K,\mathbf{H}^{n-1}, \mathbf{S}^{n-1}\right),
\end{equation}
where the transmit signal $\mathbf{X}(n)$ at time $n$, depends on the the past channel state $\left(\mathbf{H}^{n-1}\right)$ and jammer state $\left(\mathbf{S}^{n-1}\right)$ information available at the transmitter. 
%As we show in Section~\ref{Schemes}, the synergistic   % i.e., =\left(\mathbf{H}_1(1),..,\mathbf{H}_K(1),..,\mathbf{H}_1(t-1),..,\mathbf{H}_K(t-1)\right)$ and 
%and $\mathbf{S}^{t-1}$. %=\left(S(1),S(2),\ldots,S(t-1)\right)$. 
However, in the $\NP$ configuration since the transmitter does not have knowledge about the channel (as no $\CSIT$ is available), it exploits the perfect and instantaneous knowledge about the jammers' strategy $\left(S(t)\right)$ by sending information exclusively to the unjammed receivers. 
%users only when they are not jammed. 
As a result, the encoding function for the $\NP$ configuration can be represented as 
\begin{equation}
\mathbf{X}(n)=f^{(n)}\left(W_{1},W_{2},\ldots,W_K,\mathbf{S}^n\right).
\end{equation}
The encoding functions across various channel and jammer states depend on the transmission strategies used and are discussed in more detail in Sections~\ref{Schemes} and \ref{TheoremsKuser}. 
%The coding schemes that use the above encoding functions to achieve the $\DoF$ regions described in Section~\ref{Theorems} and \ref{TheoremsKuser} are explained in Section~\ref{Schemes}.
%In particular, we denote the ($\CSIT$, $\JSIT$) pair by the variables $I_{$\CSIT$}I_{$\JSIT$}$; where
%
%\begin{align}
%I_{$\CSIT$/$\JSIT$}=
%\begin{cases}
%P,& \mbox{  instantaneous $\CSIT$/$\JSIT$},\\
%D,& \mbox{ delayed $\CSIT$/$\JSIT$},\\
%N,& \mbox{ no $\CSIT$/$\JSIT$}.
%\end{cases}
%\end{align}

%For instance, 
%\begin{itemize}
%\item ($\CSIT$, $\JSIT$) configuration $PN$ corresponds to the case in which the transmitter has access to $\{\mathbf{H}(i)\}_{i=1}^{t}$ at time $t$ and has no knowledge of $S_1(t),S_2(t)$. 
%\item ($\CSIT$, $\JSIT$) configuration $ND$ corresponds to the case in which the transmitter has no knowledge of CSI but has access to $\{S_1(i),S_2(i)\}_{i=1}^{t-1}$ at time $t$. 
%\end{itemize}
%\begin{figure}[t]
%  \centering
%\includegraphics[width=13.2cm]{ComparisonJournal.pdf}
%\caption{Degrees-of-freedom regions for various $\CSIT$-$\JSIT$ configurations.}\label{Fig:Figure1}
%\label{DoF}
%\end{figure}
\subsection{Review of Known Results}
As mentioned earlier, the $\DoF$ region for the $K$-user MISO BC has been studied extensively in the absence of external interference. We briefly present some of those important results that are relevant to the work presented in this paper. 
\begin{enumerate}
\item In the absence of jamming, the $\DoF$ region with perfect $\CSIT$ is given by,
\begin{align}
d_k\leq 1,\quad k=1,2,\ldots,K,
\end{align}
and the achievable sum $\DoF$ is $K$ \cite{ACSIT2012}. 
\item With delayed $\CSIT$, the $\DoF$ region in the absence of a jammer was characterized by Maddah-Ali and Tse in \cite{MAT2012}, and is given by
\begin{align}
\sum_{k=1}^K \frac{d_{\pi(k)}}{k}\leq 1,
\end{align}
where $\pi(K)$ is a permutation of the set of numbers $\{1,2,3,\ldots,K\}$. In such a scenario, the sum $\DoF$ (henceforth referred to as 
$\DoF_{\mathsf{MAT}}$) is given by 
\begin{align}\label{DoF_MAT}
\DoF_{\mathsf{MAT}}(K)=\frac{K}{1+\frac{1}{2}+\ldots\frac{1}{K}}.
\end{align}
\item The $\DoF$ region with no $\CSIT$ is given by 
\begin{align}
\sum_{k=1}^K d_k\leq 1.
\end{align}
and the sum $\DoF$ in this case reduces to $1$ \cite{ACSIT2012}. 
\end{enumerate}
It is easy to see that the sum $\DoF$ achieved in a delayed $\CSIT$ scenario lies in between the sum $\DoF$ achieved in the perfect $\CSIT$ and no $\CSIT$ scenarios. 
%In the absence of jamming, the sum $\DoF$ with perfect $\CSIT$ is $K$ and with no $\CSIT$ is $1$ . It has been shown in \cite{MAT2012} that the sum $\DoF$ with delayed $\CSIT$ is given by 

%Similar characterization of the $\DoF$ region in the presence of jamming attacks is presented in the next section. 

\section{Main Results and Discussion}\label{Theorems}
We first present $\DoF$ results for the $2$-user MISO BC under various assumptions on the availability of $\CSIT$ and $\JSIT$ and discuss various insights arising from these results. In the $2$-user case, the jammer state $S(t)$ at time $t$ can take one out of four values: $00, 01, 10$, or $11$, where 
\begin{itemize}
\item $S(t)=00$ indicates that none of the receivers are jammed, which happens with probability $\lambda_{00}$,
\item $S(t)=01$ indicates that only receiver $1$ is not jammed, which happens with probability $\lambda_{01}$,
\item $S(t)=10$ indicates that only the $2$nd receiver is un-jammed with probability $\lambda_{01}$, and finally 
\item $S(t)=11$ indicates that both the receivers are jammed with probability $\lambda_{11}$.
\end{itemize} 
In order to compactly present the results, we define the marginal probabilities 
\begin{align}
\lambda_{1}&\triangleq \lambda_{00}+\lambda_{01},\nonumber\\
\lambda_{2}&\triangleq \lambda_{00}+\lambda_{10}\nonumber,
\end{align}
where $\lambda_{k}$, for $k=1,2$ is the total probability with which receiver $k$ is \emph{not jammed}. In the sequel, Theorems~\ref{TheoremPP}-\ref{TheoremNN} present the optimal $\DoF$ characterization for the $\left(\CSIT,\JSIT\right)$ configurations $\PP,\PD,\PN,\DP,\DD,\NP$ and $\NN$ while Theorems~\ref{TheoremDN} and \ref{TheoremND} present non-trivial achievable schemes (novel inner bounds) for the $\DN$ and $\ND$ configurations. 

\begin{Theo}\label{TheoremPP}
The $\DoF$ region of the $2$-user MISO BC for each of the $\CSIT$-$\JSIT$ configurations $\PP$, $\PD$ and $\PN$ is the same and is given by the set of non-negative pairs $(d_{1},d_{2})$ that satisfy
\begin{align}
d_{1}&\leq \lambda_{1}\\
d_{2}&\leq \lambda_{2}.
\end{align}
\end{Theo}

\begin{Theo}\label{TheoremDP}
The $\DoF$ region of the $2$-user MISO BC for the $\CSIT$-$\JSIT$ configuration $\DP$, is given by the set of non-negative pairs $(d_{1},d_{2})$ that satisfy\begin{align}
d_{1}&\leq \lambda_1\\
d_{2}&\leq \lambda_2\\
2d_{1}+d_{2}&\leq 2\lambda_1+\lambda_{10}\\
d_{1}+2d_{2}&\leq 2\lambda_2+\lambda_{01}.
\end{align}
\end{Theo}

\begin{Theo}\label{TheoremDD}
The $\DoF$ region of the $2$-user MISO BC for the $\CSIT$-$\JSIT$ configuration $\DD$, is given by the set of non-negative pairs $(d_{1},d_{2})$ that satisfy
\begin{align}
\frac{d_{1}}{\lambda_{1}}+\frac{d_{2}}{(\lambda_{1}+\lambda_{2})}&\leq 1 \label{DD1} \\
\frac{d_{1}}{(\lambda_{1}+\lambda_{2})}+\frac{d_{2}}{\lambda_{2}}&\leq 1. \label{DD2}
\end{align}
\end{Theo}

\begin{Theo}\label{TheoremNP}
The $\DoF$ region for the $2$-user MISO BC for the $\CSIT$-$\JSIT$ configuration $\NP$, is given by the set of non-negative pairs $(d_{1},d_{2})$ that satisfy
\begin{align}
d_{1}&\leq \lambda_1\\
d_{2}&\leq \lambda_2\\
d_{1}+d_{2} &\leq \lambda_{00}+\lambda_{01}+\lambda_{10}.
\end{align}
\end{Theo}

\begin{Theo}\label{TheoremNN}
The $\DoF$ region of the $2$-user MISO BC for the $\CSIT$-$\JSIT$ configuration $\NN$ is given by the set of non-negative pairs $(d_{1},d_{2})$ that satisfy
\begin{align}
\frac{d_{1}}{\lambda_{1}}+\frac{d_{2}}{\lambda_{2}}&\leq 1.
\end{align}
\end{Theo}

\begin{remark}{\em [Redundancy of $\JSIT$ with Perfect $\CSIT$]} 
{\em We note from Theorem~\ref{TheoremPP} that when Perfect $\CSIT$ is available, the $\DoF$ region remains the same regardless of availability/un-availability of jammer state information at the transmitter. This implies that with perfect $\CSIT$, only statistical knowledge about the jammer's strategy suffices to achieve the optimal $\DoF$ region (note that it is assumed that the transmitter has statistical knowledge of the jammers' strategy). The availability of perfect $\CSIT$ helps to avoid cross-interference in such a broadcast type communication system and thereby enables the receivers to decode their intended symbols whenever they are not jammed. }
\end{remark}

\begin{remark}{\em [Quantifying $\DoF$ Loss]}  
{\em When the transmitter has perfect knowledge about the jammers state i.e, perfect $\JSIT$, it is seen that the $\mathsf{Sum}\ \DoF$ for the various configurations is 
\begin{align}
\mathsf{Sum}\ \DoF \textsf{ (with Perfect $\JSIT$)}=
\begin{cases}
\lambda_1+\lambda_2,& \mbox{  perfect $\CSIT$},\\
\lambda_1+\lambda_2-\frac{2}{3}\lambda_{00},& \mbox{ delayed $\CSIT$},\\
\lambda_1+\lambda_2-\lambda_{00},& \mbox{ no $\CSIT$}.
\end{cases}
\end{align}
It is seen that the sum $\DoF$s achieved in the $\DP$ and $\NP$ configurations are less than $\left(\lambda_1+\lambda_2\right)$, the sum $\DoF$ achieved in the $\PP$ configuration. The loss in $\DoF$ due to delayed channel knowledge is $\frac{2}{3}\lambda_{00}$ and due to no channel knowledge is $\lambda_{00}$. As expected, the loss in the $\NP$ configuration is more than the corresponding $\DoF$ loss in the $\DP$ configuration due to the un-availability of $\CSIT$. Interestingly, the loss in $\DoF$ due to delayed channel state information in the absence of a jammer is $2-\frac{4}{3}=\frac{2}{3}$ (where $2 \ \left(\frac{4}{3}\right)$ is the $\DoF$ achieved in a $2$-user MISO BC with perfect (delayed) $\CSIT$ \cite{MAT2012}), which, in the presence of a jammer, corresponds to the case when $\lambda_{00}=1$ i.e, none of the receivers are jammed. Along similar lines, the $\DoF$ loss due to no $\CSIT$ is $2-1=1$ where $1$ is the $\DoF$ achieved in the $2$-user MISO BC when there is no $\CSIT$ \cite{ACSIT2012} (in the absence of jamming). The loss in $\DoF$ converges to $0$ as $\lambda_{00}\rightarrow 0$ i.e, the $\PP$, $\DP$ and $\NP$ configurations are equivalent when the jammer disrupts either one or both the receivers at any given time. }
\end{remark}

%\begin{remark}
%{\em The achievability of Theorem \ref{TheoremDD} is based on synergistic usage of delayed $\CSIT$ (d-$\CSIT$) and delayed $\JSIT$ (d-$\JSIT$) by exploiting side-information created at the un-jammed receiver in the past and transmitting linear combinations of such side-information symbols in a retroactive manner. }
%%in the future 
%\end{remark}
%\begin{remark}
%\end{remark}

\begin{remark}{\em [Separability with Perfect $\JSIT$] 
%When the transmitter has perfect knowledge about the jammers state i.e, perfect $\JSIT$, it is seen that the $\mathsf{Sum}\ \DoF$ for the various configurations is 
%\begin{align}
%\mathsf{Sum}\ \DoF \textsf{ (with Perfect $\JSIT$)}=
%\begin{cases}
%\lambda_1+\lambda_2,& \mbox{  perfect $\CSIT$},\\
%\lambda_1+\lambda_2-\frac{2}{3}\lambda_{00},& \mbox{ delayed $\CSIT$},\\
%\lambda_1+\lambda_2-\lambda_{00},& \mbox{ no $\CSIT$}.
%\end{cases}
%\end{align}
When perfect $\JSIT$ is present, i.e., in the $\PP$, $\DP$ and $\NP$ configurations, the transmitter \textbf{does not} need to code (transmit) \emph{across} different jammer states; or in other words, the jammer's states are \emph{separable}. For instance, consider the case of delayed $\CSIT$. In the absence of a jammer, the optimal $\DoF$ with delayed $\CSIT$ is $4/3$ as shown in \cite{MAT2012}. The optimal strategy in presence of a jammer and with perfect $\JSIT$ is the following: use the $00$ state to achieve $\frac{4}{3}\lambda_{00}$ $\DoF$ by employing the $\mathsf{MAT}$ scheme \cite{MAT2012} (transmission scheme to achieve the sum $\DoF$ given in \eqref{DoF_MAT}, explained in Section~\ref{Schemes}), use $01$ state to achieve $\lambda_{01}$ $\DoF$ by transmitting to receiver $1$, use $10$ state to achieve $\lambda_{10}$ $\DoF$ by transmitting to receiver $2$. The state $11$ yields $0$ $\DoF$ since both the receivers are jammed. Thus, the net achievable $\DoF$ of this separation based strategy is given as: $\frac{4}{3}\lambda_{00}+ \lambda_{01}+\lambda_{10}= \lambda_{1}+\lambda_{2}-\frac{2}{3}\lambda_{00}$. Similar interpretations hold with perfect $\CSIT$ and no $\CSIT$. The transmission schemes that achieve these $\DoF$s and make the jammers' states separable are illustrated in more detail in Section~\ref{Schemes}. }

%These results indicate that the various jammers' states are separable i.e, the transmitter can develop a transmission scheme independently across these jamming states. For example, the $\mathsf{Sum}\ \DoF$ is $2$ when perfect $\CSIT$ is available while it reduces to $\frac{4}{3}$ when the $\CSIT$ knowledge is available with a delay and to $1$ when no $\CSIT$ information is available. Using this knowledge, it can be seen that if the transmitter can transmit independently across different states of the jammer, the $\mathsf{Sum}\ \DoF$ when none of the receivers are jammed (with probability $\lambda_{00}$) is $2\lambda_{00}$, $\frac{4}{3}\lambda_{00}$, and $\lambda_{00}$ with perfect, delayed and no knowledge about $\CSIT$. However, in all the different configurations, the $\mathsf{Sum}\ \DoF$ is $1$ when only one receiver is jammed (irrespective of the user) and $0$ when both the receivers are jammed. Thus it is seen that the transmitter has the flexibility of treating the states independently given perfect knowledge of the jammers strategy, that is as if the jammers states are separable. }
\end{remark}

\begin{remark} {\em [Marginal Equivalence] The $\DoF$ regions in Theorems \ref{TheoremPP}, \ref{TheoremDD} and \ref{TheoremNN} only depend on the marginal probabilities $(\lambda_{1}, \lambda_{2})$ with which each receiver is not jammed. This implies that two different jamming strategies with statistics, $\{\lambda_{00}, \lambda_{01}, \lambda_{10}, \lambda_{11}\}$ and $\{\lambda^{'}_{00}, \lambda^{'}_{01}, \lambda^{'}_{10}, \lambda^{'}_{11}\}$ result in the same $\DoF$ regions for $\PP$, $\PD$, $\PN$, $\DD$ and $\NN$ configurations as long as $\lambda_{00}+\lambda_{01}=\lambda^{'}_{00}+\lambda^{'}_{01}=\lambda_{1}$ and $\lambda_{00}+\lambda_{10}=\lambda^{'}_{00}+\lambda^{'}_{10}=\lambda_{2}$.}
\end{remark}

In the next two Theorems, we present  achievable $\DoF$ regions for the remaining configurations $\DN$ and $\ND$ respectively. It should be noticed that ignoring the availability of delayed $\CSIT$ in the $\DN$ configuration and the availability of delayed $\JSIT$ in the $\ND$ configuration, the $\DoF$ region described by Theorem~\ref{TheoremNN} can always be achieved. However, the novel inner bounds presented in Theorems~\ref{TheoremDN},\ref{TheoremND} show that the achievable $\DoF$ can be improved by synergistically using the delayed feedback regarding $\CSIT$ and $\JSIT$. 

\begin{Theo}\label{TheoremDN}
An achievable $\DoF$ region for the $2$-user MISO BC for the $\CSIT$-$\JSIT$ configuration $\DN$, is given as follows. 

\noindent For  $\frac{|\lambda_{1}-\lambda_{2}|}{\lambda_{1}\lambda_{2}}\leq 1$,  following region is achievable
\begin{align}
d_{1} +\frac{\left(2\max(1,\lambda_{1}/\lambda_{2})-1\right)}{(1+\lambda_2)}d_{2} &\leq \lambda_{1} \\
 \frac{\left(2\max(1,\lambda_{2}/\lambda_{1})-1\right)}{(1+\lambda_1)}d_{1} + d_{2}&\leq \lambda_{2}.
\end{align}
For $\frac{|\lambda_{1}-\lambda_{2}|}{\lambda_{1}\lambda_{2}}> 1$, following region is achievable
\begin{align}
\frac{d_{1}}{\lambda_{1}}+\frac{d_{2}}{\lambda_{2}}&\leq 1.
\end{align}
\end{Theo}

Though the optimal $\DoF$ region for the $\DN$ configuration remains unknown, we propose a novel inner bound (achievable scheme) to the $\DoF$ region as specified in Theorem~\ref{TheoremDN}. This scheme is based on a coding scheme (alternative to the original transmission scheme proposed in \cite{MAT2012}) to achieve $\DoF$ of $\frac{4}{3}$ for the $2$-user MISO BC in the absence of jamming attacks. This alternative scheme is discussed in Section~\ref{Schemes}. 
%We would also like to mention here that $\DoF$ gains over the naive TDMA scheme is achievable in this configuration only when $\lambda_1$ and $\lambda_2$ (the probabilities with which the receivers are not jammed)
%satisfy the condition 
%\begin{equation}
%\frac{|\lambda_{1}-\lambda_{2}|}{\lambda_{1}\lambda_{2}}\leq 1.
%\end{equation} 

\begin{Theo}\label{TheoremND}
An achievable $\DoF$ region for the $2$-user MISO BC in the $\CSIT$-$\JSIT$ configuration $\ND$, is given by the set of non-negative pairs $(d_{1},d_{2})$ that satisfy
\begin{align}
\frac{d_1}{\lambda_1} + \frac{d_2}{\lambda_{00}+\lambda_{01}+\lambda_{10}} &\leq 1 \\
\frac{d_1}{\lambda_{00}+\lambda_{01}+\lambda_{10}} + \frac{d_2}{\lambda_2} &\leq 1.
\end{align}
\end{Theo}

By noticing that $\lambda_{00}+\lambda_{01}+\lambda_{10} \geq  \mathsf{max}\left(\lambda_1,\lambda_2\right)$, it can be seen that the $\DoF$ region described by Theorem~\ref{TheoremND} is better than the region described by Theorem~\ref{TheoremNN} i.e., the region achieved in the $\NN$ configuration can be improved by utilizing the delayed $\JSIT$ information. Also, the $\DoF$ achievable in the $\ND$ configuration is a subset of the $\DoF$ achieved in the $\DD$ configuration. This is because $\lambda_1+\lambda_2\geq\lambda_{00}+\lambda_{01}+\lambda_{10}$. However, in scenarios where $\lambda_{00}=0$, the $\DoF$ region achieved by these two configurations is the same. Thus the converse proof in the Appendix that shows the optimality of the $\DoF$ region achieved in the $\DD$ configuration also holds true for the $\ND$ scenario when $\lambda_{00}=0$. This equivalence will be explained further in Section~\ref{Schemes}. 

Table~\ref{Theorem_table} summarizes the mapping between the $(\CSIT,\JSIT)$ configurations and the
theorems that specify their $\DoF$. The coding schemes that achieve the corresponding degrees of freedom regions are detailed in Section \ref{Schemes} and the corresponding converse proofs are presented in the Appendix. 
%We next present a series of remarks that highlight various properties and insights offered by the results in Theorems \ref{TheoremPP}-\ref{TheoremNN}.

\begin{table}[t]
\centering
\begin{tabular}{|c|c|c|c|l|l|l|l|}\hline
$\CSIT$&$\JSIT$&Configuration ($I_{\CSIT} I_{\JSIT}$) & Theorem \\ \hline
&Perfect&$\PP$& \\
Perfect&Delayed&$\PD$& Theorem~\ref{TheoremPP}\\
&None&$\PN$& \\ \hline
&Perfect&$\DP$& Theorem~\ref{TheoremDP}\\ 
Delayed&Delayed&$\DD$& Theorem~\ref{TheoremDD}\\ 
&None&$\DN$& Theorem~\ref{TheoremDN} [inner bound]\\ \hline
&Perfect&$\NP$& Theorem~\ref{TheoremNP}\\ 
None&Delayed&$\ND$& Theorem~\ref{TheoremND} [inner bound]\\ 
&None&$\NN$& Theorem~\ref{TheoremNN}\\ \hline
\end{tabular}
\caption{$\CSIT,\JSIT$ configurations and corresponding theorems.}
\label{Theorem_table}
\end{table}

\section{Achievability Proofs}\label{Schemes}
Here, we present the transmission schemes achieving the bounds mentioned in Theorems~\ref{TheoremPP}-\ref{TheoremND}. 

\subsection{Perfect $\CSIT$}
In this sub-section schemes achieving the $\DoF$ for $\PP$, $\PD$ and $\PN$ configurations are discussed.
It is clear that the following ordering holds: 
\begin{align}\label{DoF_PCSIT_Compare}
\DoF_{\PN} \subseteq \DoF_{\PD} \subseteq \DoF_{\PP},
\end{align}  
i.e, the $\DoF$ is never reduced when $\mathsf{JSI}$ $(i.e., S(t))$ is available at the transmitter. 
%Therefore, we show the achievability of the pair $(d_{1},d_{2})=(\lambda_{1}, \lambda_{2})$ with PP and $\PN$ configurations. 

\subsubsection{Perfect $\CSIT$, Perfect $\JSIT$ ($\PP$):}
In this configuration the transmitter has perfect and instantaneous knowledge of $\CSIT$ and $\JSIT$. Further, since the jammers' states ($4$ in this case) 
are i.i.d across time, the transmitter's strategy in this configuration is also independent across time. This is further explained below. 
\begin{itemize}
\item When $S(t)=11$, i.e., when both the receivers are jammed, the transmitter does not send any information symbols to the receivers 
as they are completely disrupted by the jamming signals. 
\item When $S(t)=01$, i.e., the case when only the $2$nd receiver is jammed and the $1$st receiver is un-jammed, the transmitter sends 
\begin{align}
\mathbf{X}(t)=\left[\begin{matrix}a \\ 0\end{matrix} \right], 
\end{align}
where $a$ is an information symbol intended for the $1$st receiver. In this case, the receiver $1$ gets 
\begin{align}
Y_1(t)=\mathbf{H}_1(t)\mathbf{X}(t)+N_1(t)\equiv h_{11}(t)a+N_1(t),
\end{align}
and the $2$nd receiver gets 
\begin{align}
Y_2(t)=\mathbf{H}_2(t)\mathbf{X}(t)+\mathbf{G}_2(t)\mathbf{J}(t)+N_2(t).
\end{align}
The $2$nd receiver cannot recover its symbols because it is disrupted by the jamming signals. However, since the $1$st receiver is un-jammed, it can recover the intended symbols within noise distortion\footnote{Throughout the paper, it is assumed that the receivers are capable of recovering their symbols within noise distortion whenever they are not jammed (a valid assumption given that the $\DoF$ characterization is done for $P_T\rightarrow \infty$). }. 
\item $S(t)=10$, i.e., the case when only the $1$st receiver is jammed and the $2$nd receiver is un-jammed. This is the converse case of the jammers' state $S(t)=01$. In this scenario, the transmitter sends 
\begin{align}
\mathbf{X}(t)=\left[\begin{matrix}0 \\ b\end{matrix} \right],
\end{align}
where $b$ is an information symbol intended for the $2$nd receiver. The $2$nd receiver can recover the symbol $b$ within noise distortion. 
\item Finally, for the jammer state $S(t)=00$, i.e., none of the receivers are jammed, the transmitter can increase the $\DoF$ by sending symbols to both the receivers. This is achieved by using the knowledge of the perfect and instantaneous channel state information. In such a scenario, the transmitter employs a pre-coding based zero-forcing transmission strategy as illustrated below. The transmitter sends 
\begin{align}
\mathbf{X}(t)=\mathbf{B}_1(t)a+\mathbf{B}_2(t)b,%\left[\begin{matrix}B_1(t)a \\ B_2(t)b\end{matrix} \right] 
\end{align}
where $\mathbf{B}_1(t)$ and $\mathbf{B}_2(t)$ are $2\times 1$ auxiliary pre-coding vectors such that $\mathbf{H}_{1}(t)\mathbf{B}_2(t)=0$ and 
$\mathbf{H}_{2}(t)\mathbf{B}_1(t)=0$ (i.e, there is no interference caused at a user due to the un-intended information symbols). Thus, the received signals at the users are given by 
\begin{align}
Y_{1}(t)&= \mathbf{H}_{1}(t)\mathbf{B}_1(t)a + N_{1}(t)\\
Y_{2}(t)&= \mathbf{H}_{2}(t)\mathbf{B}_2(t)b + N_{2}(t)\\
\end{align}
which are decoded at the receivers using available $\CSIR$ (jamming signal $\mathbf{J}(t)$ is not present in the received signal since $S_1(t)=S_2(t)=0$). 
%When both the receivers are jammed, the transmitter does not transmit to any receiver. 
\end{itemize}
Based on the above transmission scheme, it is seen that each receiver can decode the intended information symbols whenever they are not jammed. Since, the $1$st receiver is not jammed in the states $S(t)=00$ and $S(t)=01$, which happen with probabilities $\lambda_{00},\lambda_{01}$ respectively (i.e., it can recover symbols for $\lambda_{00}+\lambda_{01}$ fraction of the total transmission time), the $\DoF$ achieved is $\lambda_1=\lambda_{00}+\lambda_{01}$. Similarly, the $\DoF$ achieved by the $2$nd receiver is $\lambda_2=\lambda_{00}+\lambda_{10}$. Thus the $\DoF$ pair $(\lambda_1,\lambda_2)$ described by Theorem~\ref{TheoremPP} is achieved using this transmission scheme.

\subsubsection{Perfect $\CSIT$, Delayed $\JSIT$ ($\PD$):}
Unlike in the $\PP$ configuration, the transmitters' strategy in the $\PD$ configuration is not independent (or not separable) across various time instants due to the unavailability of instantaneous $\JSIT$. However, we show that using the knowledge of perfect and instantaneous $\CSIT$ and the delayed knowledge of $\JSIT$, the $\DoF$ pair  $(d_{1},d_{2})=(\lambda_{1}, \lambda_{2})$ can still be achieved. Since the transmitter has delayed knowledge about the jammers strategy, it adapts its transmission scheme at time $t$ based on the feedback it receives about the jammers' strategy at time $t-1$ i.e., $S(t-1)$. This transmission scheme is briefly explained here.%Hence, owing to the perfect $\JSIT$/$\CSIT$, transmit power can be conserved in the PP configuration.

Let $\{a_1,a_2\}$ denote the symbols to be sent to the $1$st receiver and $\{b_1,b_2\}$ to the $2$nd receiver. Since the transmitter has perfect knowledge about the channel or $\CSIT$, it creates pre-coding vectors $\mathbf{B}_1(t)$ and $\mathbf{B}_2(t)$ such that $\mathbf{H}_{1}(t)\mathbf{B}_2(t)=0$ and $\mathbf{H}_{2}(t)\mathbf{B}_1(t)=0$ (similar to the $\PP$ configuration).  For example, at $t=1$, it sends 
\begin{align}
\mathbf{X}(1)=\mathbf{B}_1(1)a_1+\mathbf{B}_2(1)b_1.
\end{align}
\begin{itemize}
\item If the d-$\JSIT$ about the jammer's state at $t=1$ indicates that none of the receivers were jammed i.e., $S(1)=00$, then the transmitter sends new symbols $a_2$ and $b_2$ as 
\begin{align}
\mathbf{X}(2)=\mathbf{B}_1(2)a_2+\mathbf{B}_2(2)b_2,
\end{align}
at time $t=2$ because both the receivers can decode their intended symbols $a_1$ and $b_1$ within noise distortion in the absence of jamming signals. 
\item If the jammer's state at $t=1$ suggests that only the $1$st receiver was jammed i.e., $S(t)=10$, then the transmitter sends 
\begin{align}
\mathbf{X}(2)=\mathbf{B}_1(2)a_1+\mathbf{B}_2(2)b_2,
\end{align}
in order to deliver the undelivered symbol to the $1$st receiver and a new symbol for the $2$nd receiver (since it was not jammed at $t=1$). 
\item When the feedback about the jammers' state at $t=1$ indicates that $S(1)=01$, the coding scheme used when $S(t)=10$ is reversed (roles of the receivers are flipped) and the transmitter sends a new symbol to the $1$st receiver and the undelivered symbol to the $2$nd receiver as
\begin{align}
\mathbf{X}(2)=\mathbf{B}_1(2)a_2+\mathbf{B}_2(2)b_1.
\end{align}
\item If both the receivers were jammed i.e., $S(1)=11$, then the transmitter re-transmits the symbols for the both the receivers as 
\begin{align}
\mathbf{X}(2)=\mathbf{B}_1(2)a_1+\mathbf{B}_2(2)b_1.
\end{align}
\end{itemize}
By extending this transmission scheme to multiple time instants, the $\DoF$ described by Theorem~\ref{TheoremPP} is also achieved in the $\PD$ configuration (since the receivers $1$ and $2$ get jamming free symbols whenever they are not jammed which happen with probabilities $\lambda_1$ and $\lambda_2$ respectively).

\subsubsection{Perfect $\CSIT$, No $\JSIT$ ($\PN$):}
In this section, we sketch the achievability of the pair $(d_{1},d_{2})=(\lambda_{1}, \lambda_{2})$ for the $\PN$ configuration. We first note that for a scheme of block length $n$, for sufficiently large $n$, only $\lambda_k n$ symbols will be received cleanly (i.e., not-jammed) at receiver $k$, since at each time instant the $k$th receiver gets a jamming free signal with probability $\lambda_k$. As the transmitter is statistically aware of  jammers' strategy, it only sends $\lambda_k n$ symbols for receiver $k$ over the entire transmission period. It overcomes the problem of no feedback by sending pre-coded random linear combinations (LC) of these $\{\lambda_k n\}_{k=1,2}$ symbols at each time instant. Notice here the difference between the schemes suggested for the $\PD$ and $\PN$ configurations. Due to the availability of $\JSIT$, albeit in a delayed manner in the $\PD$ configuration, the transmitter can deliver information symbols to the receivers in a timely fashion without combining the symbols. This is not the case in the $\PN$ configuration. The proposed scheme for $\PN$ configuration is illustrated below. 

Let $\{a_{j}\}_{j=1}^{\lambda_{1}n}$ and $\{b_{j}\}_{j=1}^{\lambda_{2}n}$ denote the information symbols intended to be sent to receiver $1$ and $2$ respectively.
Having the knowledge of $\{\mathbf{H}_{1}(t), \mathbf{H}_{2}(t)\}$, the transmitter sends the following input at time $t$:
\begin{align}
\mathbf{X}(t)=\mathbf{B}_1(t)f_{t}(a_{1},\ldots,a_{\lambda_{1}n})+ \mathbf{B}_2(t)g_{t}(b_{1},\ldots,b_{\lambda_{2}n}), 
\end{align}
where $f_{t}(\cdot), g_{t}(\cdot)$ are \textit{random} linear combinations\footnote{The random coefficients are assumed to be known at the receivers. The characterization of the overhead involved in this process is beyond the scope of this paper. } of the respective $\lambda_{1}n$ and $\lambda_{2}n$ symbols; and the $\mathbf{B}_1(t)$, $\mathbf{B}_2(t)$ are $2\times 1$ precoding vectors (similar to the ones used in $\PP$ and $\PD$ configurations). Thus, the received signals at time $t$ are given as 
\begin{align}
Y_{1}(t)\hspace{-0.1cm}&= \hspace{-0.1cm}\mathbf{H}_{1}(t)\mathbf{B}_1(t)f_{t}(a_{1},..,a_{\lambda_{1}n})\hspace{-0.05cm}+\hspace{-0.05cm} S_1(t)\mathbf{G}_{1}(t)\mathbf{J}(t) +N_{1}(t)\nonumber\\
Y_{2}(t)\hspace{-0.1cm}&= \hspace{-0.1cm}\mathbf{H}_{2}(t)\mathbf{B}_2(t)g_{t}(b_{1},..,b_{\lambda_{2}n})\hspace{-0.05cm} +\hspace{-0.05cm} S_2(t)\mathbf{G}_{2}(t)\mathbf{J}(t) +N_{2}(t).\nonumber
\end{align}
Each receiver can decode all these symbols upon successfully receiving $\lambda_k n$ linearly independent combinations\footnote{ Note here that in order to be able to decode all $\lambda_{1}n$ symbols, we need $\lambda_{1}n$ linearly independent combinations of $\lambda_{1}n$ symbols. For example to be able to decode $a_1,a_2,a_3$ , 3 LCs say $f_1(a_1,a_2, a_3),f_2(a_1,a_2,a_3),f_3(a_1,a_2, a_3)$ are sufficient.} transmitted using the zero-forcing strategy. 
Using this scheme, each receiver can decode $\lambda_k n$ symbols over $n$ time instants using the received $\lambda_k n$ LCs. 
Hence $(d_{1}, d_{2})=(\lambda_1, \lambda_2)$ is achievable. The proposed scheme is in similar spirit to the random network coding used in broadcast packet erasure channels where the receivers collect sufficient number of packets before being able to decode their intended information (see \cite{ChihChunWang2012}, \cite{Erasure} and references therein).

\begin{remark}{\em For all possible ($\CSIT$, $\JSIT$) configurations, the $\DoF$ pairs: $(d_{1}, d_{2})=(\lambda_1,0)$ and $(d_{1}, d_{2})=(0, \lambda_2)$ 
are achievable. This is possible via a simple scheme in which the transmitter sends random LC's of $\lambda_{k}n$ symbols to only the $k$th receiver throughout the transmission interval. The $k$th receiver can decode $\lambda_{k}n$ symbols in $n$ time slots given the fact that it receives jamming free LCs with probability $\lambda_k$.  As Theorem \ref{TheoremNN} suggests, for the case in which the transmitter has neither $\mathsf{CSI}$ nor $\mathsf{JSI}$ (i.e., in the $\NN$ configuration), the optimal strategy is to alternate between transmitting symbols exclusively to only one receiver. }
\end{remark}

\begin{remark}
{\em Although the $\PP$, $\PD$ and $\PN$ configurations are equivalent in terms of the achievable $\DoF$ region, they may not be equivalent in terms of the achievable capacity region. For instance, it can be seen in the $\PN$ configuration that the intended symbols can be decoded only after sufficient linear combinations of the intended symbols are received. However, this is not the case in the other configurations. In $\PP$ and $\PD$ configurations, the receivers can decode their intended symbols instantaneously whenever they are not jammed. Thus with respect to the receivers, the decoding delay is maximum in the case of $\PN$ configuration while it is the least in the $\PP$ and $\PD$ configurations. In addition, with respect to the transmitter, re-transmissions are not required in the $\PP$ configuration while they are necessary in the case of the $\PD$ and $\PN$ configurations to ensure that the receivers get their intended symbols. Thus it must not be confused that the $\PP$, $\PD$ and $\PN$ configurations are equivalent. }
\end{remark}

\subsection{Delayed $\CSIT$}\label{DCSIT}
The $\DoF$ region of a 2-user MISO BC using delayed-$\CSIT$ has been studied in the absence of a jammer \cite{MAT2012}. A 3-stage scheme was proposed by the authors in \cite{MAT2012} to increase the optimal $\DoF$ from 1 (no $\CSIT$) to $\frac{4}{3}$. We briefly explain this scheme here. 
\subsubsection{Scheme achieving $\DoF =\frac{4}{3}$ in the absence of jamming} 
At $t=1$, the transmitter sends 
\begin{equation}
\mathbf{X}(1)=\left[\begin{matrix}a_1 \\ a_2\end{matrix} \right],
\end{equation}
where $a_1,a_2$ are symbols intended for the $1$st receiver.
The outputs at the receivers (within noise distortion) at $t=1$ are given as 
\begin{align}
Y_1(1)&=\mathbf{H}_1(1)\left[\begin{matrix}a_1 \\ a_2\end{matrix} \right] = h_{11}(1)a_1+h_{21}(1)a_2\triangleq \mathcal{F}_1(a_1,a_2) \\
Y_2(1)&=\mathbf{H}_2(1)\left[\begin{matrix}a_1 \\ a_2\end{matrix} \right] = h_{12}(1)a_1+h_{22}(1)a_2\triangleq \mathcal{F}_2(a_1,a_2),
\end{align}
where $\mathbf{H}_{k}(t)=[h_{1k}(t)\quad h_{2k}(t)]$ for $k=1,2$ and $h_{1k}(t)$, $h_{2k}(t)$ represent the channel between the $2$ transmit antennas and the $k$th receive antenna. The LC at $2$nd receiver is not discarded, instead it is used as side information in Stage $3$. In Stage $2$ the transmitter creates a symmetric situation at the $2$nd receiver by transmitting $b_1,b_2$, the symbols intended for the $2$nd receiver. 
\begin{equation}
\mathbf{X}(2)=\left[\begin{matrix}b_1 \\ b_2\end{matrix} \right].
\end{equation}
The outputs at the receivers at $t=2$ are given as 
\begin{align}
Y_1(2)&=\mathbf{H}_1(2)\left[\begin{matrix}b_1 \\ b_2\end{matrix} \right] = h_{11}(2)b_1+h_{21}(2)b_2\triangleq \mathcal{G}_1(b_1,b_2) \\
%&\triangleq \mathcal{G}_1(b_1,b_2) \\
Y_2(2)&=\mathbf{H}_2(2)\left[\begin{matrix}b_1 \\ b_2\end{matrix} \right] = h_{12}(2)b_1+h_{22}(2)b_2\triangleq \mathcal{G}_2(b_1,b_2).
%&\triangleq \mathcal{G}_2(b_1,b_2).
\end{align}
Similar to stage $1$, the undesired LC at receiver $1$ is not discarded. The transmitter is aware of the LCs $\mathcal{F}_1,\mathcal{F}_2,\mathcal{G}_1,\mathcal{G}_2$ via delayed $\CSIT$. At this point, each receiver has one LC that is not intended for them, but is useful if it is delivered at the other receiver. Having access to $\mathcal{F}_2$ along with $\mathcal{F}_1$ will enable the $1$st receiver to decode its intended symbols. Similarly, the $2$nd receiver can decode its $b$-symbols using $\mathcal{G}_1$ and $\mathcal{G}_2$. 
To achieve this, the transmitter multicasts 
\begin{equation}
\mathbf{X}(3)=\left[\begin{matrix}\mathcal{F}_2(a_1,a_2)+\mathcal{G}_1(b_1,b_2) \\ 0 \end{matrix} \right]
\end{equation}
at $t=3$ to the receivers. Upon successfully receiving this symbol within noise distortion, 
the receivers can recover $\mathcal{F}_2(a_1,a_2)$ and $\mathcal{G}_1(b_1,b_2)$ using the available side information (the side information can be cancelled from the new LC). Thus each receiver has $2$ LCs of $2$ intended symbols. Using this transmission scheme the receivers can decode $2$ symbols each in $3$ time slots. Thus the optimal $\DoF$ $(\frac{2}{3},\frac{2}{3})$ is achieved using this transmit strategy. Hereafter, this scheme is referred to as the ``$\mathsf{MAT}$ scheme''. 

%Having seen this scheme, it should be noted by the reader that this transmission scheme cannot be directly extended to the current scenario when delayed $\CSIT$ alone is available in the presence of jamming attacks. 
Below we present transmission schemes to achieve optimal $\DoF$ in the presence of jamming signals, specifically in scenarios where the jamming state information ($\JSIT$) is either available instantaneously or with a delay or is not available i.e., for the $\DP$, $\DD$ and $\DN$ configurations. The following relationship holds true,
\begin{align}\label{DoF_DCSIT_Compare}
\DoF_{\DN} \subseteq \DoF_{\DD} \subseteq \DoF_{\DP}.
\end{align}
%In this section we present the achievability schemes for the $\DP$, $\DD$ and $\DN$ configurations.

\subsubsection{Delayed $\CSIT$, Perfect $\JSIT$ ($\DP$):}
\begin{figure}[t]
  \centering
\includegraphics[width=11.0cm]{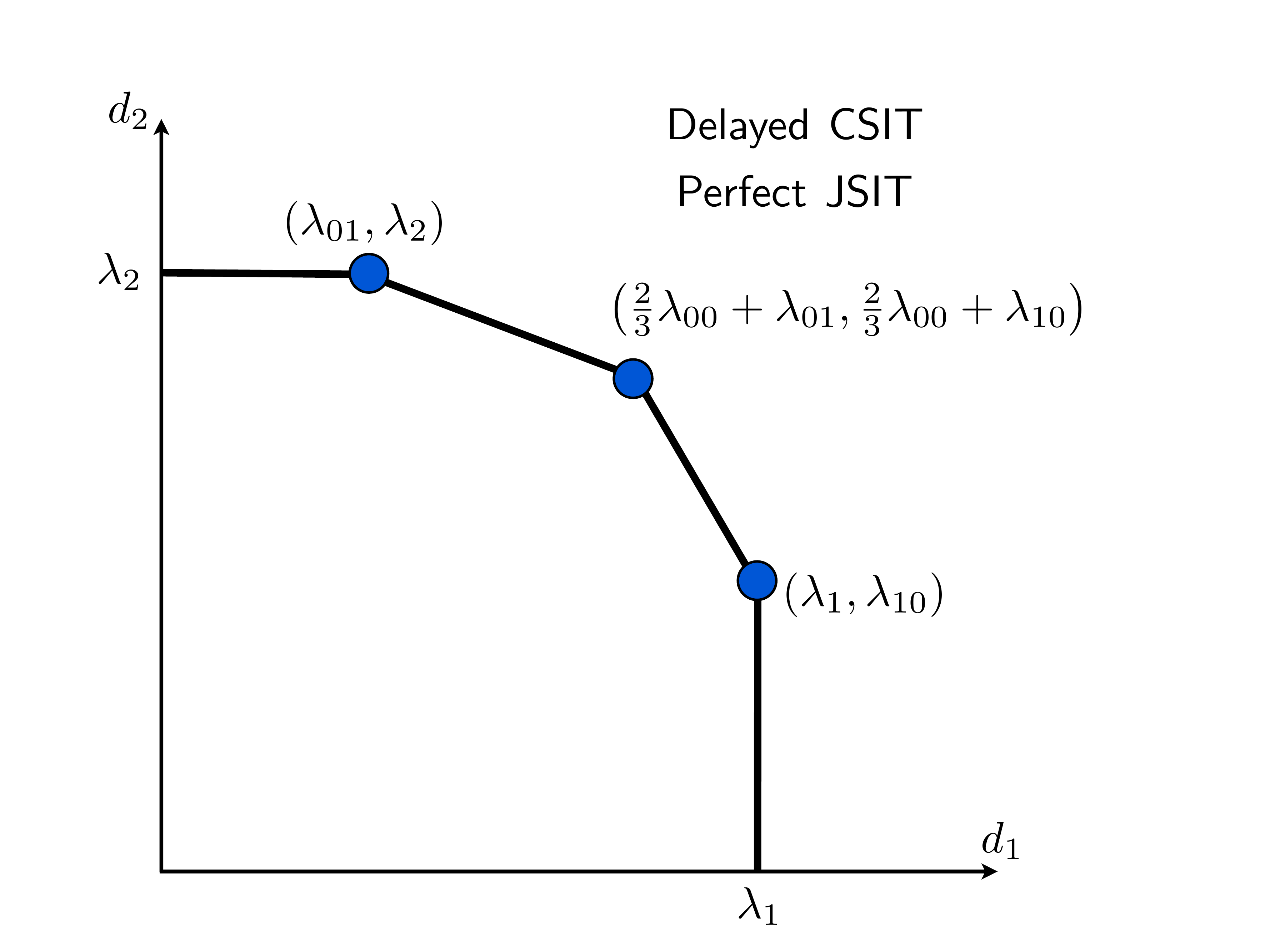}
\caption{$\DoF$ region with delayed $\CSIT$ and perfect $\JSIT$.}\label{Fig:FigureDP}
\end{figure}

As seen in Fig.~\ref{Fig:FigureDP}, the following $\DoF$ pairs $(d_1,d_2)=(\lambda_1,0)$,
$(\lambda_1,\lambda_{10}), \ (\frac{2}{3}\lambda_{00}+\lambda_{01},\frac{2}{3}\lambda_{00}+\lambda_{10})$ and $ (\lambda_{01},\lambda_2), \ (0,\lambda_2)$ are achievable in the $\DP$ configuration. The $\DoF$ pairs $(\lambda_1,0)$ and $(0,\lambda_2)$ are readily achievable by transmitting to only receiver $1$ (resp. receiver $2$). Here, we present transmission schemes to achieve the $\DoF$ pairs $(\frac{2}{3}\lambda_{00}+\lambda_{01},\frac{2}{3}\lambda_{00}+\lambda_{10})$, $(\lambda_1,\lambda_{10})$ and $(\lambda_{01},\lambda_2)$. 

Due to the availability of perfect $\JSIT$, the transmitters strategy is independent across time i.e., the transmitter uses a different strategy based on the jammers' state. Thus the transmission scheme can be divided into $4$ different strategies based on the jammers' state $S(t)$ which is detailed below. 
\begin{itemize}
\item When the jammers' state $S(t)=00$, the transmitter uses the $\mathsf{MAT}$ scheme which was described earlier. Since this state is seen with 
probability $\lambda_{00}$ and the $\DoF$ achieved by the $\mathsf{MAT}$ scheme in the presence of delayed $\CSIT$ is $\left(\frac{2}{3},\frac{2}{3}\right)$, the overall $\DoF$ achieved whenever this jammer state is seen is given by $\left(\frac{2}{3}\lambda_{00},\frac{2}{3}\lambda_{00}\right)$. 

Instead of using the $\mathsf{MAT}$ scheme, if the transmitter chooses to send symbols exclusively to only one receiver, then the $\DoF$ pair $\left(\lambda_{00},0\right)$ or $\left(0,\lambda_{00}\right)$ is achieved depending on whether it chooses the $1$st or the $2$nd receiver (notice the $\DoF$ loss by using this strategy). 
\item When $S(t)=01$, the jammer transmits symbols only to the $1$st receiver (since the $2$nd receiver cannot recover its symbols due to jamming) which can recover the intended symbol within noise distortion. Since this state is seen with probability $\lambda_{01}$, the $\DoF$ achievable in this state is given by $\left(\lambda_{01},0\right)$.
\item The state $S(t)=10$ is the converse of the previous state $S(t)=01$ with the roles of the two receivers flipped. Thus the $\DoF$ achieved in this state is 
$\left(0,\lambda_{10}\right)$. 
\item When the jammers' state is $S(t)=11$, none of the receivers can recover the symbols as their received signals are completely disrupted by the jamming signals. Thus the transmitter does not send symbols whenever this jamming state occurs. 
\end{itemize}
Since the jammers' states are disjoint, the overall $\DoF$ achieved in the $\DP$ configuration is given by the pair $(d_1,d_2)=(\frac{2}{3}\lambda_{00}+\lambda_{01},\frac{2}{3}\lambda_{00}+\lambda_{10})$ if it chooses to use the $\mathsf{MAT}$ scheme. Else the $\DoF$ pairs, 
$(d_1,d_2)=(\lambda_{1},\lambda_{10})$ or  $(d_1,d_2)=(\lambda_{01},\lambda_2)$ are achievable. This completes the achievability scheme for the $\DP$ configuration. Hence, the $\DoF$ region mentioned by Theorem~\ref{TheoremDP} is achieved. 

As mentioned earlier, if perfect $\JSIT$ 	is available, the transmitter \textbf{does not} have to transmit/ code across different jammers' states in order to achieve $\DoF$ gains. In other words, the jammers' states are separable due to availability of perfect $\JSIT$. As will be seen next, this separability no longer holds true in the case of $\DD$ and $\DN$ configurations and hence necessitate transmitting across various jamming states. These transmission schemes thereby introduce decoding delays at the intended receivers.

\subsubsection{Delayed $\CSIT$, Delayed $\JSIT$ ($\DD$):}
\begin{figure}[t]
  \centering
\includegraphics[width=12.0cm]{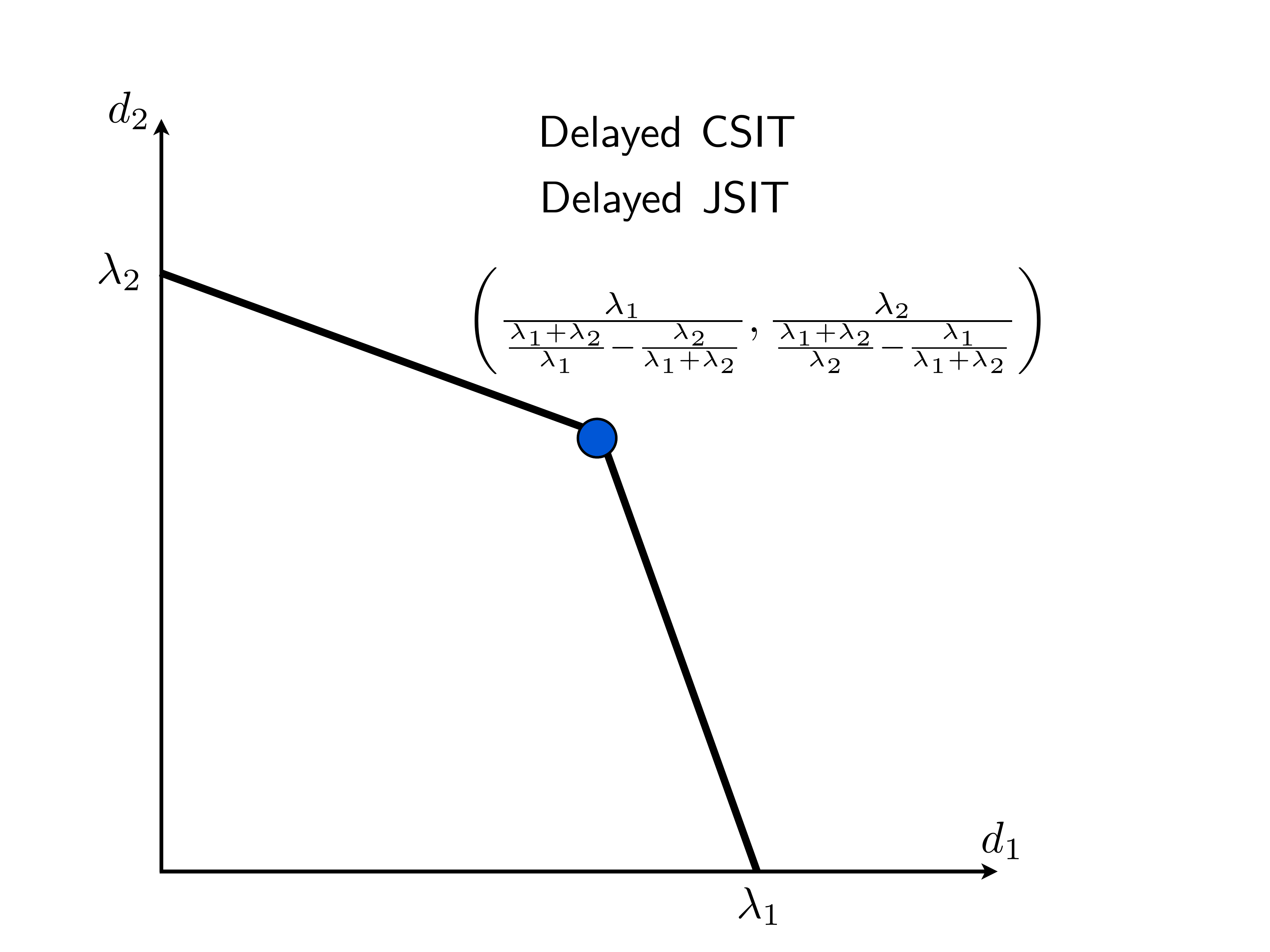}
\vspace{-10pt}
\caption{$\DoF$ region with delayed $\CSIT$ and delayed $\JSIT$ i.e. $\DD$ configuration.}\label{Fig:FigureDD}
\end{figure}

In this subsection, we propose a transmission scheme that achieves the following $(d_{1}, d_{2})$ pair (which corresponds to intersection of (\ref{DD1}) and (\ref{DD2}), see Fig. \ref{Fig:FigureDD}):
\begin{align}\label{DoFDD_opt_point}
(d_{1}, d_{2})&=\left(\frac{\lambda_{1}}{\frac{\lambda_{1}+\lambda_{2}}{\lambda_{1}}-\frac{\lambda_{2}}{\lambda_{1}+\lambda_{2}}}, \frac{\lambda_{2}}{\frac{\lambda_{1}+\lambda_{2}}{\lambda_{2}}-\frac{\lambda_{1}}{\lambda_{1}+\lambda_{2}}}\right).
\end{align}
In this scheme, the decoding process follows once the transmission of the symbols has finished and the receivers have all required linear combinations of the symbols which are used to decode the symbols. %We are specifically not interested in the decoding delays involved in such a scheme. 
The decoding process using the linear combinations is explicitly mentioned in the transmission schemes below. 
%\begin{enumerate}
%\item {\textbf{Algorithm 1}}\\
This algorithm operates in three stages. In stage $1$, the transmitter sends symbols intended only for receiver $1$ and keeps re-transmitting them until they are received within noise distortion (or uncorrupted by the jamming signal) at at least one receiver. In stage $2$, the transmitter sends symbols intended only for receiver $2$ in the same manner. Stage $3$ consists of transmitting the undelivered symbols to the intended receivers. The specific LCs to be transmitted in stage $3$ are determined by the feedback (i.e., d-$\CSIT$ and d-$\JSIT$) received from the stages $1$ and $2$. The eventual goal of the scheme is to deliver $n_{1}$ symbols
(denoted by $\{a_{j}\}_{j=1}^{n_{1}}$; or $a$-symbols) to receiver $1$ and  $n_{2}$ symbols (denoted by $\{b_{j}\}_{j=1}^{n_{2}}$; or $b$-symbols) to receiver $2$.

Below we explain the 3-stages involved in the proposed transmission scheme. 

%
%At the end of stages 1 and 2, the transmitter intends to deliver sufficient linear combinations that enable the receivers to decode their intended symbols. However, due to jamming, the receivers do not have access to all the linear combinations necessary to decode their intended symbols. Hence, in order that the receivers are capable of decoding all the symbols, an additional stage is necessary. 

\textbf{\textit{Stage 1}}--In this stage, the transmitter intends to deliver $n_1$ $a$-symbols, in a manner such that each $a$-symbol is received at \textit{at least} one of the receivers (either $1$st or $2$nd receiver). At every time instant the transmitter sends two symbols on two transmit antennas. A pair of symbols (say $a_1$ and $a_2$) are re-transmitted until they are received at at least one receiver (this knowledge is available via d-$\JSIT$). Any one of the following four scenarios can arise: 
\begin{enumerate}
\item \textit{Event $00$}: none of the receivers are jammed (which happens with probability $\lambda_{00}$). As an example, suppose that at time $t$, if the transmitter sends $(a_{1}, a_{2})$: then receiver $1$ gets $\mathcal{F}_{1}(a_{1},a_{2})$ and receiver $2$ gets $\mathcal{F}_{2}(a_{1},a_{2})$. The fact that the event $00$ occurred at time $t$ is known at time $t+1$ via d-$\JSIT$; and the LCs $(\mathcal{F}_{1}(a_{1},a_{2}), \mathcal{F}_{2}(a_{1},a_{2}))$ can be obtained at the transmitter at time $t+1$ via d-$\CSIT$. The goal of stage $3$ would be to deliver $\mathcal{F}_{2}(a_{1},a_{2})$ to receiver $1$ by exploiting the fact that it is already received at receiver $2$.  Thus, at time $t+1$, the transmitter sends two new symbols $(a_{3}, a_{4})$. 

\item \textit{Event $01$}: receiver $1$ is not jammed, while receiver $2$ is jammed (which happens with probability $\lambda_{01}$). As an example, suppose that at time $t$, if the transmitter sends $(a_{1}, a_{2})$: then receiver $1$ gets $\mathcal{F}_{1}(a_{1},a_{2})$ and receiver $2$'s signal is drowned in the jamming signal. The fact that the event $01$ occurred at time $t$ is known at time $t+1$ via d-$\JSIT$; and the LC $\mathcal{F}_{1}(a_{1},a_{2})$ can be obtained at the transmitter at time $t+1$ via d-$\CSIT$. Thus, at time $t+1$, the transmitter sends a fresh symbol $a_{3}$ on one antenna; and a LC of $(a_{1},a_{2})$; say $\tilde{\mathcal{F}}_{1}(a_{1},a_{2})$; such that $\mathcal{F}_{1}(a_{1},a_{2})$ and $\tilde{\mathcal{F}}_{1}(a_{1},a_{2})$ constitute two linearly independent combinations of $(a_{1}, a_{2})$. In summary, at time $t+1$, the transmitter sends $(a_{3}, \tilde{\mathcal{F}}_{1}(a_{1},a_{2}))$. 

\item \textit{Event $10$}: receiver $2$ is not jammed, while receiver $1$ is jammed (which happens with probability $\lambda_{10}$). As an example, suppose that at time $t$, if the transmitter sends $(a_{1}, a_{2})$: then receiver $1$'s signal is drowned in the jamming signal, whereas receiver $2$ gets $\mathcal{F}_{2}(a_{1},a_{2})$. The fact that the event $10$ occurred at time $t$ is known at time $t+1$ via d-$\JSIT$; and the LC $\mathcal{F}_{2}(a_{1},a_{2})$ can be obtained at the transmitter at time $t+1$ via d-$\CSIT$. The goal of stage $3$ would be to deliver $\mathcal{F}_{2}(a_{1},a_{2})$ to receiver $1$ by exploiting the fact that it is already received at receiver $2$. Thus, at time $t+1$, the transmitter sends a fresh symbol $a_{3}$ on one antenna; and a LC of $(a_{1},a_{2})$; say $\tilde{\mathcal{F}}_{2}(a_{1},a_{2})$; such that $\mathcal{F}_{2}(a_{1},a_{2})$ and $\tilde{\mathcal{F}}_{2}(a_{1},a_{2})$ constitute two linearly independent combinations of $(a_{1}, a_{2})$. In summary, at time $t+1$, the transmitter sends $(a_{3}, \tilde{\mathcal{F}}_{2}(a_{1},a_{2}))$. 

\item \textit{Event $11$}: both receivers are jammed (which happens with probability $\lambda_{11}$). Using d-$\JSIT$, transmitter knows at time $t+1$ that the event $11$ occurred and hence at time $t+1$, it re-transmits $(a_{1}, a_{2})$ on the two transmit antennas.
\end{enumerate}
The above events are \emph{disjoint}, so in one time slot, the average number of useful LCs 
delivered to \textbf{at least one receiver} is given by 
\begin{equation}\label{expected_symbols_DD}
E[\textnormal{$\#$ of LC's delivered}] =2\lambda_{00}+\lambda_{01}+\lambda_{10}\triangleq \phi. \nonumber
\end{equation}
Hence, the expected time to deliver one LC is 
\begin{equation}
\frac{1}{\phi}=\frac{1}{2\lambda_{00}+\lambda_{01}+\lambda_{10}}\triangleq \frac{1}{\lambda_1+\lambda_2}.
\end{equation}
The time spent in this stage to deliver $n_1$ LCs is 
\begin{align}\label{N_1}
N_1=\frac{n_1}{\lambda_1+\lambda_2}.
\end{align}
Since receiver $1$ is not jammed in events $00$ and $01$, i.e., for $\lambda_1$ fraction of the time, it receives only $\lambda_1 N_1$ LCs. The number of undelivered LCs is $n_1-\lambda_1N_1=\frac{\lambda_2n_1}{\lambda_1+\lambda_2}$. These LCs are available at receiver $2$ (corresponding to events $00$ and $10$) and are known to the transmitter via d-$\CSIT$. This side information created at receiver $2$ is not discarded, instead it is used in Stage 3 of the transmission scheme. 

{\textbf{\textit{Stage 2}}}--
In this stage, the transmitter intends to deliver $n_2$ $b$-symbols, in a manner such that each symbol is received at \textit{at least} one of the receivers. Stage 1 is repeated here with the roles of the receivers 1 and 2 interchanged. On similar lines to Stage 1, the time spent in this stage is 
\begin{align}\label{N_2}
N_2=\frac{n_2}{\lambda_1+\lambda_2}.
\end{align} 
The number of LCs received at receiver $2$ is $\lambda_2N_2$ and the number of LCs not delivered to receiver $2$ but are available as side information at receiver $1$ is $n_2-\lambda_2N_2=\frac{\lambda_1n_2}{\lambda_1+\lambda_2}$. 

\begin{remark} {\em At the end of these $2$ stages, following typical situation arises: $\mathcal{F}(a_1,a_2)$ (resp. $\mathcal{G}(b_1,b_2)$) is a LC intended for receiver $1$ (resp. $2$) but is available as side information at receiver $2$ (resp. $1$)\footnote{Such situations correspond to events $00$ and $01$ in Stage $1$; and events $00$, $10$ in Stage $2$.}. Notice that these LCs must be transmitted to the complementary receivers so that the desired symbols can be decoded. In Stage $3$, the transmitter sends a random LC of these symbols, say $\mathcal{L}=l_1\mathcal{F}(a_1,a_2)+l_2\mathcal{G}(b_1,b_2)$ where $l_1, l_2$ that form the new LC are known to the transmitter and receivers \emph{a priori}. Now, assuming that only receiver $2$ (resp. $1$) is jammed, $\mathcal{L}$ is received at receiver $1$ (resp. $2$) within noise distortion. Using this LC, it can recover $\mathcal{F}(a_1,a_2)$ (resp. $\mathcal{G}(b_1,b_2)$) from $\mathcal{L}$ since it already has $\mathcal{G}(b_1,b_2)$ (resp. $\mathcal{F}(a_1,a_2)$) as side information. When no receiver is jammed, both the receivers are capable of recovering $\mathcal{F}(a_1,a_2)$,  $\mathcal{G}(b_1,b_2)$ simultaneously.}
\end{remark}
%This technique of transmitting random LCs of previously received symbols, to deliver the side information, as mentioned earlier in Section~\ref{Theorems}, is similar to the transmission schemes used in network coding and packet erasure channels with delayed information about the channel.

{\textbf{\textit{Stage 3}}}--In this stage, the undelivered LCs to each receiver are transmitted using the technique mentioned above. 
Let us assume that $\mathcal{F}_1(a_1,a_2)$ and $\mathcal{G}_1(b_1,b_2)$ are LCs available as side information at receivers 2 and 1 respectively. The transmitter sends $\mathcal{L}(\mathcal{F}_1,\mathcal{G}_1)$, a LC of these symbols on one  transmit antenna, with the eventual goal of multicasting this LC (i.e., send it to \textit{both} receivers). The following events, as specified earlier in Stages $1$ and $2$, are also possible while in this stage. 

\textit{Event $00$}: Suppose at time $t$, if the transmitter sends $\mathcal{L}(\mathcal{F}_1,\mathcal{G}_1)$, then both the receivers get this LC within noise distortion. With the capability to recover $\mathcal{L}(\mathcal{F}_1,\mathcal{G}_1)$ within a scaling factor, the receivers 1 and 2 decode their intended LCs $\mathcal{F}_1$ and $\mathcal{G}_1$ respectively using the side informations $\mathcal{G}_1$ and $\mathcal{F}_1$ that are available with them. Since the intended LCs are delivered at the intended receivers, the transmitter, at time $t+1$, sends a new LC of two new symbols $\tilde{\mathcal{L}}(\tilde{\mathcal{F}}_1,\tilde{\mathcal{G}}_1)$. 

\textit{Event $01$}: Since receiver $2$ is jammed, its signal is drowned in the jamming signal while receiver $1$ gets $\mathcal{L}(\mathcal{F}_1,\mathcal{G}_1)$ and is capable of recovering $\mathcal{F}_1$ using $\mathcal{G}_1$ available as side information. The fact that event $01$ occurred is known to the transmitter at time $t+1$ via d-$\JSIT$. Thus, at time $t+1$, the transmitter sends a new LC $\tilde{\mathcal{L}}(\tilde{\mathcal{F}}_1,\mathcal{G}_1)$ since $\mathcal{G}_1$ has not yet been delivered to receiver $2$. 

\textit{Event $10$}: This event is similar to event $01$, with the roles of the receivers 1 and 2 interchanged. Hence, receiver $2$ is capable of recovering $\mathcal{G}_1$ from 
$\mathcal{L}(\mathcal{F}_1,\mathcal{G}_1)$ while receiver $1$'s signal is drowned in the jamming signal. Thus at time $t+1$, the transmitter sends a new LC $\tilde{\mathcal{L}}(\mathcal{F}_1,\tilde{\mathcal{G}}_1)$ since $\mathcal{F}_1$ has not yet been delivered to receiver $1$. 

\textit{Event $11$}: Using d-$\JSIT$, transmitter knows at time $t+1$ that the event $11$ occurred and hence at time $t+1$, it re-transmits $\mathcal{L}(\mathcal{F}_1,\mathcal{G}_1)$ on one of its transmit antennas. 
\begin{figure}[t]
  \centering
\includegraphics[width=13.0cm]{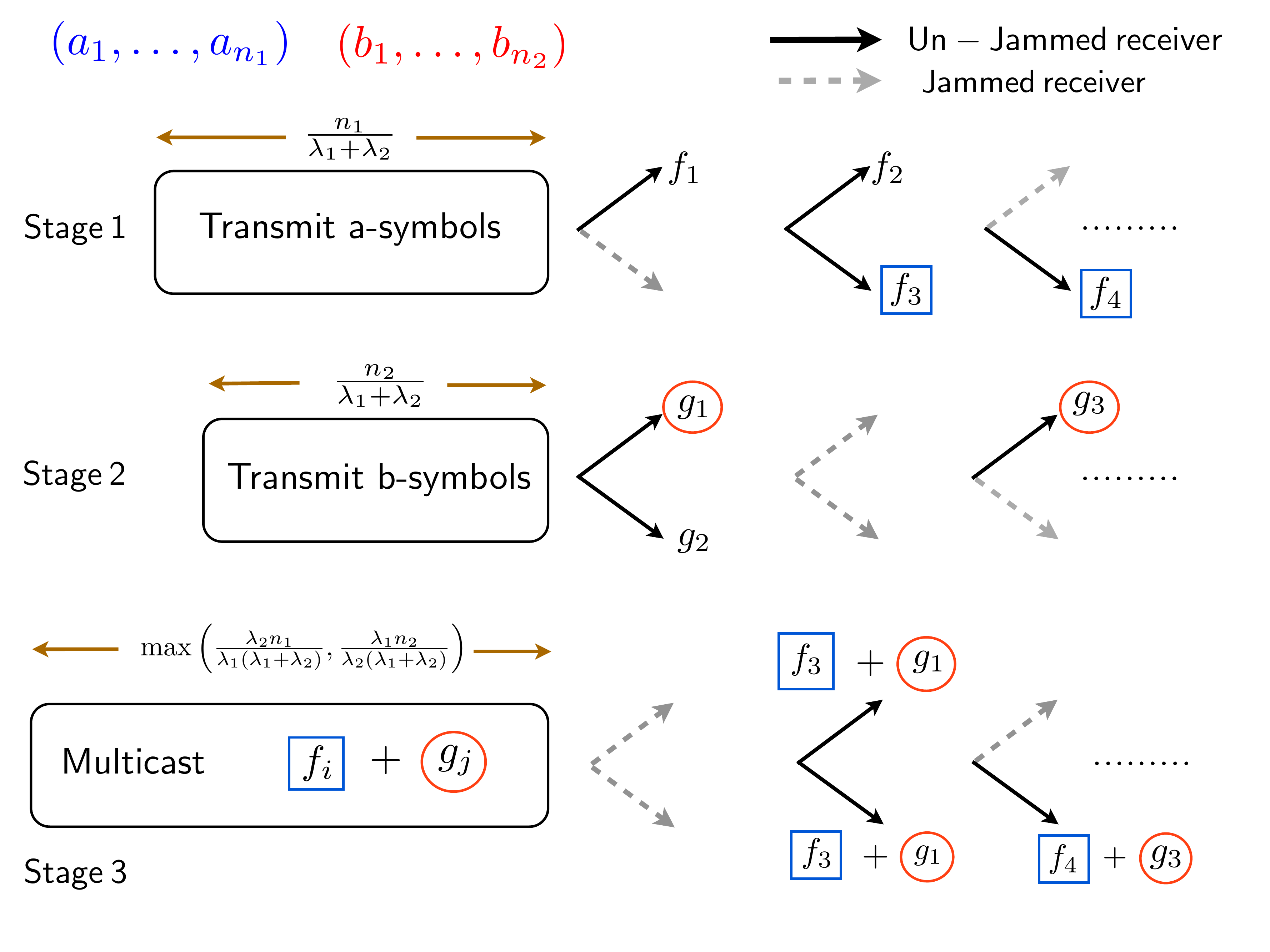}
\caption{Coding with  delayed $\CSIT$ and delayed $\JSIT$.}\label{CodingDD}
\end{figure}

Since, all the events are disjoint, in one time slot, the average number of LCs delivered to receiver $1$ is given by 
\begin{equation}
E[\textnormal{$\#$ of LC's delivered to user 1}] =\lambda_{00}+\lambda_{01}\triangleq \lambda_1. \nonumber
\end{equation}
Hence, the expected time to deliver one LC to receiver $1$ in this stage is $\frac{1}{\lambda_1}$. Given that $\frac{\lambda_2 n_1}{\lambda_1+\lambda_2}$ LCs are to be delivered to receiver $1$ in this stage, the time taken to achieve this is 
$\frac{\lambda_2 n_1}{\lambda_1(\lambda_1+\lambda_2)}$. Interchanging the roles of the users, the time taken to deliver $\frac{\lambda_1 n_2}{\lambda_1+\lambda_2}$ LCs to receiver $2$ is $\frac{\lambda_1 n_2}{\lambda_2(\lambda_1+\lambda_2)}$. Thus the total time required to satisfy the requirements of both the receivers in Stage 3 is given by 
\begin{equation}\label{N_3}
N_3=\mathrm{max}\left(\frac{\lambda_2n_1}{\lambda_1(\lambda_1+\lambda_2)},\frac{\lambda_1n_2}{\lambda_2(\lambda_1+\lambda_2)}\right).
\end{equation}
The optimal $\DoF$ achieved in the $\DD$ configuration is readily evaluated as 
\begin{eqnarray}
d_1=\frac{n_1}{N_1+N_2+N_3}, \ d_2=\frac{n_2}{N_1+N_2+N_3}.
\end{eqnarray}
Substituting for $\{N_i\}_{i=1,2,3}$ from \eqref{N_1}--\eqref{N_3}, we have,
\begin{align}
d_k&=\frac{n_k}{\frac{n_1}{\lambda_1+\lambda_2}+\frac{n_2}{\lambda_1+\lambda_2}+\mathrm{max}\left(\frac{\lambda_2 n_1}{\lambda_1(\lambda_1+\lambda_2)},\frac{\lambda_1n_2}{\lambda_2(\lambda_1+\lambda_2)}\right)},\ k=1,2. \nonumber \\
\end{align}
Using $\eta=\frac{n_1}{n_1+n_2}$, we have 
\begin{align}
d_1&=\frac{\eta}{\frac{1}{\lambda_1+\lambda_2}+\mathrm{max}\left(\frac{\lambda_2\eta}{\lambda_1(\lambda_1+\lambda_2)},\frac{\lambda_1(1-\eta)}{\lambda_2(\lambda_1+\lambda_2)}\right)} \nonumber \\
d_2&=\frac{1-\eta}{\frac{1}{\lambda_1+\lambda_2}+\mathrm{max}\left(\frac{\lambda_2\eta}{\lambda_1(\lambda_1+\lambda_2)},\frac{\lambda_1(1-\eta)}{\lambda_2(\lambda_1+\lambda_2)}\right)}. 
\end{align}
Eliminating $\eta$ from the above two equations, yields the $(d_{1}, d_{2})$ pair given in \eqref{DoFDD_opt_point}. 

\begin{remark}{\em It is seen that only JSI at time $t$ is necessary for the transmitter to make a decision on the LCs to be transmitted at time $t+1$ in Stage 3. Also, it is worth noting that the outer most points on the $\DoF$ region described by Theorem~\ref{TheoremDD} (for a given $\lambda_1$, $\lambda_2$) are obtained for different values of $\eta \in [0,1]$. Another interesting point to note here is that if $\lambda_1=\lambda_2=1$,(which is possible only if $\lambda_{00}=1$) i.e none of the receivers are jammed, the $\DoF$ achieved is $\frac{4}{3}$ which is the optimum $\DoF$ achieved in a d-$\CSIT$ scenario for the 2-user MISO broadcast channel as shown by Maddah-Ali and Tse in \cite{MAT2012}. }
\end{remark}

\subsubsection{Delayed $\CSIT$, No $\JSIT$ ($\DN$):}
One of the novel contributions of this paper is developing a new coding/transmission scheme for the $\DN$ configuration. 
Before we explain the proposed scheme, we first present a modified $\mathsf{MAT}$ scheme (original $\mathsf{MAT}$ scheme proposed in  \cite{MAT2012}) that achieves a $\DoF$ of $\frac{4}{3}$ in a 2-user MISO BC (in the absence of jamming). 

\paragraph{Modified $\mathsf{MAT}$ Scheme:} 
Consider a 2-user MISO BC where the transmitter intends to deliver $a$-symbols ($a_1,a_2$) to the $1$st receiver and $b$-symbols ($b_1,b_2$) to the $2$nd receiver respectively. The $\mathsf{MAT}$ scheme proposed in \cite{MAT2012} was illustrated earlier in Section~\ref{DCSIT}. Here we first revise the modified $\mathsf{MAT}$ scheme to achieve the same results. 

At $t=1$, the transmitter sends 
\begin{equation}
\mathbf{X}(1)=\left[\begin{matrix}a_1+b_1 \\ a_2+b_2\end{matrix} \right],
\end{equation}
on its two transmit antennas. 
The outputs (within noise distortion) at the $2$ receivers are given as (ignoring noise)
\begin{align}
Y_1(1)&=\mathbf{H}_1(1)\left[\begin{matrix}a_1+b_1 \\ a_2+b_2\end{matrix} \right] = h_{11}(1)(a_1+b_1)+h_{21}(1)(a_2+b_2) \nonumber \\
&=\underbrace{(h_{11}(1)a_1+h_{21}a_2)}_{\mathcal{F}_1(a_1,a_2)}+\underbrace{(h_{11}(1)b_1+h_{21}b_2)}_{\mathcal{G}_1(b_1,b_2)} \nonumber \\
&\triangleq \mathcal{F}_1(a_1,a_2)+\mathcal{G}_1(b_1,b_2) \\
Y_2(1)&=\mathbf{H}_2(1)\left[\begin{matrix}a_1+b_1 \\ a_2+b_2\end{matrix} \right] = h_{21}(1)(a_1+b_1)+h_{22}(1)(a_2+b_2) \nonumber \\
&=\underbrace{(h_{21}(1)a_1+h_{21}a_2)}_{\mathcal{F}_2(a_1,a_2)}+\underbrace{(h_{22}(1)b_1+h_{21}b_2)}_{\mathcal{G}_2(b_1,b_2)} \nonumber \\
&\triangleq \mathcal{F}_2(a_1,a_2)+\mathcal{G}_2(b_1,b_2),
\end{align}
where $\mathcal{F}_1,\mathcal{F}_2$ represent LCs of the symbols $a_1,a_2$ and similarly $\mathcal{G}_1,\mathcal{G}_2$ are LCs of the symbols $b_1,b_2$ (the received symbols can be grouped in this manner as the receivers have $\CSIR$.). These LCs are known to the  transmitter at time $t=2$ via d-$\CSIT$. The $1$st receiver requires $\mathcal{F}_2$ (apart from $\mathcal{F}_1$) to decode its symbols and the $2$nd receiver needs $\mathcal{G}_1$ (apart from $\mathcal{G}_2$) for its symbols. Thus at time $t=2$, the transmitter multicasts $\mathcal{G}_1$ to both the receivers on one of its transmit antennas as 
\begin{equation}
\mathbf{X}(2)=\left[\begin{matrix}\mathcal{G}_1(b_1,b_2) \\ 0\end{matrix} \right].
\end{equation}
which is received within noise distortion at both the receivers. Using the recovered $\mathcal{G}_1$ (within noise distortion), the $1$st receiver can recover $\mathcal{F}_1$ by removing it from the symbol $Y_1(1)$ that it received at time $t=1$. At this point, receiver $1$ has one LC of intended symbols $\mathcal{F}_1$ and also needs $\mathcal{F}_2$ to recover its symbols. Thus the transmitter multicasts $\mathcal{F}_2$ to both the receivers at time $t=3$ as 
\begin{equation}
\mathbf{X}(3)=\left[\begin{matrix}\mathcal{F}_2(a_1,a_2)\\ 0\end{matrix} \right].
\end{equation}
Using the same technique as receiver $1$, the $2$nd receiver can recover $\mathcal{G}_2$ by removing $\mathcal{F}_2$ from the symbol  $Y_2(1)$ that it received at time $t=1$. Thus at the end of $3$ time instants, the receivers $1$ and $2$ have $\mathcal{F}_1,\mathcal{F}_2$ and $\mathcal{G}_1,\mathcal{G}_2$ respectively, that help them decode their intended symbols. Thus using this transmission scheme, $4$ symbols are decoded at the receivers in $3$ time slots that leads to a sum $\DoF$ of $\frac{4}{3}$ which is also the $\DoF$ achieved by the $\mathsf{MAT}$ scheme in the 2-user MISO BC with delayed $\CSIT$.

\begin{figure}[t]
  \centering
\includegraphics[width=11.0cm]{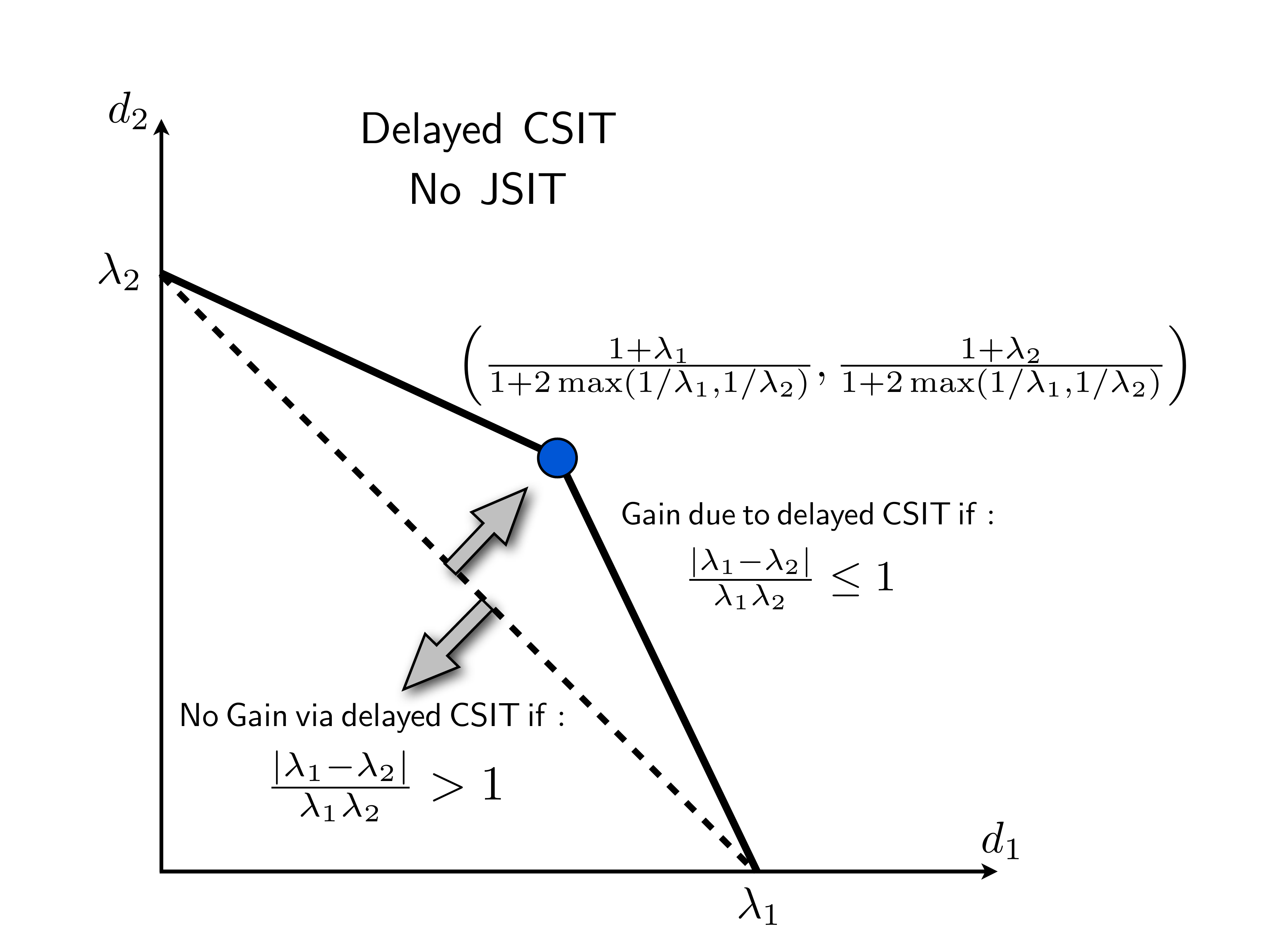}
\caption{Achievable $\DoF$ region with delayed $\CSIT$ and no $\JSIT$.}\label{Fig:FigureDN}
\end{figure}

\paragraph{Proposed Scheme for $\DN$:}It is clearly seen that the modified $\mathsf{MAT}$ scheme presented above cannot be directly extended to the case where the jammer disrupts the receivers. Below, we present a novel 3-stage transmission strategy to achieve the $\DoF$ described by Theorem~\ref{TheoremDN}. The transmitter uses the statistical knowledge of the jammers strategy to deliver symbols to both the receivers in this configuration (as feedback information about the undelivered symbols is not available at the transmitter). Similar to the $\PN$ configuration, the transmitter sends random LCs of the intended symbols to both the users to overcome the unavailability of $\JSIT$. 

Let $(1+\lambda_1)n$ and $(1+\lambda_2)n$ (the reason for choosing $(1+\lambda_k)n$, $k=1,2$, as the length of symbol sequence will be clear as we proceed through the algorithm) denote the total number of symbols the transmitter intends to deliver to receivers $1$ and $2$ respectively, where $\lambda_1,\lambda_2$ indicate the probability with which the receivers are not disrupted by the jammer. In this scheme, we assume that the decoding process follows once the transmission of the symbols has finished and the receivers have all required linear combinations of the symbols which are used to decode the symbols. 
%We are specifically not interested in the decoding delays involved in such a scheme. 
So each receiver needs $(1+\lambda_1)n$, $(1+\lambda_2)n$ LCs respectively to completely decode their symbols. 
\begin{itemize}
\item \textbf{Stage 1}: The transmitter forms random LCs of the $(1+\lambda_1)n$ $a$-symbols and $(1+\lambda_2)n$ $b$-symbols symbols intended for both the receivers. Let us denote these LCs by $(a_1,a_2,\ldots,a_{\left(1+\lambda_1\right)n})$ and $(b_1,b_2,\ldots,b_{\left(1+\lambda_2\right)n})$ respectively (these are the actual transmitted symbols and are similar to the $a$-symbols and $b$-symbols mentioned earlier in the modified $\mathsf{MAT}$ scheme). In Stage 1, the transmitter combines these $a$-symbols and $b$-symbols and sends them over $n$ time instants (please refer to the modified $\mathsf{MAT}$ scheme to see how combination of $a$-symbols and $b$-symbols are sent). Since the receivers $1,2$ are not jammed with a probability $\lambda_1,\lambda_2$ respectively, they receive 
$\lambda_1n$ and $\lambda_2n$ combinations of $a$-symbols and $b$-symbols over $\tau_1=n$ time instants. 
\begin{figure}[t]
  \centering
\includegraphics[width=13.0cm]{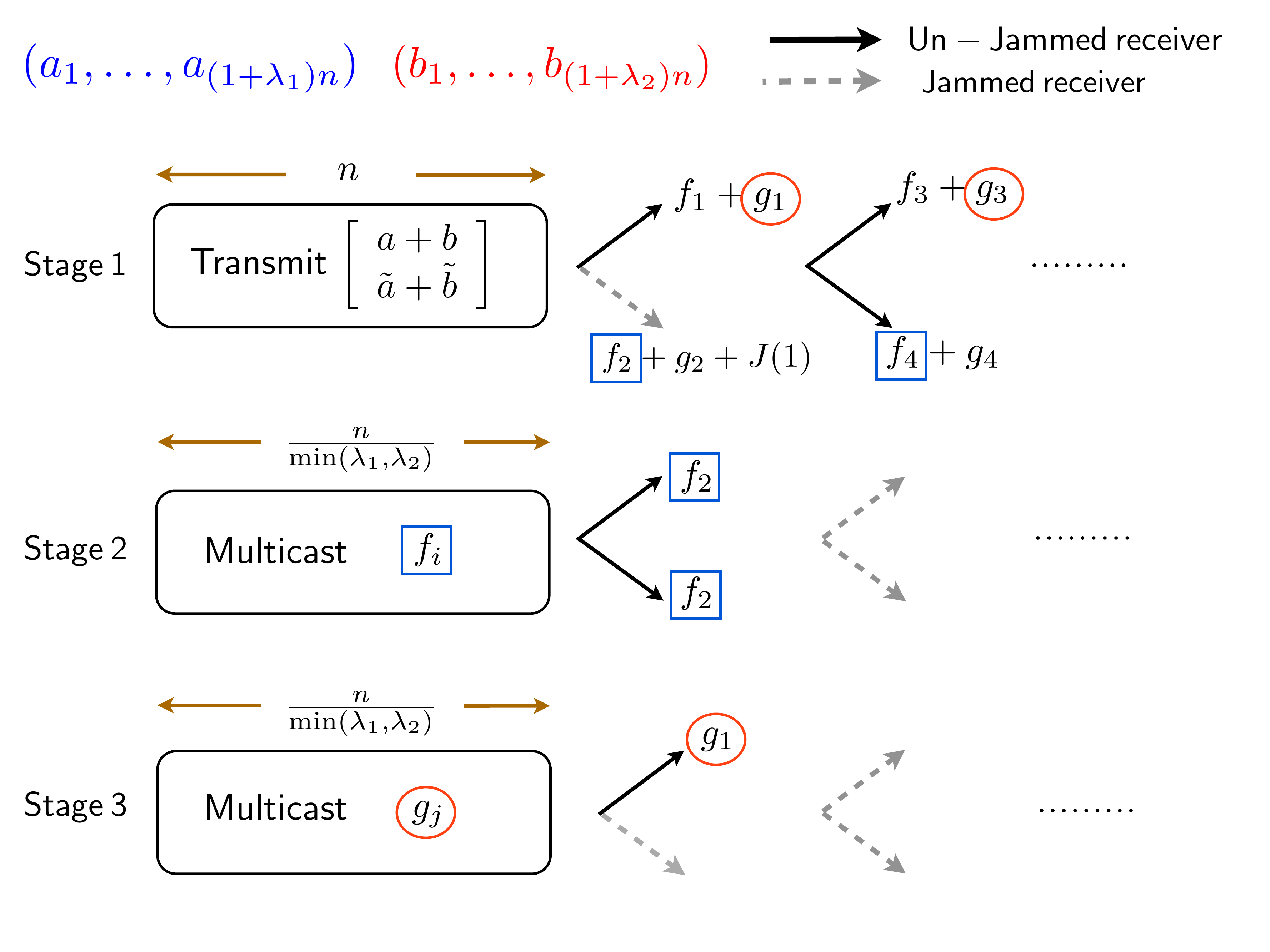}
\caption{Coding with  delayed $\CSIT$ and no $\JSIT$.}\label{CodingDN}
\end{figure}

As mentioned, the transmitter does not have knowledge about the LCs undelivered to the receivers. However, using d-$\CSIT$, it can reconstruct the LCs that would have been received at each receiver irrespective of whether they are jammed or not. For example, let us denote these LCs by $\mathcal{F}_1,\mathcal{F}_2$,  $\mathcal{G}_1,\mathcal{G}_2$ that correspond to combinations of $a_1,a_2,b_1$ and $b_2$ (refer to modified $\mathsf{MAT}$ scheme). Irrespective of whether  $\mathcal{F}_1+\mathcal{G}_1$ is received at receiver $1$ or not, the LC $\mathcal{F}_2$ is useful for it as it will act as an additional LC that helps decode its intended symbols. Similar reasoning holds for receiver $2$ with respect to the symbol $\mathcal{G}_1$. But because these LCs have been received at the un-intended receiver, these act as side information which are used in the stages $2$ and $3$ of the algorithm. 

\item \textbf{Stage 2}: In this stage, the transmitter multicasts $\mathcal{F}$-type LCs that would have been received at receiver $2$ (irrespective of whether it is jammed or not, the transmitter can reconstruct them using d-$\CSIT$). This is now available at the $1$st receiver with a probability $\lambda_1$ and with probability $\lambda_2$ at the $2$nd receiver.  This is useful for both the receivers as it is a useful LC of intended symbols for the $1$st receiver while it can be used to remove the side information at receiver $2$ to recover its intended LC if at all it was received in the past (note that this is not useful for the $2$nd receiver, if a LC consisting of this $\mathcal{F}$- symbol was never received in the past). Thus the total time taken to deliver one such $\mathcal{F}$-symbol at both the receivers is given by 
\begin{equation}
\mathrm{max}\left(\frac{1}{\lambda_1},\frac{1}{\lambda_2}\right),
\end{equation}
as they are un-jammed with probabilities $\lambda_1,\lambda_2$ respectively. Since there are $n$ such $\mathcal{F}$-type LCs (created over $n$ time instants in stage $1$), 
the total time necessary to deliver them is given by 
\begin{equation}
\tau_2=\mathrm{max}\left(\frac{1}{\lambda_1},\frac{1}{\lambda_2}\right)n.
\end{equation}

\item \textbf{Stage 3}: This stage is the complement of the Stage $2$, where the transmitter sends the $\mathcal{G}-$type LCs that would have been received at the $1$st receiver, but are useful to both of them. Thus the total time spent in Stage $3$ is given by 
\begin{equation}
\tau_3=\mathrm{max}\left(\frac{1}{\lambda_1},\frac{1}{\lambda_2}\right)n.
\end{equation}
\end{itemize}

\subparagraph{$\DoF$ analysis:} At the end of the proposed $3-$stage algorithm, notice that both the receivers have $(1+\lambda_1)n$ and $(1+\lambda_2)n$ intended LCs. Since $(1+\lambda_1)n$ random LCs are sufficient to decode $(1+\lambda_1)n$ symbols, at the end of this stage $3$, both the receivers have successfully decoded all intended symbols. Thus the $\DoF$ is given by 
\begin{eqnarray}
d_1&=&\frac{(1+\lambda_1)n}{\tau_1+\tau_2+\tau_3} \\
&=&\frac{(1+\lambda_1)n}{n+\mathrm{max}\left(\frac{n}{\lambda_1},\frac{n}{\lambda_2}\right)+\mathrm{max}\left(\frac{n}{\lambda_1},\frac{n}{\lambda_2}\right)} \\
&=&\frac{(1+\lambda_1)}{1+2\mathrm{max}\left(\frac{1}{\lambda_1},\frac{1}{\lambda_2}\right)}
\end{eqnarray}
On similar lines, we have 
\begin{eqnarray}
d_2&=&\frac{(1+\lambda_2)}{1+2\mathrm{max}\left(\frac{1}{\lambda_1},\frac{1}{\lambda_2}\right)},
\end{eqnarray}
which is the $\DoF$ region given by Theorem~\ref{TheoremDN}.

\begin{figure}[t]
  \centering
\includegraphics[width=12.0cm]{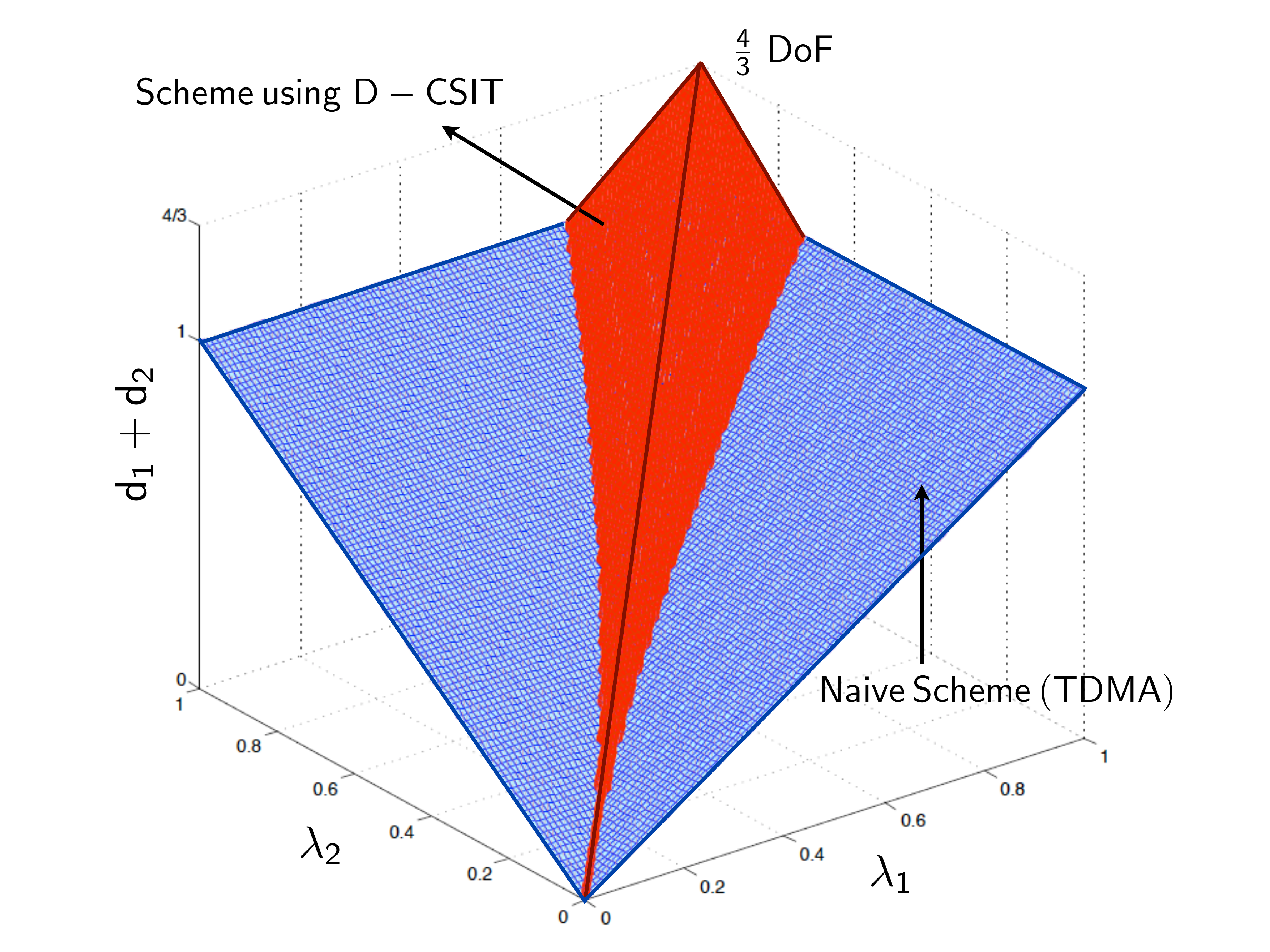}
\caption{Achievable sum $\DoF$ in the $\DN$ configuration}
\label{Fig:3DPlotDN}
\end{figure}
Theorems~\ref{TheoremNN} and \ref{TheoremDN}, suggest that the $\DoF$ in the $\DN$ configuration can be increased only when the region described in Theorem~\ref{TheoremNN} is a subset of the region described in Theorem~\ref{TheoremDN}. This is possible only when 
\begin{align}
\frac{(1+\lambda_2)\lambda_1}{2\mathsf{max}\left(1,\lambda_1/\lambda_2-1\right)} \geq \lambda_2 \nonumber \\
\frac{(1+\lambda_1)\lambda_2}{2\mathsf{max}\left(1,\lambda_2/\lambda_1-1\right)} \geq \lambda_1. 
\end{align}
In other words, the proposed scheme for the $\DN$ configuration can achieve $\DoF$ gains over the naive TDMA-based scheme if and only if $\lambda_1,\lambda_2$ satisfy (obtained by solving the above two equations)
\begin{align}
\frac{|\lambda_1-\lambda_2|}{\lambda_1\lambda_2}\leq 1. 
\end{align}
Fig.~\ref{Fig:3DPlotDN} shows the $\mathsf{Sum}\ \DoF$ achieved using the naive TDMA scheme and the proposed scheme for the $\DN$ configuration. 
%It is seen that when $\lambda_1,\lambda_2$ satisfy 
%\begin{align}
%\frac{|\lambda_1-\lambda_2|}{\lambda_1\lambda_2}\leq 1,
%\end{align}
%the proposed transmission scheme helps increase the $\mathsf{Sum} \ \DoF$ from what is achievable using the naive TDMA scheme. 
Since the transmitter has statistical knowledge about the jammers strategy, it can choose to use the naive scheme or the novel scheme based on the values of $\lambda_1,\lambda_2$.

\subsection{No $\CSIT$}
The following relationship holds true, 
\begin{align}\label{DoF_NCSIT_Compare}
\DoF_{\NN} \subseteq \DoF_{\ND} \subseteq \DoF_{\NP}.
\end{align}
i.e, the $\DoF$ is never reduced when JSI is available at the transmitter. 
\subsubsection{No $\CSIT$, Perfect $\JSIT$ ($\NP$) :}
\begin{figure}[t]
  \centering
\includegraphics[width=12.0cm]{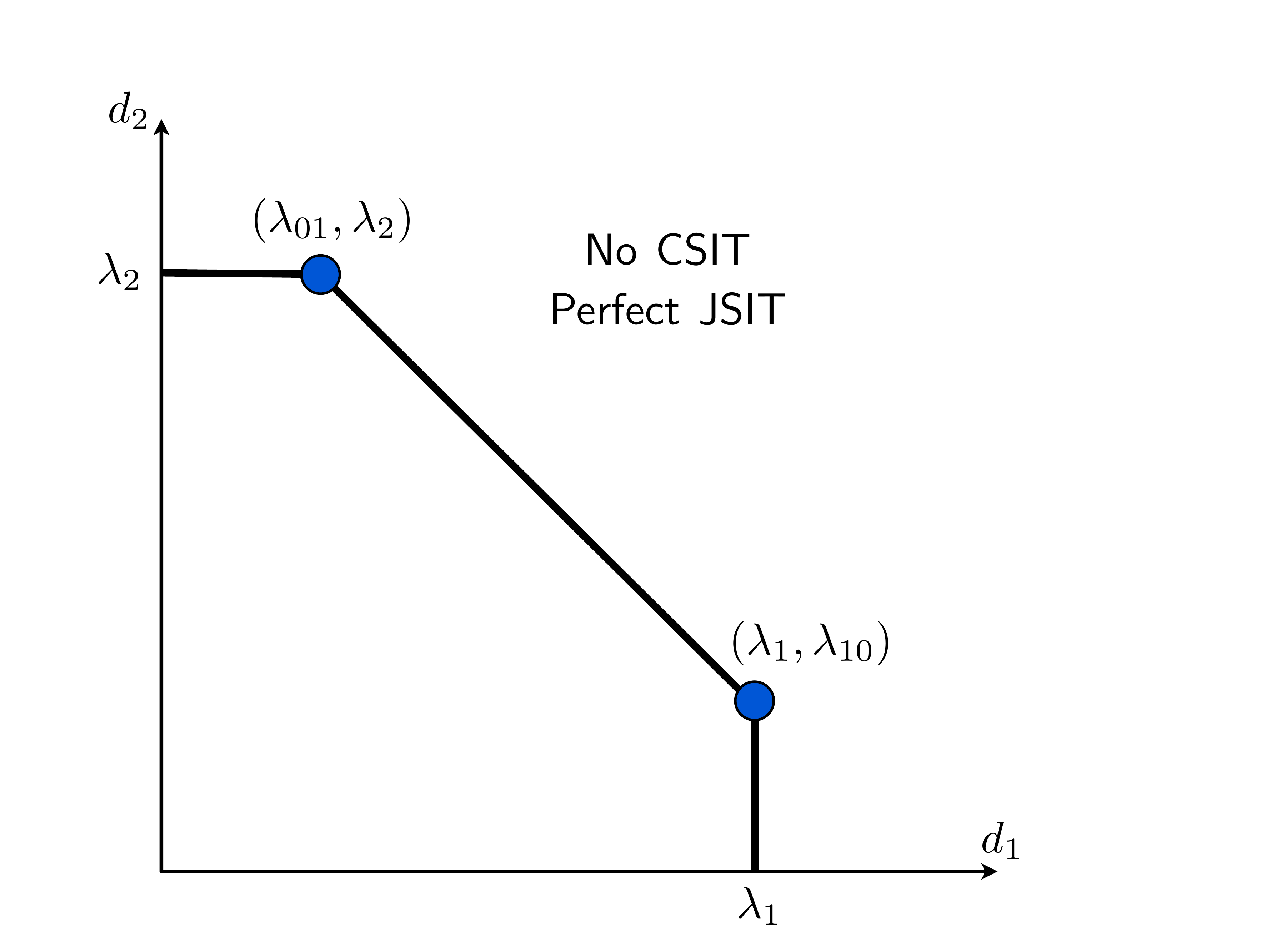}
\vspace{-10pt}
\caption{$\DoF$ region with no $\CSIT$ and perfect $\JSIT$.}\label{Fig:FigureNP}
\end{figure}

As seen in Fig.~\ref{Fig:FigureNP}, the following $\DoF$ pairs $(d_1,d_2)=(\lambda_1,0)$,
$(\lambda_1,\lambda_{10})$, $(\lambda_{01},\lambda_2)$, and $(0,\lambda_2)$ are achievable in the $\NP$ configuration. 
The $\DoF$ pairs $(\lambda_1,0)$ and $(0,\lambda_2)$ are readily achievable using the naive scheme mentioned before where the transmitter sends symbols exclusively to the receiver that is not jammed (when the transmitter sends $n$ symbols to the $k$th receiver using the knowledge of perfect $\JSIT$, it receives $\lambda_kn$ symbols since it is not jammed with probability $\lambda_k$). The remaining $\DoF$ pairs, $(\lambda_1,\lambda_{10})$ and $(\lambda_{01},\lambda_2)$ are achieved via the transmission schemes suggested in the $\DP$ configuration for the corresponding $\DoF$ pairs. 

\subsubsection{No $\CSIT$, Delayed $\JSIT$ ($\ND$):}
Here, we present a 3-stage scheme that achieves the $\DoF$ region given by Theorem~\ref{TheoremND}. This scheme is similar to the algorithm proposed for the $\DD$ configuration. In stage $1$, the transmitter sends symbols intended for receiver $1$ alone and keeps re-transmitting them until it is received (jamming free signal) at at least one receiver. On similar lines, the transmitter sends symbols intended only for receiver 2 in Stage 2. Stage 3 consists of transmitting the undelivered symbols to the intended receiver. However, since there is no CSI available at the transmitter, the algorithm proposed for the $\DD$ configuration cannot be applied here. The modified 3-stage algorithm is presented henceforth. 
\begin{figure}[t]
  \centering
\includegraphics[width=12.0cm]{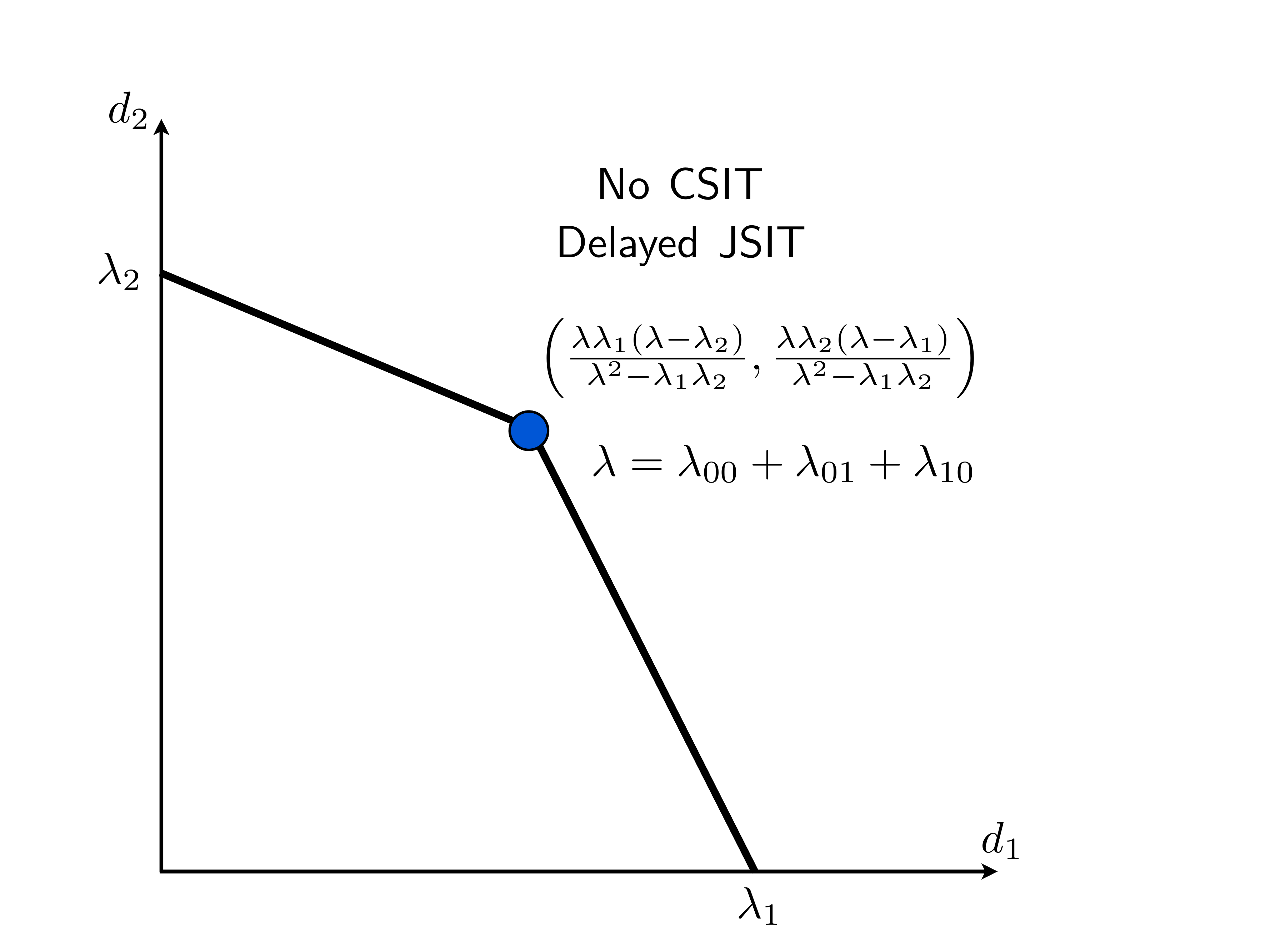}
\vspace{-10pt}
\caption{An achievable $\DoF$ region with no $\CSIT$ and delayed $\JSIT$.}\label{Fig:FigureND}
\end{figure}

\textbf{\textit{Stage 1}}--In this stage, the transmitter intends to deliver $n_1$ $a$-symbols, in a manner such that each symbol is received at \textit{at least} one of the receivers. At every time instant the transmitter sends one symbol on one of its transmit antennas. This message is re-transmitted until it is received at at least one receiver. Any one of the following four scenarios can arise: \\
\textit{Event $00$}: none of the receivers are jammed (which happens with probability $\lambda_{00}$). As an example, suppose that at time $t$, if the transmitter sends $a_{1}$: then receiver $1$ gets $\mathcal{F}_{1}(a_{1})$ and receiver $2$ gets $\mathcal{F}_{2}(a_{1})$ (note that these are scaled versions of the transmit signal corrupted by white Gaussian noise and are recovered by the receivers within noise distortion). The fact that the event $00$ occurred at time $t$ is known at time $t+1$ via d-$\JSIT$. The transmitter ignores the side information created at receiver $2$, since the intended symbol is delivered to receiver $1$. The transmitter sends a new symbol $a_2$ at time $t+1$. 

\textit{Event $01$}: receiver $1$ is not jammed, while receiver $2$ is jammed (which happens with probability $\lambda_{01}$). As an example, suppose that at time $t$, if the transmitter sends $a_{1}$: then receiver $1$ gets $\mathcal{F}_{1}(a_{1})$ and receiver $2$'s signal is drowned in the jamming signal. The fact that the event $01$ occurred at time $t$ is known at time $t+1$ via d-$\JSIT$. Since the intended $a-$ symbol is delivered to receiver $1$, at time $t+1$, the transmitter sends a new symbol $a_2$ from the message queue of symbols intended for receiver $1$. 

\textit{Event $10$}: receiver $2$ is not jammed, while receiver $1$ is jammed (which happens with probability $\lambda_{10}$). As an example, suppose that at time $t$, if the transmitter sends $a_{1}$: then receiver $1$'s signal is drowned in the jamming signal, whereas receiver $2$ gets $\mathcal{F}_{2}(a_{1})$. The fact that the event $10$ occurred at time $t$ is known at time $t+1$ via d-$\JSIT$. Since the receivers have CSI and JSI, receiver $1$ is aware of the message received at receiver $2$ within noise distortion. This message is not discarded, but instead used as side information and is delivered to the receiver $1$ in Stage $3$. 

\textit{Event $11$}: both receivers are jammed (which happens with probability $\lambda_{11}$). Using d-$\JSIT$, transmitter knows at time $t+1$ that the event $11$ occurred and hence at time $t+1$, it re-transmits $a_{1}$ on one of its transmit antennas.

The above events are \emph{disjoint}, so in one time slot, the average number of useful messages 
delivered to at least one receiver is given by 
\begin{equation}
E[\textnormal{$\#$ of symbols delivered}] =\lambda_{00}+\lambda_{01}+\lambda_{10}\triangleq \phi. \nonumber
\end{equation}
Hence, the expected time to deliver one LC is 
\begin{equation}
\frac{1}{\phi}=\frac{1}{\lambda_{00}+\lambda_{01}+\lambda_{10}}.
\end{equation}
\emph{Summary of Stage $1$:} 
\begin{itemize}
\item The time spent in this stage to deliver $n_1$ LCs is 
\begin{align}\label{N_1_ND}
N_1=\frac{n_1}{\phi}.
\end{align} 
\item Since receiver $1$ is not jammed in events $00$ and $01$, i.e., with probability $\lambda_1$, it receives only $\lambda_1 N_1$ symbols. 
\item The number of undelivered symbols is $n_1-\lambda_1N_1=\frac{\lambda_{10} n_1}{\phi}$. These symbols are available at receiver $2$ (corresponding to the event $10$) and are known to the transmitter via d-$\JSIT$. This side information created at receiver $2$ is not discarded, instead it is used in Stage 3 of the transmission scheme. 
\item The loss in $\DoF$ in this configuration due to the unavailability of $\CSIT$ is observed by noticing the expected number of symbols delivered in the $\ND$ configuration which is given by $\lambda_{00}+\lambda_{01}+\lambda_{10}$ while it is $2\lambda_{00}+\lambda_{01}+\lambda_{10}$  in the $\DD$ configuration as seen in \eqref{expected_symbols_DD}. 

\end{itemize}
{\textbf{\textit{Stage 2}}}--
In this stage, the transmitter intends to deliver $n_2$ $b$-symbols, in a manner such that each symbol is received at \textit{at least} one of the receivers. Stage 1 is repeated here with the roles of the receivers 1 and 2 interchanged. On similar lines to Stage 1, the time spent in this stage is 
\begin{align}\label{N_2_ND}
N_2=\frac{n_2}{\phi}.
\end{align} The number of symbols received at receiver $2$ is $\lambda_2N_2$ and the number of symbols not delivered to receiver $2$ but are available as side information at receiver $1$ is $n_2-\lambda_2N_2=\frac{\lambda_{01}n_2}{\phi}$. 

\begin{remark}{\em At the end of these $2$ stages, following typical situation arises: $\mathcal{F}(a_1)$ (resp. $\mathcal{G}(b_1)$) is a symbol intended for receiver $1$ (resp. $2$) but is available as side information at receiver $2$ (resp. $1$)\footnote{Such situations correspond to the event $10$ in Stage $1$; and the event $01$ in Stage $2$.}. Notice that these symbols must be transmitted to the complementary receivers so that the desired symbols can be decoded. The transmitter, via delayed-$\JSIT$, is aware of the symbols that that are not delivered to the receivers (however the transmitter is not required to be aware of $\mathcal{F}$ (resp. $\mathcal{G}$) since the receivers have this knowledge and that $\mathcal{F}$ (resp. $\mathcal{G}$) is the noise corrupted version of one symbol $a_1$ (resp. $b_1$)). In Stage $3$, the transmitter sends a random LC of these symbols, say $\mathcal{L}=h_1\mathcal{F}(a_1)+h_2\mathcal{G}(b_1)$ where $h_1, h_2$ that form the new LC are known to the transmitter and receivers \emph{a priori}. Now, assuming that only receiver $2$ (resp. $1$) is jammed, $\mathcal{L}$ is received at receiver $1$ (resp. $2$) within noise distortion. Using this LC, it can recover $\mathcal{F}(a_1)$ (resp. $\mathcal{G}(b_1)$) from $\mathcal{L}$ since it already has $\mathcal{G}(b_1)$ (resp. $\mathcal{F}(a_1)$) as side information. When no receiver is jammed, both the receivers are capable of recovering $\mathcal{F}(a_1)$,  $\mathcal{G}(b_1)$ simultaneously.}
\end{remark}

{\textbf{\textit{Stage 3}}}--In this stage, the undelivered symbols to each receiver are transmitted using the technique mentioned above. Let us assume that $\mathcal{F}_1(a_1)$ and $\mathcal{G}_1(b_1)$ are symbols available as side information at receivers $2$ and $1$ respectively. 
The transmitter sends $\mathcal{L}(\mathcal{F}_1,\mathcal{G}_1)$, a LC of these symbols on one transmit antenna, with the eventual goal of multicasting this LC (i.e., send it to \textit{both} receivers). The following events, as specified earlier in Stages $1$ and $2$, are also possible while in this stage. 

\textit{Event $00$}: Suppose at time $t$, if the transmitter sends $\mathcal{L}(\mathcal{F}_1,\mathcal{G}_1)$, then both the receivers get this LC within noise distortion. With the capability to recover $\mathcal{L}(\mathcal{F}_1,\mathcal{G}_1)$ within a scaling factor, the receivers $1$ and $2$ decode their intended messages $\mathcal{F}_1$ and $\mathcal{G}_1$ respectively using the side informations $\mathcal{G}_1$ and $\mathcal{F}_1$ that are available with them. Since the intended messages are delivered at the intended receivers, the transmitter, at time $t+1$, sends a new LC of two new symbols $\tilde{\mathcal{L}}(\tilde{\mathcal{F}}_1,\tilde{\mathcal{G}}_1)$. 

\textit{Event $01$}: Since receiver $2$ is jammed, its signal is drowned in the jamming signal while receiver $1$ gets $\mathcal{L}(\mathcal{F}_1,\mathcal{G}_1)$ and is capable of recovering $\mathcal{F}_1$ using $\mathcal{G}_1$ available as side information. The fact that event $01$ occurred is known to the transmitter at time $t+1$ via d-$\JSIT$. Thus, at time $t+1$, the transmitter sends a new LC $\tilde{\mathcal{L}}(\tilde{\mathcal{F}}_1,\mathcal{G}_1)$ since $\mathcal{G}_1$ has not yet been delivered to receiver $2$. 

\textit{Event $10$}: This event is similar to event $01$, with the roles of the receivers 1 and 2 interchanged. Hence, receiver $2$ is capable of recovering $g_1$ from $\mathcal{L}(\mathcal{F}_1,\mathcal{G}_1)$ while receiver $1$'s signal is drowned in the jamming signal. Thus at time $t+1$, the transmitter sends a new LC $\tilde{\mathcal{L}}(\mathcal{F}_1,\tilde{\mathcal{G}}_1)$ since $\mathcal{F}_1$ has not yet been delivered to receiver $1$. 

\textit{Event $11$}: Using d-$\JSIT$, transmitter knows at time $t+1$ that the event $11$ occurred and hence at time $t+1$, it re-transmits $\mathcal{L}(\mathcal{F}_1,\mathcal{G}_1)$ on one of its transmit antennas. 

Since, all the events are disjoint, in one time slot, the average number of LCs delivered to receiver $1$ is given by 
\begin{equation}
E[\textnormal{$\#$ of symbols delivered to user 1}] =\lambda_{00}+\lambda_{01}\triangleq \lambda_1. \nonumber
\end{equation}
Hence, the expected time to deliver one symbol to receiver $1$ in this stage is $\frac{1}{\lambda_1}$. Given that $\frac{\lambda_{10} n_1}{\phi}$ symbols are to be delivered to receiver $1$ in this stage, the time taken to achieve this is 
$\frac{\lambda_{10} n_1}{\lambda_1\phi}$. Interchanging the roles of the users, the time taken to deliver $\frac{\lambda_{01} n_2}{\phi}$ symbols to receiver $2$ is $\frac{\lambda_{01} n_2}{\lambda_2\phi}$. Thus the total time required to satisfy the requirements of both the receivers in Stage $3$ is given by 
\begin{equation}\label{N_3_ND}
N_3=\mathrm{max}\left(\frac{\lambda_{10} n_1}{\lambda_1\phi}, \frac{\lambda_{01} n_2}{\lambda_2\phi}\right).
\end{equation}
The optimal $\DoF$ achieved in the $\DD$ configuration is readily evaluated as 
\begin{eqnarray}
d_1=\frac{n_1}{N_1+N_2+N_3}, \ d_2=\frac{n_2}{N_1+N_2+N_3}.
\end{eqnarray}
Substituting $\{N_i\}_{i=1,2,3}$ from \eqref{N_1_ND}--\eqref{N_3_ND}, we have,
\begin{eqnarray}
d_1&=&\frac{\eta}{\frac{1}{\phi}+\mathrm{max}\left(\frac{\lambda_{10} \eta}{\lambda_1\phi},\frac{\lambda_{01} (1-\eta)}{\lambda_2\phi}\right)} \nonumber \\
d_2&=&\frac{1-\eta}{\frac{1}{\phi}+\mathrm{max}\left(\frac{\lambda_{10} \eta}{\lambda_1\phi},\frac{\lambda_{01} (1-\eta)}{\lambda_2\phi}\right)},
\end{eqnarray}
where $\eta=\frac{n_1}{n_1+n_2}$. Eliminating $\eta$ from the above two equations, yields the $\DoF$ region given by Theorem~\ref{TheoremND}. 
The $\DoF$ pairs $(\lambda_1,0)$ and $(0,\lambda_2)$ are achieved by using the transmission strategy proposed for the $\NN$ configuration below. 
\subsubsection{No $\CSIT$, No $\JSIT$ ($\NN$) :}
The $\DoF$ for the $\NN$ configuration is given by Theorem~\ref{TheoremNN} and the simple time sharing scheme achieves $\DoF_{\NN}$. For completeness, we briefly explain the transmission scheme used in this configuration. 
We first explain the achievability of the $\DoF$ pair: $(d_{1}, d_{2})=(\lambda_1,0)$. To this end, note that receiver $1$ is jammed in an i.i.d. manner with probability $(1-\lambda_{1})$. This implies that for a scheme of sufficiently large duration $n$, it will receive $\lambda_{1}n$ jamming free information symbols (corresponding to those instants in which $S_{1}(t)=0$). However, in the $\NN$ configuration (no $\CSIT$ and no $\JSIT$), the transmitter is not aware of the symbols which are received without being jammed. In order to compensate for the lack of this knowledge, it sends random linear combinations (LCs) (the random coefficients are assumed to be known at the receivers \cite{Erasure}) of $\lambda_{1}n$ symbols over $n$ time slots. For sufficiently large $n$, receiver $1$ obtains $\lambda_{1}n$ jamming free LCs and hence it can decode these symbols. Thus the $\DoF$ pair $(\lambda_{1}, 0)$ is achievable. Similarly, by switching the role of the receivers, the pair $(0, \lambda_{2})$ is also achievable. Finally, the entire region in Theorem \ref{TheoremNN} is achievable by time sharing between these two strategies. 

\section{Extensions to Multi-receiver MISO Broadcast Channel}\label{TheoremsKuser}
We present extensions of the $2$-user case to that of a multi-user broadcast channel. In particular, for the $K$-user scenario, 
the total number of possible jammer states is $2^K$, which can be interpreted as:
\begin{align}
2^{K}&= \underbrace{{K \choose 0}}_{\mbox{None jammed}} + \underbrace{{K \choose 1}}_{\mbox{One receiver jammed}}+\ldots+ \underbrace{{K \choose K}}_{\mbox{All receivers jammed}}.
\end{align}
In such a scenario, the jammer state $S(t)$ at time $t$ is a length $K$ vector with each element taking values $0$ or $1$. 
%We next present the Theorems that describe the $\DoF$ regions for the $K$-user case. For most of the configurations, the achievability schemes are straight forward extensions of the coding schemes presented for the the $2$-user case. Hence, in the interest of space, we do not outline these schemes again. 
We present the optimal $\DoF$ regions for the $\PP,\PD$ and $\PN$ configurations in Theorem~\ref{TheoremPP-KUser} and for the $\NN$ configuration in Theorem~\ref{TheoremNN-KUser}. For the $\DP$ and $\DD$ configurations, we present lower bounds on the sum $\DoF$ under a class of symmetric jamming strategies. Furthermore, we illustrate the impact of jamming and the availability of $\JSIT$ (either instantaneous or delayed) by comparing the $\DoF$ achievable in these configurations with the $\DoF$ achieved in the absence of jamming (with delayed $\CSIT$) i.e., $\DoF_{\mathsf{MAT}}(K)$ (defined in Section~\ref{system_model}) \cite{MAT2012}. For most of the configurations, the achievability schemes are straight forward extensions of the coding schemes presented in the $2$-user case. Hence, in the interest of space, we do not outline these schemes again. 
%We first present extensions of Theorem \ref{TheoremPP} and Theorem \ref{TheoremNN} to the $K$-user case for arbitrary jamming strategies. 
\begin{Theo}\label{TheoremPP-KUser}
The $\DoF$ region of the $K$-user MISO BC for each of the ($\CSIT$, $\JSIT$) configurations $\PP$, $\PD$ and $\PN$ is the same and is given by the set of non-negative pairs $(d_{1},\ldots, d_{K})$ that satisfy
\begin{align}
d_{k}&\leq \lambda_{k}, \quad k=1,\ldots,K,
\end{align}
where $\lambda_k$ is the probability with which the $k$th receiver is not jammed. 
\end{Theo}
The achievability of this $\DoF$ region is a straightforward extension of the scheme proposed in Section~\ref{Schemes} for the $2$-user MISO BC for the corresponding $I_{\CSIT}I_{\JSIT}$ configurations. 

\begin{Theo}\label{TheoremNN-KUser}
The  $\DoF$ region of the $K$-user MISO BC for  the ($\CSIT$, $\JSIT$) configuration $\NN$ is given as
\begin{align}
\sum_{k=1}^{K}\frac{d_{k}}{\lambda_{k}}&\leq 1.
\end{align}
\end{Theo}
The achievability of this $\DoF$ region is also an extension of the transmission scheme proposed for the $\NN$ configuration in Section~\ref{Schemes} for the $2$-user MISO BC. This is a simple time sharing scheme (TDMA) where the transmitter sends information to only one receiver among the $K$ receivers at any given time instant. 

For the $\DP$ and $\DD$ configurations,
 we consider a symmetric scenario in which any subset of receivers are jammed symmetrically i.e, 
\begin{align}\label{invariant_lambda}
\lambda_{\mathbf{s}}=\lambda_{\pi(\mathbf{s})}, 
\end{align}
where $\lambda_{\mathbf{s}}$ is the probability that $S(t)=\mathbf{s}$ at any given time $t$ and $\pi(\mathbf{s})$ denotes any permutation of the $K$ length jamming state vector $S(t)=\mathbf{s}$. In particular, for $K=3$, this assumption corresponds to
\begin{align}\label{equal_lambda}
\lambda_{001}=\lambda_{010}=\lambda_{100}, \mbox{ and } \lambda_{011}=\lambda_{101}=\lambda_{110}. 
\end{align}
From \eqref{lambda_receiver_1} and \eqref{equal_lambda}, it is seen that
\begin{align}
\lambda_1=\lambda_2=\lambda_3= \lambda_{000}+ \lambda_{001}+\lambda_{010}+\lambda_{011},
\end{align}
i.e., the marginal probabilities of the receivers being jammed (un-jammed) are the same. For the $K$-user case, we have 
\begin{align}
\lambda_1=\lambda_2\ldots=\lambda_K.
\end{align}

Let $||\mathbf{s}||_1$ denote the $1$-norm of the $K$-length vector $\mathbf{s}$. In other words, $||\mathbf{s}||_1$ indicates the total number of $1$'s seen in the vector $\mathbf{s}$ and hence $0\leq ||\mathbf{s}||_1\leq K$. We denote $\eta_{j}$ as the total probability with which any $j$ receivers are jammed i.e.,
\begin{align}
\eta_j=Pr\left(||\mathbf{s}||_1=j\right),\ j=0,1,2,\ldots,K,
\end{align}
where $Pr(\mathcal{E})$ indicates the probability of occurrence of event $\mathcal{E}$. By definition, we have $\sum_{j=0}^K \eta_j = 1$ and we collectively define these probabilities as the $\left(K+1\right)\times1$ vector $\eta=[\eta_0,\eta_1,\ldots,\eta_K]^T$. For instance, $\eta_0=1$ corresponds to the no jamming scenario i.e., none of the receivers are jammed. For $K=3$, we have
\begin{align}\label{eta_values_3_user}
 \eta_{0}=\lambda_{000}, \quad \eta_{1}=\lambda_{001}+\lambda_{010}+\lambda_{100}\nonumber \\
 \eta_{2}=\lambda_{011}+\lambda_{101}+\lambda_{110}, \quad \eta_{3}=\lambda_{111}.
\end{align}
It is easily verified that $\eta_{0}+\eta_{1}+\eta_{2}+\eta_{3}=1$. From \eqref{lambda_receiver_1}, \eqref{equal_lambda}-\eqref{eta_values_3_user}, it is seen that
%\begin{align}
$\lambda_i=\eta_0+\frac{2}{3}\eta_1+\frac{1}{3}\eta_2$, for $i=1,2,3.$
%\end{align}
In general, it can be shown that 
\begin{align}\label{final_lambda}
\lambda_1=\lambda_2=\cdots=\lambda_K= \left(\sum_{j=0}^{K}\left(\frac{K-j}{K}\right)\eta_{j}\right)\triangleq \lambda_{\eta}.
\end{align}

\begin{Theo}\label{TheoremDP-KUser}
An achievable sum $\DoF$ of the $K$-user MISO BC for  the ($\CSIT$, $\JSIT$) configuration $\DP$ is given as\footnote{$\DoF_{\mathsf{DP}}^{\mathsf{Ach}}({\bf{\eta}}, K)$ and $\DoF_{\mathsf{DD}}^{\mathsf{Ach}}({\bf{\eta}}, K)$ denote the lower bound (achievable) on the $\DoF$  obtained in the $\DP$ and $\DD$ configurations in the $K$-user scenario.}
%To differentiate between the optimal $\DoF$ achieved Theorems~\ref{TheoremPP-KUser}-Theorem~\ref{TheoremNN-KUser} and the lower bounds presented for the $\DD$ and $\DP$ configurations, we use }
\begin{align}\label{TheoremDP-KUser-DoF}
\DoF_{\mathsf{DP}}^{\mathsf{Ach}}({\bf{\eta}}, K)&= \sum_{j=0}^{K} \eta_{j}\DoF_{\mathsf{MAT}}(K-j).
\end{align}
\end{Theo}
We note from Theorem~\ref{TheoremDP-KUser} that when perfect $\JSIT$ is available, the sum $\DoF$ in \eqref{TheoremDP-KUser-DoF} is achieved by transmitting only to the unjammed receivers. The transmission scheme that achieves this sum $\DoF$ is the $K$-user extension of the scheme presented for the $\DP$ configuration in Section~\ref{Schemes}. 

\begin{Theo}\label{TheoremDD-KUser}
An achievable sum $\DoF$ of the $K$-user MISO BC for the ($\CSIT$, $\JSIT$) configuration $\DD$ is given as
\begin{align}\label{DoF-TheoremDD-KUser}
\DoF_{\mathsf{DD}}^{\mathsf{Ach}}({\bf{\eta}}, K)&= \left(\sum_{j=0}^{K}\left(\frac{K-j}{K}\right)\eta_{j}\right)\DoF_{\mathsf{MAT}}(K)\triangleq \lambda_{\eta}\DoF_{\mathsf{MAT}}(K).
\end{align}
\end{Theo}
\begin{remark} {\em The $\DoF$ result in \eqref{DoF-TheoremDD-KUser} has the following interesting interpretation: consider a simpler problem in which only two jamming states are present: $S(t)=00\cdots0$ (none of the receivers are jammed) with probability $\lambda_{\eta}$  and  $S(t)=11\cdots1$ (all receivers are jammed) with probability $1-\lambda_{\eta}$. In addition, assume that the transmitter has perfect $\JSIT$. In such a scenario, the transmitter can use the $\textsf{MAT}$ scheme (for the $K$-user case) for $\lambda_{\eta}$ fraction of time to achieve $\lambda_{\eta}\DoF_{\textsf{MAT}}$ degrees-of-freedom (this scenario is equivalent to  jamming state $S(t)=00$  in the $\DP$ configuration for a $2$-user scenario which is discussed in Section~\ref{Schemes}) which is precisely as shown in  \eqref{DoF-TheoremDD-KUser}. Even though equivalence of these distinct problems is not evident \emph{a priori},  the $\DoF$ result indicates the  benefits of using $\JSIT$, although it is completely delayed.}
%An interesting conclusion arises from the $\DoF$ result in \eqref{DoF_K_User_Result}. 
%This result indicates that the achieved $\DoF$ in the $\DD$ configuration is equal to the $\DoF$ achieved using delayed-$\CSIT$ in a scenario where the jammer disrupts the receivers collectively with probability $1-\lambda_{\eta}$ and hence the receivers are jamming free with probability $\lambda_{\eta}$. In such a scenario, the transmitter uses the $\mathsf{MAT}$ scheme for $\lambda_{\eta}$ fraction of the time where the receivers are not jammed. Thus the $\DoF$ is given by \eqref{DoF_K_User_Result}. Though such a reasoning is not evident from the problem formulation, it is seen that the $\DoF$ result indicates the synergistic benefits of transmitting across various jammer states in a joint manner in the long run. 
\end{remark}
%The achievability of Theorem~\ref{TheoremDD-KUser} is based on the synergistic usage of delayed $\CSIT$ and delayed
%$\JSIT$ by exploiting side-information created at the un-jammed receivers in the past and transmitting linear combinations of such side-information symbols in the future. The sum $\DoF$ in \eqref{DoF-TheoremDD-KUser} is equal to the sum $\DoF$ achieved by using delayed $\CSIT$ and the transmission scheme proposed in \cite{MAT2012} (briefly explained in Section~\ref{Schemes}) in a scenario where all the receivers are collectively jammed with probability $\lambda_{\eta}$. 
It is reasonable to expect that the $\DoF$ achievable in the $\DP$ configuration will be higher than the $\DoF$ that can be achieved in the $\DD$ configuration. This can be readily shown as 
%It is worth noting that the bounds in Theorems~\ref{TheoremDP-KUser} and \ref{TheoremDD-KUser} satisfy $\DoF_{\mathsf{DP}}({\bf{\eta}}, K)\geq \DoF_{\mathsf{DD}}({\bf{\eta}}, K)$, which can be readily shown as follows:
\begin{align}
\DoF_{\mathsf{DP}}^{\mathsf{Ach}}({\bf{\eta}}, K)&= \sum_{i=0}^{K} \eta_{i}\DoF_{\mathsf{MAT}}(K-i)\nonumber \\
&= \sum_{i=0}^{K} \Bigg[\left(\frac{K-i}{K}\right)\eta_{i}\left(\frac{K}{K-i}\right)\DoF_{\mathsf{MAT}}(K-i)\Bigg]\nonumber \\
&= \sum_{i=0}^{K} \Bigg[\left(\frac{K-i}{K}\right)\eta_{i}\left(\frac{K}{K-i}\right)\frac{K-i}{1+\frac{1}{2}+\cdots+\frac{1}{K-i}}\Bigg]\nonumber \\
&= \sum_{i=0}^{K} \Bigg[\left(\frac{K-i}{K}\right)\eta_{i}\frac{K}{1+\frac{1}{2}+\cdots+\frac{1}{K-i}}\Bigg]\nonumber \\
&\geq \sum_{i=0}^{K} \Bigg[\left(\frac{K-i}{K}\right)\eta_{i}\frac{K}{1+\frac{1}{2}+\cdots+\frac{1}{K}}\Bigg]\nonumber \\
&= \sum_{i=0}^{K} \Bigg[\left(\frac{K-i}{K}\right)\eta_{i}\DoF_{\mathsf{MAT}}(K)\Bigg]\nonumber \\
&=\DoF_{\mathsf{DD}}^{\mathsf{Ach}}({\bf{\eta}}, K).
\end{align}
\begin{figure}[t]
  \centering
\includegraphics[width=0.8\textwidth]{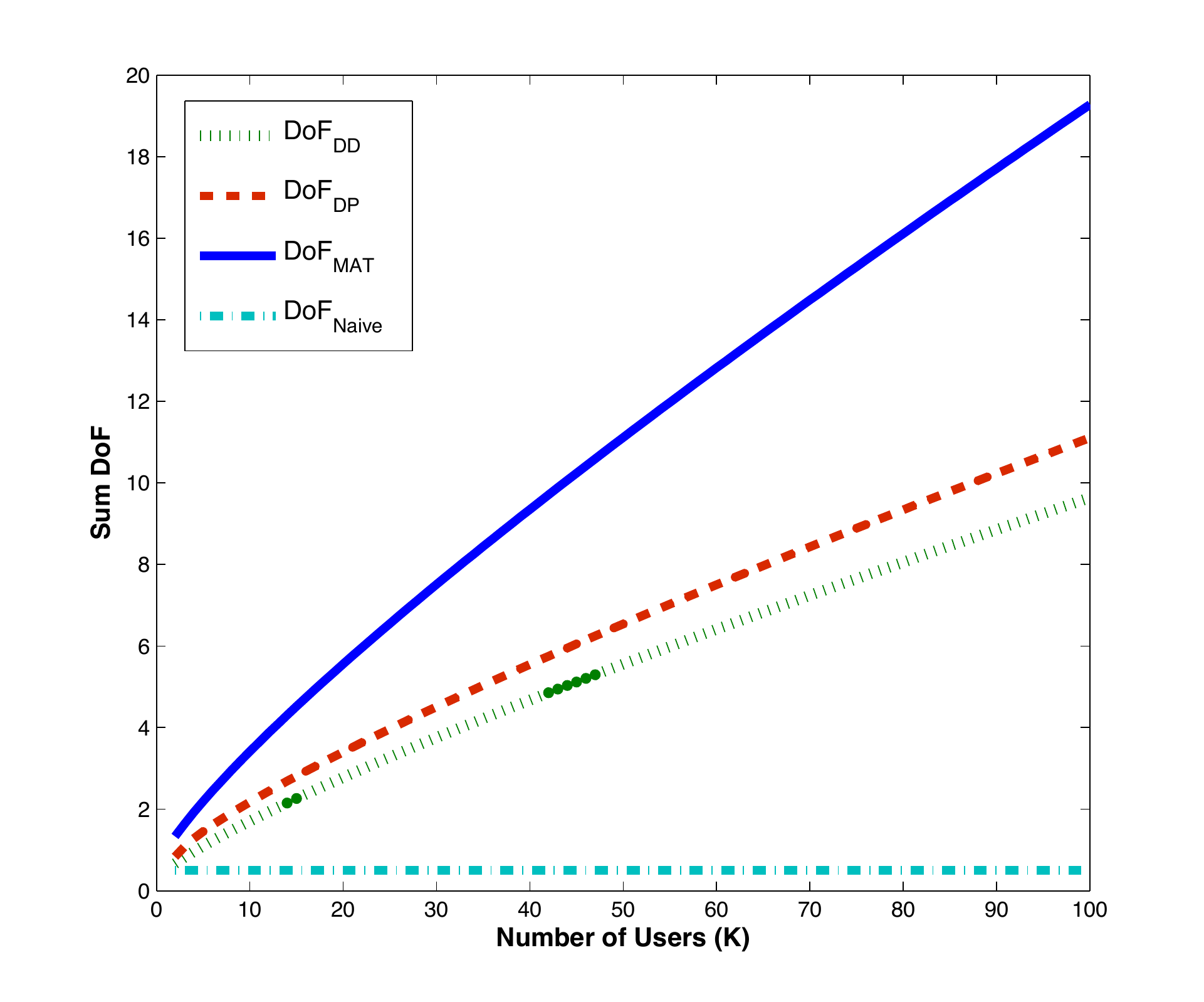}
\vspace{-18pt}
\caption{$\DoF$ comparison of $\mathsf{MAT}$ scheme, $\DP$ and $\DD$ configurations.}
\label{K_User_DP_DD}
\end{figure}

\noindent Fig.~\ref{K_User_DP_DD} shows the $\DoF$ comparison between $\DP$ and the $\DD$ configurations for a special case in which 
any subset of receivers is jammed with probability $\lambda_s=\frac{1}{2^K},\forall s$ i.e.,
\begin{align}\label{eta_j}
\eta_j=\frac{{K \choose j}}{2^K}.
\end{align}
It is seen that the sum $\DoF$ achieved in these configurations increases with the number of users, $K$. The additional $\DoF$ achievable in the $\DP$ configuration compared to the $\DD$ configuration increases with $K$ and is lower bounded by\footnote{For large values of $K$, the expression 
$1+\frac{1}{2}+\ldots+\frac{1}{K}\rightarrow \mathsf{log}(K)$. Hence the right side expression of \eqref{DoF_DD_DP_LB} behaves as $\frac{K}{\left(\mathsf{log}(K)\right)^2}\underset{K\rightarrow \infty}{\longrightarrow}\infty$.} %\eqref{DoF_DD_DP_LB}.
\begin{align}\label{DoF_DD_DP_LB}
\DoF_{\DP}^{\mathsf{Ach}}(\eta,K)-\DoF_{\DD}^{\mathsf{Ach}}(&\eta,K)\geq\frac{K-1}{4\left(1+\frac{1}{2}+\ldots+\frac{1}{K}\right)^2} \underset{K\rightarrow \infty}{\longrightarrow}\infty.
\end{align}
Also, it can be shown that the $\DoF$ gap between $\DoF_{\mathsf{MAT}}(K)$ and $\DoF_{\DP}^{\mathsf{Ach}}(\eta,K)$ is lower bounded by\footnote{For large $K$, the expression on the right side of \eqref{DoF_MAT_DP_LB} behaves as $\frac{K}{\mathsf{log}(K)}-\frac{K}{\left(\mathsf{log}(K)\right)^2}$ which tends to $\infty$ as $K\rightarrow \infty$.} 
\begin{align}\label{DoF_MAT_DP_LB}
\DoF_{\mathsf{MAT}}(K)-\DoF_{\DP}^{\mathsf{Ach}}(\eta,K)\geq\frac{K}{2\left(1+\frac{1}{2}+\ldots+\frac{1}{K}\right)}
-\frac{K\left(2^K-1\right)}{2^K\left(1+\frac{1}{2}+\ldots+\frac{1}{K}\right)^2} \underset{K\rightarrow \infty}{\longrightarrow}\infty.
\end{align}
These bounds illustrate the dependence of the sum $\DoF$ on the availability of perfect $\JSIT$ in a multi-user MISO BC in the presence of jamming attacks. For example, since the transmitter has instantaneous knowledge of the users that are jammed (at any given instant) in the $\DP$ configuration, it can conserve energy by only transmitting to the un-jammed receivers. However since no such information is available in the $\DD$ configuration, the transmitter has to transmit across different jamming scenarios (different subsets of receivers jammed) in such a configuration to realize $\DoF$ gains over naive transmission schemes. 
The sum $\DoF$ achieved in these configurations is much larger than the $\DoF$ achieved using a naive  transmission scheme ($\DoF=\lambda_{\eta}$) where the transmitter sends information to only one user at any given time instant without using $\CSIT$ or $\JSIT$. The coding schemes that achieve the sum $\DoF$ in \eqref{TheoremDP-KUser-DoF} and \eqref{DoF-TheoremDD-KUser} are detailed in Section~\ref{Schemes}.

\subsection{Achievability Scheme for $\DD$ configuration in $K$-user scenario}
Before we explain the $\DoF$ achievability scheme for the $K$-user $\DD$ configuration, we briefly explain the $\DD$ configuration for the $2$-user MISO BC for a special case in which the users are un-jammed with equal probability i.e,
\begin{align}
\lambda=\lambda_1=\lambda_2 \triangleq \lambda_{01}=\lambda_{10}.
\end{align}
In such a scenario, a simple $2$-phase scheme can be developed to achieve the optimal sum $\DoF$ of $\frac{4\lambda}{3}$ (this is seen by substituting $\lambda_1=\lambda_2=\lambda$ and $n_1=n_2$ in \eqref{DoFDD_opt_point}). We define order $1$ symbols as the set of symbols intended to only $1$ receiver while order $2$ symbols as the ones that are intended at both the receivers. Phase $1$ of the algorithm only uses order $1$ symbols while the order $2$ symbols are used in the $2$nd phase. We define $\DoF_1(2,\lambda)$ as the $\DoF$ of the $2$-user MISO BC to deliver order $1$ symbols in the case where the receivers are un-jammed with probability $\lambda$. On similar lines, $\DoF_2(2,\lambda)$ is the $\DoF$ of the system in delivering the order $2$ symbols to both the receivers. 

\begin{itemize}
\item {Phase 1:} Phase $1$ consists of $2$-stages one each for both the users. In each of these stages, symbols intended for a particular user are transmitted such that they are received at either receiver. Since each receiver is un-jammed with a probability $\lambda$, it receives $\lambda d$ symbols intended for itself and $\lambda d$ symbols of the other user which is used as side information in the $2$nd phase of this algorithm. Here $d$ is the time duration of each stage of this phase. Since a total of $n$ symbols are transmitted in each stage, we have
\begin{align}
2\lambda d=n \implies d=\frac{n}{2\lambda}.
\end{align}
The total time spent in this phase is $2d=\frac{n}{\lambda}$. At the end of this phase, each user has $\lambda d=\frac{n}{2}$ intended symbols and $\frac{n}{2}$ symbols intended for the other user. Using these $\frac{n}{2}$ side information symbols available at both the users, the transmitter can form $\frac{n}{2}$ LCs of these symbols which are transmitted in the $2$nd phase of the algorithm. These LCs are required by both the users that help them decode their intended symbols. Thus we have
\begin{align}
\DoF_1(2,\lambda)=\frac{2n}{\frac{n}{\lambda}+\frac{\frac{n}{2}}{\DoF_2(2,\lambda)}}.
\end{align}

\item Phase 2: The $\frac{n}{2}$ LCs of the side information symbols created at the transmitter are multicasted in this phase until both the receivers receive all the LCs. These LCs help the receivers decode their intended symbols using the available $\CSIR$ and the side information created in the $1$st phase of the algorithm. Since each receiver is jammed with probability $(1-\lambda)$, the expected time taken to deliver a order $2$ symbol to any receiver is $\frac{1}{\lambda}$. Hence the total time spent in this stage is 
\begin{align}
\frac{n}{2}\mathrm{max}\left(\frac{1}{\lambda},\frac{1}{\lambda}\right)=\frac{n}{2\lambda}.
\end{align}
Using the above result we can calculate $\DoF_2(2,\lambda)$ as 
\begin{align}
\DoF_2(2,\lambda)=\frac{\frac{n}{2}}{\frac{n}{2\lambda}}=\lambda.
\end{align}
\end{itemize}
Hence the sum $\DoF$ of the $2$-user MISO BC is given by 
\begin{align}
\DoF_1(2,\lambda)&=\frac{2n}{\frac{n}{\lambda}+\frac{\frac{n}{2}}{\lambda}}\\
&=\frac{4\lambda}{3},
\end{align}
which is also the sum $\DoF$ obtained from \eqref{DoFDD_opt_point} for the specified scenario. This algorithm also builds up the platform for developing the transmission scheme for the $K$-receiver MISO BC whose $\DoF$ is given by \eqref{DoF-TheoremDD-KUser}. 

\begin{figure}
\hspace{-20pt}\includegraphics[width=1.1\textwidth]{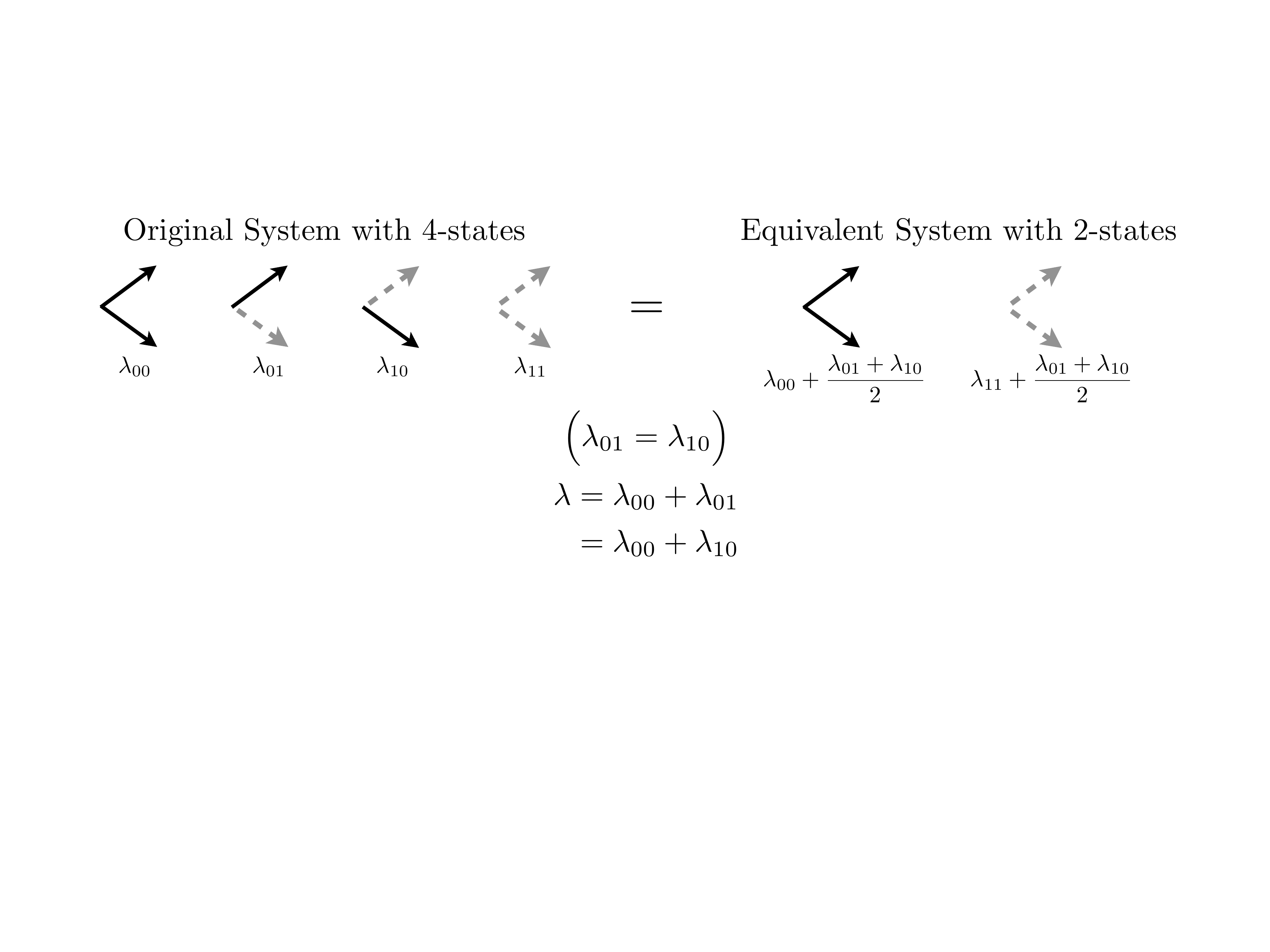}
\vspace{-140pt}\caption{State Equivalence when $\lambda_{01}=\lambda_{10}$.}
\label{state_equivalence}
\end{figure}

An interesting observation can be made from this result. If the jammer attacks either both or none of the receivers at any given time (i.e., $\lambda_{01}=\lambda_{10}=0$) such that the total probability with which the receivers are jammed together is $(1-\lambda)$ (and hence the probability with which they are not jammed is $\lambda$), the $\DoF$ achievable is $\frac{4}{3}\lambda$ ( $\frac{4}{3}$ is the optimal $\DoF$ achieved in a $2$-receiver MISO BC with d-$\CSIT$ \cite{MAT2012}). This is shown in Fig.~\ref{state_equivalence}. Though such an equivalence is not seen \emph{apriori}, the sum $\DoF$ achieved by this transmission scheme shows that a synergistic benefit is achievable over a long duration of time if all the possible jammer states are used jointly. 
\subsubsection{$K$-User:}
In this subsection, we present a $K$-phase transmission scheme that achieves the $\DoF$ described in Theorem~\ref{TheoremDD-KUser}.  
The achievability of Theorem~\ref{TheoremDD-KUser} is based on the synergistic usage of delayed $\CSIT$ and delayed
$\JSIT$ by exploiting side-information created at the un-jammed receivers in the past and transmitting linear combinations of such side-information symbols in the future. Before we explain the scheme for this configuration, we first give a brief description of the transmission scheme that achieves $\DoF_{\mathsf{MAT}}(K)$ for the $K$-user MISO BC with delayed $\CSIT$ and in the absence of any jamming attacks \cite{MAT2012}. Hereafter this scheme is referred to as the $\mathsf{MAT}$ scheme. 

A $K$-phase transmission scheme is presented in \cite{MAT2012} to achieve $\DoF_{\mathsf{MAT}}(K)$. The transmitter has information about the symbols (or linear combinations of the transmitted symbols) available at the receivers via delayed-$\CSIT$. The first phase of the algorithm sends symbols intended for each receiver. The side information  (symbols that are desired at a user but are available at other users) created at the receivers are used in the subsequent phases of the algorithm to create higher order symbols (symbols required by $>1$ receivers) \cite{MAT2012}, thereby increasing the $\DoF$. 

Specifically, $(K-j+1){K\choose j}$ order $j$ symbols  (symbols intended for $j\leq K$ receivers) are chosen in the $j$th phase to create $j{K\choose j+1}$ order $(j+1)$ symbols that are necessary for $(j+1)\leq K$ receivers and are used in the $(j+1)$th phase of the algorithm. Using this, a recursive relationship between 
$\DoF$ of the $j$th and $(j+1)$th phases is obtained as \cite[eq. (28)]{MAT2012}.\vspace{-10pt}
\begin{align}
\DoF_j(K)=\frac{(K-j+1){K\choose j}}{{K\choose j}+\frac{j {K\choose j+1}}{\DoF_{j+1}(K)}},\label{OrderjMAT}
\end{align}
where $\DoF_j(K)$ is the $\DoF$ of the $K$-user MISO BC to deliver order $j$ symbols. This recursive relationship then leads to the $\DoF$ for a $K$-user MISO BC given by $\DoF_{\mathsf{MAT}}(K)$. See \cite{MAT2012} for a complete description of the coding scheme. 

It is assumed that the decoding process takes place when the receivers have received sufficient linear combinations (LCs) of the intended symbols required to decode their symbols. For example, $n$ \textit{jamming free} LCs are sufficient to decode $n$ symbols at a receiver. The synergistic benefits of transmitting over different jamming states in these configurations is achievable in the long run by exploiting the knowledge about the present and past jamming states. 

%In this subsection, we present a $K$-phase transmission scheme that achieves the $\DoF$ described in Theorem~\ref{TheoremDD-KUser}.  
%For ease of analysis we assume that all the receivers are un-jammed or jammed with equal probabilities $\{\lambda_i\}_{i=1}^K=\lambda$ and $(1-\lambda)$ respectively (see (3) for definition of $\lambda_i$). Note that this assumption does not imply that all possible jamming states are equiprobable. It tells that the probability that any subset $\{j\}_{j=1}^K$ of receivers being jammed is equiprobable with probabilty $\eta_j$ given by \eqref{eta_j}.
Before we present the proposed scheme, notations necessary for the proposed multi-phase transmission scheme are presented. Let $\DoF_{j}(\eta,K)$ denote the $\DoF$ of the $K$-user MISO BC to deliver order $j$ symbols to the users in a scenario where the receivers are jamming free with equal probability $\lambda_{\eta}$ given by \eqref{final_lambda} which is a function of $\eta=[\eta_{0}, \eta_{1}, \ldots, \eta_{K}]$.

We show that in the presence of a jammer, the following relationship (analogous to (\ref{OrderjMAT}))  holds: 
\begin{align}\label{DoF_Iterative_K_user}
\DoF_{j}(\eta,K)=\frac{(K-j+1){K \choose j}}{\frac{{K \choose j}}{\lambda_{\eta}}+\frac{j{K\choose {j+1}}}{\DoF_{j+1}(\eta,K)}}.
\end{align}
Using \eqref{DoF_Iterative_K_user}, it can be shown that the $\DoF$ of a $K$-user MISO BC in the presence of such a jamming attack  is given by
\begin{align}\label{DoF_Iterative_K_user_compare_MAT}
\DoF_{\DD}(\eta,K)\triangleq\DoF_1(\eta,K)=\lambda_{\eta}\DoF_{\mathsf{MAT}}(K),
\end{align}
where $\DoF_{\mathsf{MAT}}(K)$ is given by \eqref{DoF_MAT}.
We initially present the transmission scheme for the $1$st phase and later generalize it for the $j$th $(j\leq K)$ phase.

\paragraph{Phase $1$:} 
Phase $1$ of the coding scheme consists of $K$-stages, one for each receiver. In these stages, symbols intended for each user are transmitted in their respective stages. For instance, let $\left(a_1,a_2,\ldots,a_K\right)$ represent the symbols to be delivered to the $1$st receiver. The transmitter sends these symbols on its $K$ transmit antennas during the $1$st stage. The receivers get \textit{jamming free} LCs of these symbols when they are not jammed. 
% ends when $K$ jammming free random LCs, say $\{f_i(a_1,a_2,\ldots,a_K)\}_{i=1}^K$, of the intended symbols are available at any of the $K$ receivers i.e., 
Each of these $K$-stages end when the LCs intended for a particular receiver are received \textit{jamming free} by at least one of the $K$ receivers. This information (i.e., which LC was received and whether it was received unjammed or not) is available at the transmitter using d-$\CSIT$ and d-$\JSIT$. 

Let $d$ denote the duration of one such stage. A particular receiver is not jammed with probability $\lambda_{\eta}$, and hence 
$\lambda_{\eta} d$ jamming free LCs are available at each of the $K$ receivers. Since $K$ jamming free LCs suffice to decode $K$ symbols, we enforce %\vspace{-10pt} 
%\begin{align}
$K\times (\lambda_{\eta} d)=K \Rightarrow d=\frac{1}{\lambda_{\eta}}$. 
%\end{align}
Since there are $K$ such stages in the $1$st phase, the total time duration of this phase is 
%\begin{align}
$\tau_1=\frac{K}{\lambda_{\eta}}$.
%\end{align}
At the end of this phase, each receiver requires $(K-1)\lambda_{\eta} d=(K-1)$ additional jamming free LCs that are available at the other receivers to decode its symbols. Each receiver has order $1$ LCs (side information) that are required by the other receivers. These order $1$ LCs are used to create order $2$ LCs which are subsequently used in the $2$nd phase of the algorithm. 
Notice that the total number of $(K-1)K$ order $1$ LCs available at the end of this phase can be used to create $\frac{(K-1)K}{2}$ order $2$ LCs that are used in the $2$nd phase of the algorithm. Thus the $\DoF$ can be represented as
\begin{align}
\DoF_1(\eta,K)=\frac{K^{2}}{\tau_1+\tau_2},
\end{align}
where $\tau_2$ is the total time taken to deliver $\frac{(K-1)K}{2}$ order $2$ LCs to the receivers and is given by
\begin{align}
\tau_2=\frac{\frac{(K-1)K}{2}}{\DoF_2(\eta,K)}.
\end{align}
Thus the $\DoF_1(\eta,K)$ is given by 
\begin{align}
\DoF_1(\eta,K)&=\frac{K^2}{\frac{K}{\lambda_{\eta}}+\frac{\frac{(K-1)K}{2}}{\DoF_2(\eta,K)}}=\frac{K}{\frac{1}{\lambda_{\eta}}+\frac{\frac{(K-1)}{2}}{\DoF_2(\eta,K)}}.
\end{align}
Notice here that this conforms with the recursion  given in \eqref{DoF_Iterative_K_user}. 
\paragraph{Phase $j$ :} 
In the $j$th phase the transmitter sends $(K-j+1)$ order $j$-symbols on its $(K-j+1)$ transmit antennas. The $j$th phase has ${K \choose j}$ such stages one each for the ${K \choose j}$ different subsets of $j\leq K$ receivers. It can be shown that $(j+1)$ order $j$ jamming free symbols (LCs) can be used to create $j$ symbols (LCs) of order $(j+1)$. Equivalently, $1$ order $j$ symbol helps to create $\frac{j}{j+1}$ order $(j+1)$ symbols. Hence, $(K-j+1)$ order $j$ jamming free symbols transmitted in the $j$th phase, help to create $(K-j)\frac{j}{j+1}$ order $(j+1)$ symbols which are subsequently transmitted in the $(j+1)$th phase of the algorithm. 
%For any subset of $j$ receivers, there exist $(K-j+1)$ order $j$ symbols necessary for all the receivers in that subset.
%Based on these definitions, it is seen that the $K$th phase of the algorithm does not create any more symbols (LCs). 
%In the $j$th phase the transmitter sends the $(K-j+1)$ order $j$-symbols on $(K-j+1)$ antennas. The $j$th phase has ${K \choose j}$ stages one each for the ${K \choose j}$ different subsets of $j\leq K$ receivers. 

Since each receiver is not jammed with probability $\lambda_{\eta}$, the average time required to deliver an order $j$ symbol (LC) is $\frac{1}{\lambda_{\eta}}$. The total time duration of this phase is $\frac{{K \choose j}}{\lambda_{\eta}}$ since we have ${K \choose j}$ stages. Thus the $j$th phase transmits $(K-j+1){K \choose j}$ \textit{jamming free} symbols of order $j$ in $\frac{{K \choose j}}{\lambda_{\eta}}$ time slots and generates $j{K \choose j+1}$ order $(j+1)$ symbols which are delivered to the receivers in the subsequent phases. The $K$th phase transmits symbols of order $K$ and does not create any new symbols (LCs). %It is seen that in such a scheme, the $K$th phase does not create any new symbols (LCs).  
%Assume there are $n$ such common symbols for a given subset of $j$ receivers. For every receiver that is in this subset of $j$ receivers, receives one LC of these n symbols. Apart from these 
%In any subset of $j\leq K$ receivers, each receiver has a LC that is useful for all the receivers in that corresponding subset \cite{MAT2012}. Hence $(j+1)$ order $j$ symbols generate $j$ order $(j+1)$ symbols. 
%Notice that each of these subsets have $(K-j+1)$ common order $j$ jamming free symbols (LCs) that are sought by all the receivers in that subset. 
%Thus the total number of order $j$ symbols is $(K-j+1){K \choose j}$. At the end of these stages, notice that every receiver in a subset of $(j+1)$ receivers has a LC that is simultaneously useful for the remaining $j$ users in that subset. Thus given a subset of $(j+1)$ receivers, we can form $j$ random LCs of order $(j+1)$. 
%In general, the $j$th phase takes $(K-j+1){K \choose j}$ \textit{jamming free} symbols of order $j$ and generates $j{K \choose j+1}$ order $(j+1)$ symbols \cite{MAT2012}. 
Thus we have\vspace{-10pt}
\begin{align}
\DoF_{j}(\eta,K)=\frac{(K-j+1){K \choose j}}{\frac{{K \choose j}}{\lambda_{\eta}}+\frac{j{K\choose {j+1}}}{\DoF_{j+1}(\eta,K)}},
\end{align}
Using this recurrence relation we can show that 
\begin{align}\label{DoF_K_User_Result}
\DoF_{1}(\eta,K)=\lambda_{\eta} \frac{K}{1+\frac{1}{2}\ldots+\frac{1}{K}}.
\end{align}

\section{Conclusions}\label{Conclusions}
In this paper, the MISO broadcast channel has been studied
in the presence of a time-varying jammer. We introduced a new variable $\JSIT$ to indicate the presence or absence of information regarding the jammer. 
From our results, the interplay between $\CSIT$ and $\JSIT$ and associated impact on the $\DoF$ regions are illuminated. 
For the case in which there is perfect $\CSIT$, by employing a randomized zero-forcing precoding scheme, the $\DoF$ region remains the same irrespective of the availability/un-availability of $\JSIT$. 
%In such a scenario, only the statistical knowledge of the jammer is sufficient to achieve the optimal $\DoF$ region. 
%The transmitter in such a scenario can use randomized pre-coding schemes to achieve $\DoF$ gains compared to naive scehemes.  
On the other hand, for the case of delayed $\CSIT$ and $\JSIT$, our results show that both the jammer and channel state information must be synergistically used in order 
to provide $\DoF$ gains. Whenever there is perfect $\JSIT$, it is seen that the jammers' states are separable and the optimal strategy is to send information symbols independently across different jamming states. The result for the $\NN$ configuration quantifies the $\DoF$ loss  in case of unavailability of $\JSIT$ and $\CSIT$. The results for the $K$-user MISO BC indicate the scaling of the sum $\DoF$  with the number of users in  the presence of jamming attacks. Finally, several interesting open questions and directions emerge out of this work. We outline some of these below. 
\begin{enumerate}
\item It remains unclear if the inner bounds to the $\DoF$ region for the $\DN$ and $\ND$ configurations are optimal. The exact $\DoF$ region for these configurations remains an interesting open problem. The $\DoF$ region achieved by the $\DD$ configuration in the $2$-user MISO BC serves as an outer bound for both $\DN$ and $\ND$ configurations. Improving both these inner and outer bounds for the $\DN$ and $\ND$ configurations is a challenging problem.  
%While it may be difficult to treat these two states independently, a joint analysis of these two states i.e., the case where the availability of $\CSIT$ and $\JSIT$ at the transmitter can vary between $\DN$ and $\ND$ configurations would be feasible. This is similar in spirit to the $\DoF$ analysis done in the presence of alternating $\CSIT$ (in the absence of jamming) in \cite{ACSIT2012}. 
\item For the $\DD$ configuration, a $3$-stage scheme is proposed to achieve the optimal $\DoF$ region. In the $3$rd stage of this coding scheme, the transmitter did not require any $\CSIT$  or $\JSIT$. This raises an interesting question: what is the minimum fraction of time over which $\CSIT$ and $\JSIT$ must be acquired in order to achieve the optimal $\DoF$. A similar problem has been considered in the absence of a jammer  \cite{ACSIT2012}, in which the minimum amount of $\CSIT$ required to achieve a particular $\DoF$ value is characterized. 

\item Finally, the results presented in this paper can possibly be extended to scenarios where the jammers' statistics are not stationary. While the analysis presented in this paper assumes that the jammers' states are i.i.d and that its statistics are constant across time, it would be interesting to understand the behavior of $\DoF$ regions in a scenario where the jammers' states are correlated across time and also possibly correlated with the transmit signals.  
\end{enumerate}

\section{Appendix}\label{Appendix}
\subsection{Converse Proof for Theorem \ref{TheoremPP}}
We first present the proof for the bounds $d_{1}\leq \lambda_{1}$ and $d_{2}\leq \lambda_{2}$ for the ($\CSIT$, $\JSIT$) configuration PP. Clearly, these bounds would also continue to serve as valid outer bounds for the worse configurations $\PD$ and $\PN$.  Since these bounds are symmetric, it suffices to prove that $d_{1}\leq \lambda_{1}$. We have the following sequence of bounds for receiver $1$:
\begin{align}
nR_{1}&= H(W_{1})= H(W_{1}|\mathbf{H}^{n}, S_1^{n}, S_{2}^{n})\\
&= I(W_{1}; Y_{1}^{n}| \mathbf{H}^{n}, S_1^{n}, S_{2}^{n}) + H(W_{1}|Y_{1}^{n}, \mathbf{H}^{n}, S_1^{n}, S_{2}^{n})\\
&\leq  I(W_{1}; Y_{1}^{n}| \mathbf{H}^{n}, S_1^{n}, S_{2}^{n}) + n\epsilon_{n}\label{Fano1}\\
&= h(Y_{1}^{n}| \mathbf{H}^{n}, S_1^{n}, S_{2}^{n}) - h(Y_{1}^{n}| W_{1}, \mathbf{H}^{n}, S_1^{n}, S_{2}^{n}) +n\epsilon_{n}\\
&\leq n\log(P_{T})- h(Y_{1}^{n}| W_{1}, \mathbf{H}^{n}, S_1^{n}, S_{2}^{n}) +n\epsilon_{n}\\
&\leq n\log(P_{T})- h(Y_{1}^{n}| \mathbf{X}^{n}, W_{1}, \mathbf{H}^{n}, S_1^{n}, S_{2}^{n}) +n\epsilon_{n}\\
&= n\log(P_{T})- h(S_{1}^{n}\mathbf{G}_{1}^{n}\mathbf{J}_{1}^{n}+ N_{1}^{n}| \mathbf{X}^{n}, W_{1}, \mathbf{H}^{n}, S_1^{n}, S_{2}^{n}) +n\epsilon_{n}\\
&= n\log(P_{T})- h(S_{1}^{n}\mathbf{G}_{1}^{n}\mathbf{J}_{1}^{n}+ N_{1}^{n}| S_1^{n},S_{2}^{n}) +n\epsilon_{n}\\
&\leq n\log(P_{T})- n(\lambda_{10}+\lambda_{11})\log(P_{T})+n\epsilon_{n}\label{eq2}\\
&= n(1-\lambda_{10}-\lambda_{11})\log(P_{T}) + n\epsilon_{n}\\
&= n(\lambda_{00}+\lambda_{01})\log(P_{T})+ n\epsilon_{n}\\
&=n\lambda_{1}\log(P_{T})+ n\epsilon_{n},\label{eq3}
\end{align}
where (\ref{Fano1}) follows from Fano's inequality, \eqref{eq2} is obtained from the fact that $\mbox{Pr}(S_1(t)=1)=(\lambda_{11}+\lambda_{10})$ and the assumption that the jammer's signal is AWGN with power $P_{T}$. Normalizing \eqref{eq3} by $n\log(P_{T})$, and taking the limit $n\rightarrow \infty$ and then $P_{T}\rightarrow \infty$, we obtain 
\begin{align}
d_{1}&\leq \lambda_{1}.
\end{align}
On similar lines since user $2$ is jammed with probability $(\lambda_{11}+\lambda_{01})$, it can be readily proved that
\begin{align}
d_{2}&\leq(\lambda_{00}+\lambda_{10})=\lambda_{2}.
\end{align}
\subsection{Converse Proof for Theorem \ref{TheoremDD}}
We next provide the proof for the ($\CSIT$, $\JSIT$) configuration $\DD$, in which the transmitter has delayed $\CSIT$ and delayed $\JSIT$. In this case, we prove the bound: 
\begin{align}
\frac{d_{1}}{\lambda_{1}}+\frac{d_{2}}{(\lambda_{1}+\lambda_{2})}&\leq 1.
\end{align}

Let  $\Omega=(\mathbf{H}^{n}, S_{1}^{n}, S_{2}^{n})$ denote the global $\CSIT$ and $\JSIT$ for the entire block length $n$. 
We next enhance the original MISO broadcast channel and make it physically degraded by letting a genie provide 
the output of receiver $1$ to receiver $2$. Formally, in the new MISO BC, receiver $1$ has $(Y_{1}^{n}, \Omega)$
and receiver $2$ has $(Y_{1}^{n}, Y_{2}^{n}, \Omega)$. We next note that for a physically degraded BC, it is known 
from \cite{ElGamalFB} that feedback from the receivers \textit{does not} increase the capacity region. 
We can therefore remove delayed $\CSIT$ and delayed $\JSIT$ from the transmitter without decreasing the capacity region of the enhanced MISO BC. 
The capacity region for this model serves as an outer bound to the capacity region of the original MISO BC. 

Henceforth, we will focus on the model in which receiver $1$ has $(Y_{1}^{n}, \Omega)$, receiver $2$ has $(Y_{1}^{n}, Y_{2}^{n}, \Omega)$ and most importantly, the transmitter has \textit{no} $\CSIT$ and \textit{no} $\JSIT$. 

For such a model, we next state the following key property, which we call as the statistical equivalence property (denoted in short by SEP):
\begin{align}
h(\mathbf{H}_{1}(t)\mathbf{X}(t)+ N_{1}(t))  &= h(\mathbf{H}_{2}(t)\mathbf{X}(t)+ N_{2}(t)).\label{SEP}
\end{align}
This property follows from the following facts:
\begin{enumerate}
\item $\mathbf{H}_{1}(t)$ and $\mathbf{H}_{2}(t)$ are drawn from the same distribution.
\item $N_{1}(t)$ and $N_{2}(t)$ are statistically equivalent, i.e., drawn from the same distribution.
\item $\mathbf{X}(t)$ is independent of $(\mathbf{H}_{1}^{n}, \mathbf{H}_{2}^{n}, N_{1}^{n}, N_{2}^{n})$. 
\end{enumerate}

With these in place, we have the following sequence of bounds for receiver $1$:
\begin{align}
nR_{1}&= H(W_{1})= H(W_{1}|\Omega)\\
&\leq I(W_{1}; Y_{1}^{n} | \Omega) + n\epsilon_{n}\\
&= h(Y_{1}^{n}|\Omega) - h(Y_{1}^{n}|W_{1}, \Omega) + n\epsilon_{n}\\
&\leq n\log(P_{T})- h(Y_{1}^{n}|W_{1}, \Omega) + n\epsilon_{n}.\label{Term1}
\end{align}

We now focus on the second term appearing in (\ref{Term1}):
\begin{align}
h(Y_{1}^{n}|W_{1}, \Omega)&=\sum_{t=1}^{n}h(Y_{1t} | W_{1}, \Omega, Y_{1}^{t-1})\geq \sum_{t=1}^{n}h(Y_{1t} | W_{1}, \Omega, Y_{1}^{t-1}, Y_{2}^{t-1})\\
&= \sum_{t=1}^{n}h(Y_{1t} | S_{1}(t), S_{2}(t), \underbrace{W_{1}, \Omega\setminus\{S_{1}(t), S_{2}(t)\}, Y_{1}^{t-1}, Y_{2}^{t-1}}_{\triangleq U_{t}})\\
&= \sum_{t=1}^{n}h(Y_{1t} | S_{1}(t), S_{2}(t), U_{t})\\
&= \sum_{t=1}^{n}\Big[\lambda_{00}h(Y_{1t} | S_{1}(t)=0, S_{2}(t)=0, U_{t}) \nonumber\\
&\hspace{1.3cm} + \lambda_{01}h(Y_{1t} | S_{1}(t)=0, S_{2}(t)=1, U_{t}) \nonumber\\
&\hspace{1.3cm} + \lambda_{10}h(Y_{1t} | S_{1}(t)=1, S_{2}(t)=0, U_{t}) \nonumber\\
&\hspace{1.3cm} + \lambda_{11}h(Y_{1t} | S_{1}(t)=1, S_{2}(t)=1, U_{t})\Big]\\
&= \sum_{t=1}^{n}\Big[\lambda_{00}h(\mathbf{H}_{1}(t)\mathbf{X}(t)+ N_{1}(t) | U_{t}) \nonumber\\
&\hspace{1.3cm} + \lambda_{01}h(\mathbf{H}_{1}(t)\mathbf{X}(t)+ N_{1}(t) | U_{t}) \nonumber\\
&\hspace{1.3cm} + \lambda_{10}h(\mathbf{H}_{1}(t)\mathbf{X}(t)+ \mathbf{G}_{1}(t)\mathbf{J}(t)+ N_{1}(t) | U_{t}) \nonumber\\
&\hspace{1.3cm} + \lambda_{11}h(\mathbf{H}_{1}(t)\mathbf{X}(t)+ \mathbf{G}_{1}(t)\mathbf{J}(t)+ N_{1}(t) | U_{t})\Big]\label{E1}\\
&= \sum_{t=1}^{n}\Big[(\lambda_{00}+\lambda_{01})h(\mathbf{H}_{1}(t)\mathbf{X}(t)+ N_{1}(t) | U_{t})\nonumber\\
&\hspace{1.3cm} + (\lambda_{10}+\lambda_{11})h(\mathbf{H}_{1}(t)\mathbf{X}(t)+ \mathbf{G}_{1}(t)\mathbf{J}(t)+ N_{1}(t) | U_{t}) \Big]\\
&\geq \sum_{t=1}^{n}\Big[(\lambda_{00}+\lambda_{01})h(\mathbf{H}_{1}(t)\mathbf{X}(t)+ N_{1}(t) | U_{t})\nonumber\\
&\hspace{1.3cm} + (\lambda_{10}+\lambda_{11})h(\mathbf{H}_{1}(t)\mathbf{X}(t)+ \mathbf{G}_{1}(t)\mathbf{J}(t)+ N_{1}(t) | \mathbf{H}_{1}(t)\mathbf{X}(t), U_{t}) \Big]\\
\end{align}
\begin{align}
&= \sum_{t=1}^{n}\Big[(\lambda_{00}+\lambda_{01})h(\mathbf{H}_{1}(t)\mathbf{X}(t)+ N_{1}(t) | U_{t})\nonumber\\
&\hspace{1.3cm} + (\lambda_{10}+\lambda_{11})h( \mathbf{G}_{1}(t)\mathbf{J}(t)+ N_{1}(t)) \Big]\\
&\geq \sum_{t=1}^{n}\Big[(\lambda_{00}+\lambda_{01})\underbrace{h(\mathbf{H}_{1}(t)\mathbf{X}(t)+ N_{1}(t) | U_{t})}_{\triangleq \eta_{t}}+ (\lambda_{10}+\lambda_{11})\log(P_{T})\Big]\\
&= (\lambda_{00}+\lambda_{01})\sum_{t=1}^{n}\eta_{t} + n(\lambda_{10}+\lambda_{11})\log(P_{T})\label{E2}
\end{align}
where (\ref{E1}) follows from the fact that the random variables $S_{1}(t), S_{2}(t)$ are i.i.d. across time, and independent of all other random variables including $(U_{t}, N_{1}(t), \mathbf{X}(t), \mathbf{H}_{1}(t))$, i.e., we have used that 
$h(\mathbf{H}_{1}(t)\mathbf{X}(t)+ N_{1}(t) | S_{1}(t)=0, S_{2}(t)=0, U_{t})= h(\mathbf{H}_{1}(t)\mathbf{X}(t)+ N_{1}(t) | U_{t})$ and similar simplifications for the remaining three terms. In (\ref{E2}), we have defined
\begin{align}
\eta_{t}\triangleq h(\mathbf{H}_{1}(t)\mathbf{X}(t)+ N_{1}(t) | U_{t}).
\end{align}

Substituting (\ref{E2}) back in (\ref{Term1}), we obtain
\begin{align}
nR_{1}&\leq n(\lambda_{00}+\lambda_{01})\log(P_{T}) - (\lambda_{00}+\lambda_{01})\sum_{t=1}^{n}\eta_{t} + n\epsilon_{n}\\
&= n\lambda_{1}\log(P_{T}) - \lambda_{1}\Bigg[\sum_{t=1}^{n}\eta_{t}\Bigg] + n\epsilon_{n}\label{Term1a}
\end{align}

We next focus on the receiver $2$ which has access to both $Y_{1}^{n}$ and $Y_{2}^{n}$:
\begin{align}
nR_{2}&= H(W_{2})= H(W_{2}|W_{1},\Omega)\\
&\leq I(W_{2}; Y_{1}^{n}, Y_{2}^{n}| W_{1}, \Omega) + n\epsilon_{n}\\
&= h(Y_{1}^{n}, Y_{2}^{n}| W_{1}, \Omega) - h(Y_{1}^{n}, Y_{2}^{n}| W_{1}, W_{2}, \Omega) + n\epsilon_{n}\\
&\leq h(Y_{1}^{n}, Y_{2}^{n}| W_{1}, \Omega) - h(Y_{1}^{n}, Y_{2}^{n}| \mathbf{X}^{n}, W_{1}, W_{2}, \Omega) + n\epsilon_{n}\\
&\leq h(Y_{1}^{n}, Y_{2}^{n}| W_{1}, \Omega) - n(\lambda_{01}+\lambda_{10}+2\lambda_{11})\log(P_{T}) + n\epsilon_{n},\label{E3}
\end{align} 
where (\ref{E3}) follows from the fact that given $(\mathbf{X}^{n}, W_{1}, W_{2}, \Omega)$, the contribution of the information bearing signals (i.e., $\mathbf{H}_{k}^{n}\mathbf{X}^{n}$ for $k=1,2$) can be removed from $(Y_{1}^{n}, Y_{2}^{n})$, and we are left only with jamming signals (which are assumed to be Gaussian with power $P_{T}$, i.i.d. across time and independent of all other random variables) and unit variance Gaussian noise, the entropy of which can be lower bounded as in (\ref{E3}). 

We next expand the first term in (\ref{E3}) as follows:
\begin{align}
h(Y_{1}^{n}, Y_{2}^{n}| W_{1}, \Omega)&=\sum_{t=1}^{n} h(Y_{1t}, Y_{2t}| W_{1}, \Omega, Y_{1}^{t-1}, Y_{2}^{t-1})\\
&=  \sum_{t=1}^{n}h(Y_{1t}, Y_{2t} | S_{1}(t), S_{2}(t), \underbrace{W_{1}, \Omega\setminus\{S_{1}(t), S_{2}(t)\}, Y_{1}^{t-1}, Y_{2}^{t-1}}_{\triangleq U_{t}})\\
&= \sum_{t=1}^{n} h(Y_{1t}, Y_{2t} | S_{1}(t), S_{2}(t), U_{t})
\end{align}
\begin{align}
&= \sum_{t=1}^{n} \Big[ \lambda_{00}h(Y_{1t}, Y_{2t} | S_{1}(t)=0, S_{2}(t)=0, U_{t})\nonumber\\
&\hspace{1.3cm} + \lambda_{01}h(Y_{1t}, Y_{2t} | S_{1}(t)=0, S_{2}(t)=1, U_{t})\nonumber\\
&\hspace{1.3cm} + \lambda_{10}h(Y_{1t}, Y_{2t} | S_{1}(t)=1, S_{2}(t)=0, U_{t})\nonumber\\
&\hspace{1.3cm} + \lambda_{11}h(Y_{1t}, Y_{2t} | S_{1}(t)=1, S_{2}(t)=1, U_{t})\Big].\label{E4}
\end{align}
We next bound each one of the four terms in (\ref{E4}) as follows:
\begin{align}
&h(Y_{1t}, Y_{2t} | S_{1}(t)=0, S_{2}(t)=0, U_{t})\nonumber\\
&= h(\mathbf{H}_{1}(t)\mathbf{X}(t)+ N_{1}(t), \mathbf{H}_{2}(t)\mathbf{X}(t)+ N_{2}(t)| S_{1}(t)=0, S_{2}(t)=0, U_{t})\\
&\leq h(\mathbf{H}_{1}(t)\mathbf{X}(t)+ N_{1}(t)| S_{1}(t)=0, S_{2}(t)=0, U_{t}) \nonumber\\
&\qquad+ h(\mathbf{H}_{2}(t)\mathbf{X}(t)+ N_{2}(t)| S_{1}(t)=0, S_{2}(t)=0, U_{t})\\
&= h(\mathbf{H}_{1}(t)\mathbf{X}(t)+ N_{1}(t)|  U_{t}) + h(\mathbf{H}_{2}(t)\mathbf{X}(t)+ N_{2}(t)| U_{t})\\
&= 2\eta_{t}\label{SEPuse12},
\end{align}
where in (\ref{SEPuse12}), we have made use of the (conditional version of) statistical equivalence property (SEP) for the two receivers as stated in (\ref{SEP}).
\begin{align}
&h(Y_{1t}, Y_{2t} | S_{1}(t)=0, S_{2}(t)=1, U_{t})\nonumber\\
&= h(\mathbf{H}_{1}(t)\mathbf{X}(t)+ N_{1}(t), \mathbf{H}_{2}(t)\mathbf{X}(t)+ \mathbf{G}_{2}(t)\mathbf{J}(t)+ N_{2}(t)| S_{1}(t)=0, S_{2}(t)=1, U_{t})\\
&\leq h(\mathbf{H}_{1}(t)\mathbf{X}(t)+ N_{1}(t)| S_{1}(t)=0, S_{2}(t)=1, U_{t}) \nonumber\\
&\qquad+ h(\mathbf{H}_{2}(t)\mathbf{X}(t)+  \mathbf{G}_{2}(t)\mathbf{J}(t)+N_{2}(t)| S_{1}(t)=0, S_{2}(t)=1, U_{t})\\
&\leq h(\mathbf{H}_{1}(t)\mathbf{X}(t)+ N_{1}(t)|  U_{t}) + \log(P_{T})\\
&= \eta_{t}+ \log(P_{T})\label{SEPuse1}.
\end{align}
In summary, we have 
\begin{align}
h(Y_{1t}, Y_{2t} | S_{1}(t)=0, S_{2}(t)=0, U_{t})&\leq 2\eta_{t}\\
h(Y_{1t}, Y_{2t} | S_{1}(t)=0, S_{2}(t)=1, U_{t})&\leq \eta_{t} + \log(P_{T})\\
h(Y_{1t}, Y_{2t} | S_{1}(t)=1, S_{2}(t)=0, U_{t})&\leq \eta_{t} + \log(P_{T})\\
h(Y_{1t}, Y_{2t} | S_{1}(t)=1, S_{2}(t)=1, U_{t})&\leq 2\log(P_{T}).
\end{align}
Substituting these back in (\ref{E4}), we obtain
\begin{align}
h(Y_{1}^{n}, Y_{2}^{n}| W_{1}, \Omega)&\leq n(\lambda_{01}+\lambda_{10}+2\lambda_{11})\log(P_{T}) + (\lambda_{01}+\lambda_{10}+2\lambda_{00})\Bigg[\sum_{t=1}^{n}\eta_{t}\Bigg]\label{E5}
\end{align}
Upon substituting (\ref{E5}) back in (\ref{E3}), we have the following bound on $R_{2}$:
\begin{align}
nR_{2}&\leq (\lambda_{01}+\lambda_{10}+2\lambda_{00})\Bigg[\sum_{t=1}^{n}\eta_{t}\Bigg] + n\epsilon_{n}\\
&= (\lambda_{1}+\lambda_{2})\Bigg[\sum_{t=1}^{n}\eta_{t}\Bigg] + n\epsilon_{n}\label{Term2a}
\end{align}
In summary, from (\ref{Term1a}) and (\ref{Term2a}), we can write
\begin{align}
nR_{1}&\leq n\lambda_{1}\log(P_{T}) - \lambda_{1}\Bigg[\sum_{t=1}^{n}\eta_{t}\Bigg] + n\epsilon_{n}\label{Term1Final}\\
nR_{2}&\leq (\lambda_{1}+\lambda_{2})\Bigg[\sum_{t=1}^{n}\eta_{t}\Bigg] + n\epsilon_{n}\label{Term2Final}
\end{align}
Eliminating the term $\Big[\sum_{t=1}^{n}\eta_{t}\Big]$, we obtain
\begin{align}
n\frac{R_{1}}{\lambda_{1}} + n\frac{R_{2}}{(\lambda_{1}+\lambda_{2})}&\leq n\log(P_{T}) + n\epsilon^{'}_{n}
\end{align}
Normalizing by $n\log(P_{T})$, and taking the limits $n\rightarrow \infty$ and then $P_{T}\rightarrow \infty$, we obtain the bound:
\begin{align}
\frac{d_{1}}{\lambda_{1}}+ \frac{d_{2}}{(\lambda_{1}+\lambda_{2})}&\leq 1.
\end{align}
Reversing the role of receivers $1$ and $2$, i.e., making receiver $2$ degraded with respect to receiver $1$, we can similarly obtain the other bound
\begin{align}
\frac{d_{1}}{(\lambda_{1}+\lambda_{2})}+ \frac{d_{2}}{\lambda_{2}}&\leq 1.
\end{align}
This completes the proof of the converse for Theorem \ref{TheoremDD}.

\subsection{Converse Proof for Theorem \ref{TheoremDP}}
We next provide the proof for the ($\CSIT$, $\JSIT$) configuration $\DP$, in which the transmitter has delayed $\CSIT$ and perfect (instantaneous) $\JSIT$. In this case, we prove the bound: 
\begin{align}
2d_{1}+d_{2}&\leq 2\lambda_{00}+2\lambda_{01}+\lambda_{10}
\end{align}

Let  $\Omega=(\mathbf{H}^{n}, S_{1}^{n}, S_{2}^{n})$ denote the global $\CSIT$ and $\JSIT$ for the entire block length $n$. 
As in the proof for Theorem \ref{TheoremDD}, we enhance the original MISO broadcast channel and make it physically degraded by letting a genie provide 
the output of receiver $1$ to receiver $2$. Formally, in the new MISO BC, receiver $1$ has $(Y_{1}^{n}, \Omega)$
and receiver $2$ has $(Y_{1}^{n}, Y_{2}^{n}, \Omega)$. We next note that for a physically degraded BC, it is known 
from \cite{ElGamalFB} that feedback from the receivers \textit{does not} increase the capacity region. 
We can therefore remove delayed $\CSIT$ from the transmitter without decreasing the capacity region of the enhanced MISO BC. 
The capacity region for this model serves as an outer bound to the capacity region of the original MISO BC. 

Henceforth, we will focus on the model in which receiver $1$ has $(Y_{1}^{n}, \Omega)$, receiver $2$ has $(Y_{1}^{n}, Y_{2}^{n}, \Omega)$ and 
most importantly, the transmitter has \textit{no} $\CSIT$. Note that unlike in proof for Theorem \ref{TheoremDD}, in this case we \textit{cannot} remove the
assumption of perfect $\JSIT$.  Recall that in the proof of Theorem \ref{TheoremDD}, we made use of the following relationships (which we called as the statistical equivalence property):
\begin{align}
&h(\mathbf{H}_{1}(t)\mathbf{X}(t)+ N_{1}(t) | S_{1}(t)=i, S_{2}(t)=j, U_{t})  \nonumber\\
&\quad= h(\mathbf{H}_{2}(t)\mathbf{X}(t)+ N_{2}(t) | S_{1}(t)=i^{'}, S_{2}(t)=j^{'}, U_{t}).\label{SEPDD}
\end{align}
for $i, i^{'}, j, j^{'}\in \{0,1\}$.
In this case, we can only use a stricter version of the statistical equivalence property:
\begin{align}
&h(\mathbf{H}_{1}(t)\mathbf{X}(t)+ N_{1}(t) | S_{1}(t)=0, S_{2}(t)=0, U_{t})  \nonumber\\
&\quad= h(\mathbf{H}_{2}(t)\mathbf{X}(t)+ N_{2}(t) | S_{1}(t)=0, S_{2}(t)=0, U_{t}).\label{SEPDP}
\end{align}
The reason is that for the $\DP$ configuration, due to the fact that the transmitter has perfect $\JSIT$, the marginal probabilities $p(X(t)| S_{1}(t)=i, S_{2}(t)=j, U_{t})$
can depend explicitly on $(i,j)$, the realization of jammer's strategies at time $t$, which was not the case in Theorem \ref{TheoremDD}.

With these in place, we have the following sequence of bounds for receiver $1$:
\begin{align}
nR_{1}&\leq n\log(P_{T})- h(Y_{1}^{n}|W_{1}, \Omega) + n\epsilon_{n}.\label{TermDP1}
\end{align}
We next focus on the second term in (\ref{TermDP1}):
\begin{align}
h(Y_{1}^{n}|W_{1}, \Omega)&=\sum_{t=1}^{n}h(Y_{1t} | W_{1}, \Omega, Y_{1}^{t-1})\geq \sum_{t=1}^{n}h(Y_{1t} | W_{1}, \Omega, Y_{1}^{t-1}, Y_{2}^{t-1})\\
&= \sum_{t=1}^{n}h(Y_{1t} | S_{1}(t), S_{2}(t), \underbrace{W_{1}, \Omega\setminus\{S_{1}(t), S_{2}(t)\}, Y_{1}^{t-1}, Y_{2}^{t-1}}_{\triangleq U_{t}})\\
&= \sum_{t=1}^{n}h(Y_{1t} | S_{1}(t), S_{2}(t), U_{t})\nonumber \\
&= \sum_{t=1}^{n}\Big[\lambda_{00}h(Y_{1t} | S_{1}(t)=0, S_{2}(t)=0, U_{t}) \nonumber\\
&\hspace{1.3cm} + \lambda_{01}h(Y_{1t} | S_{1}(t)=0, S_{2}(t)=1, U_{t}) \nonumber\\
&\hspace{1.3cm} + \lambda_{10}h(Y_{1t} | S_{1}(t)=1, S_{2}(t)=0, U_{t}) \nonumber\\
&\hspace{1.3cm} + \lambda_{11}h(Y_{1t} | S_{1}(t)=1, S_{2}(t)=1, U_{t})\Big]\nonumber\\
&= \sum_{t=1}^{n}\Big[\lambda_{00}h(\mathbf{H}_{1}(t)\mathbf{X}(t)+ N_{1}(t) | S_{1}(t)=0, S_{2}(t)=0, U_{t}) \nonumber\\
&\hspace{1.3cm} + \lambda_{01}\underbrace{h(\mathbf{H}_{1}(t)\mathbf{X}(t)+ N_{1}(t) | S_{1}(t)=0, S_{2}(t)=1, U_{t})}_{\geq 0} \nonumber\\
&\hspace{1.3cm} + \lambda_{10}\underbrace{h(\mathbf{H}_{1}(t)\mathbf{X}(t)+ \mathbf{G}_{1}(t)\mathbf{J}(t)+ N_{1}(t) | S_{1}(t)=1, S_{2}(t)=0, U_{t})}_{\geq \log(P_{T})} \nonumber\\
&\hspace{1.3cm} + \lambda_{11}\underbrace{h(\mathbf{H}_{1}(t)\mathbf{X}(t)+ \mathbf{G}_{1}(t)\mathbf{J}(t)+ N_{1}(t) | S_{1}(t)=1, S_{2}(t)=1, U_{t})}_{\geq \log(P_{T})}\Big]\label{E1DP}\nonumber \\
&\geq \sum_{t=1}^{n}\Big[\lambda_{00}\underbrace{h(\mathbf{H}_{1}(t)\mathbf{X}(t)+ N_{1}(t) | S_{1}(t)=0, S_{2}(t)=0, U_{t})}_{\triangleq \eta^{(00)}_{t}} \nonumber\\
&\hspace{1.3cm} + \lambda_{01}\underbrace{h(\mathbf{H}_{1}(t)\mathbf{X}(t)+ N_{1}(t) | S_{1}(t)=0, S_{2}(t)=1, U_{t})}_{\triangleq \eta^{(01)}_{t}} \nonumber\\
&\hspace{1.3cm} + (\lambda_{10}+\lambda_{11})\log(P_{T})\Big]%\label{E2DP}
\end{align}
\begin{align}
&= \lambda_{00}\sum_{t=1}^{n}\eta^{(00)}_{t} + \lambda_{01}\sum_{t=1}^{n}\eta^{(01)}_{t} + n(\lambda_{10}+\lambda_{11})\log(P_{T}),\label{E2DP}
\end{align}
where in (\ref{E1DP}), we used the fact that elements of $\mathbf{J}_{T}$ are i.i.d. with variance $P_{T}$, and in (\ref{E2DP}), we have defined
\begin{align}
\eta^{(00)}_{t}&\triangleq h(\mathbf{H}_{1}(t)\mathbf{X}(t)+ N_{1}(t) | S_{1}(t)=0, S_{2}(t)=0,U_{t})\\
\eta^{(01)}_{t}&\triangleq h(\mathbf{H}_{1}(t)\mathbf{X}(t)+ N_{1}(t) | S_{1}(t)=0, S_{2}(t)=1,U_{t}).
\end{align}
Substituting (\ref{E2DP}) in (\ref{TermDP1}), we obtain
\begin{align}
nR_{1}&\leq n(\lambda_{00}+\lambda_{01})\log(P_{T}) - \lambda_{00}\sum_{t=1}^{n}\eta^{(00)}_{t} - \lambda_{01}\sum_{t=1}^{n}\eta^{(01)}_{t} + n\epsilon_{n}\label{F1DP}
\end{align}
We next focus on the receiver $2$ which has access to both $Y_{1}^{n}$ and $Y_{2}^{n}$. We can obtain the following bound similar to the one obtained in the proof for Theorem \ref{TheoremDD}:
\begin{align}
nR_{2}&\leq h(Y_{1}^{n}, Y_{2}^{n}| W_{1}, \Omega) - n(\lambda_{01}+\lambda_{10}+2\lambda_{11})\log(P_{T}) + n\epsilon_{n},\label{E3DP}
\end{align} 
We next expand the first term in (\ref{E3DP}) as follows:
\begin{align}
h(Y_{1}^{n}, Y_{2}^{n}| W_{1}, \Omega)&=\sum_{t=1}^{n} h(Y_{1t}, Y_{2t}| W_{1}, \Omega, Y_{1}^{t-1}, Y_{2}^{t-1})\\
&=  \sum_{t=1}^{n}h(Y_{1t}, Y_{2t} | S_{1}(t), S_{2}(t), \underbrace{W_{1}, \Omega\setminus\{S_{1}(t), S_{2}(t)\}, Y_{1}^{t-1}, Y_{2}^{t-1}}_{\triangleq U_{t}})\\
&= \sum_{t=1}^{n} h(Y_{1t}, Y_{2t} | S_{1}(t), S_{2}(t), U_{t})\\
&= \sum_{t=1}^{n} \Big[ \lambda_{00}h(Y_{1t}, Y_{2t} | S_{1}(t)=0, S_{2}(t)=0, U_{t})\nonumber\\
&\hspace{1.3cm} + \lambda_{01}\underbrace{h(Y_{1t}, Y_{2t} | S_{1}(t)=0, S_{2}(t)=1, U_{t})}_{\leq h(Y_{1t} | S_{1}(t)=0, S_{2}(t)=1, U_{t})+ \log(P_{T})}\nonumber\\
&\hspace{1.3cm} + \lambda_{10}\underbrace{h(Y_{1t}, Y_{2t} | S_{1}(t)=1, S_{2}(t)=0, U_{t})}_{\leq 2\log(P_{T})}\nonumber\\
&\hspace{1.3cm} + \lambda_{11}\underbrace{h(Y_{1t}, Y_{2t} | S_{1}(t)=1, S_{2}(t)=1, U_{t})}_{\leq 2\log(P_{T})}\Big]\nonumber \\
&\leq \sum_{t=1}^{n} \Big[ \lambda_{00}h(Y_{1t}, Y_{2t} | S_{1}(t)=0, S_{2}(t)=0, U_{t})\nonumber\\
&\hspace{1.3cm} + \lambda_{01}h(Y_{1t}| S_{1}(t)=0, S_{2}(t)=1, U_{t})\nonumber\\
&\hspace{1.3cm} + (\lambda_{01}+2\lambda_{10}+2\lambda_{11})\log(P_{T})\Big]\nonumber
\end{align}
\begin{align}
&\leq \sum_{t=1}^{n} \Big[ \lambda_{00}h(Y_{1t}| S_{1}(t)=0, S_{2}(t)=0, U_{t})\nonumber\\
&\hspace{1.3cm} + \lambda_{00}h(Y_{2t}| S_{1}(t)=0, S_{2}(t)=0, U_{t})\nonumber\\
&\hspace{1.3cm} + \lambda_{01}h(Y_{1t}| S_{1}(t)=0, S_{2}(t)=1, U_{t})\nonumber\\
&\hspace{1.3cm} + (\lambda_{01}+2\lambda_{10}+2\lambda_{11})\log(P_{T})\Big]\\
&= \sum_{t=1}^{n} \Big[ \lambda_{00}\underbrace{h(\mathbf{H}_{1}(t)\mathbf{X}(t)+ N_{1}(t) | S_{1}(t)=0, S_{2}(t)=0, U_{t})}_{= \ \eta^{(00)}_{t}}\nonumber\\
&\hspace{1.3cm} + \lambda_{00}\underbrace{h(\mathbf{H}_{2}(t)\mathbf{X}(t)+ N_{2}(t) | S_{1}(t)=0, S_{2}(t)=0, U_{t})}_{= \ \eta^{(00)}_{t}}\nonumber\\
&\hspace{1.3cm} + \lambda_{00}\underbrace{h(\mathbf{H}_{1}(t)\mathbf{X}(t)+ N_{1}(t) | S_{1}(t)=0, S_{2}(t)=1, U_{t})}_{= \ \eta^{(01)}_{t}}\nonumber\\
&\hspace{1.3cm} + (\lambda_{01}+2\lambda_{10}+2\lambda_{11})\log(P_{T})\Big]\\
&= 2\lambda_{00}\sum_{t=1}^{n}\eta^{(00)}_{t} + \lambda_{01}\sum_{t=1}^{n}\eta^{(01)}_{t} + n(\lambda_{01}+2\lambda_{10}+2\lambda_{11})\log(P_{T}).\label{E4DP}
\end{align}
Substituting (\ref{E4DP}) in (\ref{E3DP}), we get
\begin{align}
nR_{2}&\leq 2\lambda_{00}\sum_{t=1}^{n}\eta^{(00)}_{t} + \lambda_{01}\sum_{t=1}^{n}\eta^{(01)}_{t} + n\lambda_{10}\log(P_{T}) + n\epsilon_{n}.\label{E5DP}
\end{align}
Collectively, from (\ref{F1DP}) and (\ref{E5DP}), we can then write:
\begin{align}
nR_{1}&\leq n(\lambda_{00}+\lambda_{01})\log(P_{T}) - \lambda_{00}\sum_{t=1}^{n}\eta^{(00)}_{t} - \lambda_{01}\sum_{t=1}^{n}\eta^{(01)}_{t} + n\epsilon_{n}\label{M1DP}\\
nR_{2}&\leq 2\lambda_{00}\sum_{t=1}^{n}\eta^{(00)}_{t} + \lambda_{01}\sum_{t=1}^{n}\eta^{(01)}_{t} + n\lambda_{10}\log(P_{T}) + n\epsilon_{n}\label{M2DP}
\end{align}
Taking $2\times$ (\ref{M1DP}) $+$ (\ref{M2DP}),  we obtain:
\begin{align}
n(2R_{1}+R_{2})&\leq n(2\lambda_{00}+ 2\lambda_{01} + \lambda_{10})\log(P_{T}) - \lambda_{01}\sum_{t=1}^{n}\eta^{(01)}_{t}+ n\epsilon_{n}\\
&\leq n(2\lambda_{00}+ 2\lambda_{01} + \lambda_{10})\log(P_{T})+ n\epsilon_{n},
\end{align}
where we used the fact that $\eta^{00}_{t}\geq 0$ for all $t$.
Normalizing by $n\log(P_{T})$, and taking the limits $n\rightarrow \infty$, and then $P_{T}\rightarrow \infty$, we obtain
\begin{align}
2d_{1}+d_{2}&\leq 2\lambda_{00}+2\lambda_{01}+\lambda_{10}.
\end{align}
Reversing the roles of receivers $1$ and $2$, we can obtain the other bound:
\begin{align}
d_{1}+2d_{2}&\leq 2\lambda_{00}+2\lambda_{10}+\lambda_{01}.
\end{align}

\subsection{Converse Proof for Theorem \ref{TheoremNN}}
Here, we consider the configuration in which there is no $\CSIT$ and no $\JSIT$ i.e., $\NN$ configuration and prove the bound:
\begin{align}
\frac{d_{1}}{\lambda_{1}}+\frac{d_{2}}{\lambda_{2}}&\leq 1.
\end{align}

To this end, we recall a classical result  \cite{Bergmans1973}, which states
that for memoryless broadcast channels without feedback, the capacity region only depends on marginal distributions $p(y_{k}|x)$, for $k=1,2$. 
This implies for the problem at hand, in which the jammer's strategy is memoryless, and there is no $\CSIT$ and no $\JSIT$,
the capacity region only depends on the \textit{marginal} probabilities $\lambda_{1}$ and $\lambda_{2}$, i.e., the probabilities with which 
each of the receiver is not jammed.  Without loss of generality, assume that $\lambda_{1}\geq \lambda_{2}$, i.e., receiver $2$ is jammed with higher probability than receiver $1$.

We will now show that this MISO BC falls in the class of \textit{stochastically degraded} broadcast channels. We first recall that a broadcast channel (defined by $p(y_{1},y_{2}|x)$) is stochastically degraded \cite{NITBook} if there exists a random variable $Y_{1^{'}}$ such that
\begin{enumerate}
\item $Y_{1^{'}}|\{X=x\}\sim p_{Y_{1}|X}(y_{1^{'}}|x)$, i.e., $Y_{1^{'}}$ has the same conditional distribution as $Y_{1}$ (given $X$), and
\item $X\rightarrow Y_{1^{'}}\rightarrow Y_{2}$ form a Markov chain. 
\end{enumerate}

Hence, in order to show that the MISO BC with no $\CSIT$ and no $\JSIT$ is stochastically degraded, we will show the existence of a random variable $Y_{1^{'}}$ such that $Y_{1^{'}}$ has the same conditional pdf as $Y_{1}$  and $X\rightarrow Y_{1^{'}}\rightarrow Y_{2}$ form a Markov chain.
We first note that the channel outputs for the original BC at time $t$ are:
\begin{align}
Y_{1}(t)&= \mathbf{H}_{1}(t)\mathbf{X}(t)+ S_{1}(t)\mathbf{G}_{1}(t)\mathbf{J}(t)+ N_{1}(t)\\
Y_{2}(t)&= \mathbf{H}_{2}(t)\mathbf{X}(t)+ S_{2}(t)\mathbf{G}_{2}(t)\mathbf{J}(t)+ N_{2}(t).
\end{align}
Next, we create an artificial output $Y_{1^{'}}$, defined at time $t$ as:
\begin{align}
Y_{1^{'}}(t)&= \mathbf{H}_{2}(t)\mathbf{X}(t)+ \tilde{S}(t)S_{2}(t)\mathbf{G}_{2}(t)\mathbf{J}(t)+ N_{2}(t),
\end{align}
where the random variable $\tilde{S}(t)$ is distributed i.i.d. as follows:
\begin{align}
\tilde{S}(t)=
\begin{cases}
0, & \mbox{ w.p. } \frac{\lambda_{1}-\lambda_{2}}{1-\lambda_{2}},\\
1, & \mbox{ w.p. } \frac{1-\lambda_{1}}{1-\lambda_{2}}.
\end{cases}
\end{align}
Furthermore, $\tilde{S}(t)$ is independent of all other random variables. 

It is straightforward to verify that $Y_{1^{'}}$ and $Y_{1}$ have the same marginal distribution: since \break $\mathbf{H}_{1}(t)$ and $\mathbf{H}_{2}(t)$ are identically distributed, $\mathbf{G}_{1}(t)$ and $\mathbf{G}_{2}(t)$ are identically distributed, $N_{1}(t)$ and $N_{2}(t)$ are identically distributed, and most importantly, the random variables $\tilde{S}(t)S_{2}(t)$ and $S_{1}(t)$ are identically distributed. Furthermore, note that when $\tilde{S}(t)=0$, then $Y_{2}(t)= Y_{1^{'}}(t)+ \mathbf{G}_{2}(t)\mathbf{J}(t)+ N_{2}(t)$, and when $\tilde{S}(t)=0$, then we have $Y_{2}(t)= Y_{1^{'}}(t)$, i.e., $X(t)\rightarrow Y_{1^{'}}(t)\rightarrow Y_{2}(t)$ forms a Markov chain. 

This argument proves that the original MISO broadcast channel with no $\CSIT$ falls in the class of stochastically degraded broadcast channels, for which the capacity region is given by the set of rate pairs $(R_{1}, R_{2})$ satisfying:
\begin{align}
R_{2}&\leq I(U; Y_{2}| \mathbf{H}, S_{1}, S_{2})\\
R_{1}&\leq I(\mathbf{X}; Y_{1}| U, \mathbf{H}, S_{1}, S_{2}),
\end{align}
where $U\rightarrow X\rightarrow (Y_{1}, Y_{2}, S_{1}, S_{2})$ forms a Markov chain. 
Using this, we can write
\begin{align}
R_{2}&\leq h(Y_{2}| \mathbf{H}, S_{1}, S_{2}) - h(Y_{2}| U, \mathbf{H}, S_{1}, S_{2})\\
&\leq \log(P_{T}) - (1-\lambda_{2})\log(P_{T})-\lambda_{2}h(\mathbf{H}_{2}X+N_{2}|U, \mathbf{H}) + o(\log(P_{T}))\\
&= \lambda_{2}\log(P_{T}) -\lambda_{2}h(\mathbf{H}_{2}X+N_{2}|U, \mathbf{H}) + o(\log(P_{T})).\label{T3a}
\end{align}
Similarly, the other bound can be written as:
\begin{align}
R_{1}&\leq h(Y_{1}| U, \mathbf{H}, S_{1}, S_{2}) - h(Y_{1}| \mathbf{X}, U, \mathbf{H}, S_{1}, S_{2})\\
&= (1-\lambda_{1})\log(P_{T}) + \lambda_{1}h(\mathbf{H}_{1}X+N_{1}|U, \mathbf{H}) - (1-\lambda_{2})\log(P_{T}) + o(\log(P_{T}))\\
&=  \lambda_{1}h(\mathbf{H}_{1}X+N_{1}|U, \mathbf{H}) + o(\log(P_{T}))\\
&= \lambda_{1}h(\mathbf{H}_{2}X+N_{2}|U, \mathbf{H}) + o(\log(P_{T})),\label{T3b}
\end{align}
where (\ref{T3b}) follows from the statistically equivalence property (as stated in the previous section).
Combining (\ref{T3a}) and (\ref{T3b}), we obtain:
\begin{align}
\frac{R_{1}}{\lambda_{1}}+ \frac{R_{2}}{\lambda_{2}}&\leq \log(P_{T}) + o(\log(P_{T}))
\end{align}
Normalizing by $\log(P_{T})$ and taking the limit $P_{T}\rightarrow \infty$, we have the proof for 
\begin{align}
\frac{d_{1}}{\lambda_{1}}+\frac{d_{2}}{\lambda_{2}}&\leq 1.
\end{align}

\subsection{Converse Proof for Theorem \ref{TheoremNP}}
Here, we consider the configuration in which there is no $\CSIT$ and perfect $\JSIT$ i.e., $\NP$ configuration and prove the bound:
\begin{align}
d_{1}+d_{2} &\leq \lambda_{00}+\lambda_{01}+\lambda_{10}.
\end{align}
Let  $\Omega=(S_{1}^{n}, S_{2}^{n})$ denote the global $\JSIT$ for the entire block length $n$. 
We have the following sequence of bounds 
\begin{align}
n(R_1+R_2)&=H(W_1)+H(W_2) \\
&=H(W_1,W_2) \\
&= H(W_1,W_2|\Omega) \\
&=I(W_1,W_2;Y_1^n,Y_2^n|\Omega)+H(W_1,W_2|Y_1^n,Y_2^n,\Omega) \\
&\leq I(W_1,W_2;Y_1^n,Y_2^n|\Omega)+n\epsilon_n \\
&=h(Y_1^n,Y_2^n|\Omega)-h(Y_1^n,Y_2^n|\Omega,W_1,W_2)+n\epsilon_n
\end{align}
Note here that the two receivers are statistically equivalent when they are not jammed with a probability $\lambda_{00}$. In such a scenario, the transmitter can send information to only one receiver as there is no $\CSIT$ available. Using this, we have the following 
\begin{align}
n(R_1+R_2)&\leq h(Y_1^n,Y_2^n|\Omega)-h(Y_1^n,Y_2^n|\Omega,W_1,W_2)+n\epsilon_n \\
&\leq n(\lambda_{00}\log{P_T}+\lambda_{01}2\log(P_T)+\lambda_{10}2\log(P_T)+\lambda_{11}2\log(P_T))\\
&\hspace{12pt}-n(\lambda_{01}\log(P_T)+\lambda_{10}\log(P_T)+\lambda_{11}\log(P_T)) \\
n(R_1+R_2)&=n\left(\lambda_{00}\log(P_T)+\lambda_{01}\log(P_T)+\lambda_{10}\log(P_T)\right).
\end{align}
Normalizing by $n\log(P_T)$ and then $n\rightarrow \infty$ and $P_T\rightarrow \infty$ we obtain the bound
\begin{align}
d_1+d_2&\leq (\lambda_{00}+\lambda_{01}+\lambda_{10}).
\end{align}
This completes the converse proof for Theorem~\ref{TheoremNN}.

\end{document}